\documentclass[CIT,biber]{nowfnt} 

\usepackage[utf8]{inputenc} 

\usepackage{graphicx}
\usepackage{tabularx}
\usepackage{psfrag}
\usepackage{subfigure}
\usepackage{overpic}
\usepackage{hyperref}
\usepackage{amssymb}
\usepackage{pgf}
\usepackage{pgffor}
\usepackage{tikz}
\usepackage{mathtools}
\usepackage{pst-sigsys}
\usepackage{stfloats}
\usetikzlibrary{patterns}
\usepackage{epsfig}
\usepackage{svg}
\usepackage{amsthm}  
\usepackage{amsfonts}
\usepackage{pgf}
\usepackage{pgffor}
\usepackage{tikz}
\usepackage{mathtools}
\usepackage{pst-sigsys}
\usepackage{stfloats}
\usepackage{graphicx}
\usetikzlibrary{patterns}
\usepackage{subfigure}
\usepackage{tkz-graph}
\usepackage{fp}
\usepackage{textcomp}
\usetikzlibrary{patterns,positioning,shapes,arrows,arrows.meta,graphs,svg.path,calc,fit}
\usepackage{pgfplots}
\usepackage{epsfig}
\usepackage{epstopdf}
\usepackage{booktabs}

\newcommand{\specstat}{\ensuremath{\Psi}}
\newcommand{\comment}[1]{}
\newcommand{\E}{\mathbb{E}} 
\newcommand{\R}{\mathbb{R}} 

\newcommand{\tr}[1]{\text{trace}\left(#1\right)}

\newcommand{\norm}[1]{\left|\left| #1 \right|\right|}

\newcommand{\cA}{{\cal A}}

\newcommand{\cB}{{\cal B}}

\newcommand{\cX}{{\cal X}}

\newcommand{\tX}{\tilde{X}}

\newcommand{\hX}{\hat{X}}
\newcommand{\hx}{\hat{x}}

\newcommand{\beq}[1]{\begin{equation}\label{#1}}
\newcommand{\eeq}{\end{equation}}

\newcommand{\beqn}[1]{\begin{eqnarray}\label{#1}}
\newcommand{\eeqn}{\end{eqnarray}}

\newcommand{\bE}{{\bf E}}
\newcommand{\bff}{{\bf f}}

\newcommand{\bW}{{\bf W}}

\newcommand{\bx}{{\bf x}}
\newcommand{\by}{{\bf y}}
\newcommand{\bz}{{\bf z}}

\newcommand{\bS}{{\bf S}}

\newcommand{\bX}{{\bf X}}

\newcommand{\bZ}{{\bf Z}}

\newcommand{\Ddef}{\stackrel{\Delta}{=}}
\newtheorem{prop}{Proposition}
\newtheorem{cor}{Corollary}

\title{Asymptotic Frame Theory for Analog Coding\textsuperscript{\textregistered}\\
(final manuscript October 2021)
}

\maintitleauthorlist{
Marina Haikin \\
Amazon, Tel Aviv, Israel \\
mkokotov@gmail.com
\and
Matan Gavish \\
The Hebrew University of Jerusalem, Israel \\
gavish@cs.huji.ac.il
\and
Dustin G. Mixon \\
The Ohio State University, Colombus, OH, USA \\
mixon.23@osu.edu 
\and
Ram Zamir \\
Tel Aviv University, Israel \\
zamir@eng.tau.ac.il
}

\issuesetup
{%
 copyrightowner={A.~Heezemans and M.~Casey},
 volume        = xx,
 issue         = xx,
 pubyear       = 2021,
 isbn          = xxx-x-xxxxx-xxx-x,
 eisbn         = xxx-x-xxxxx-xxx-x,
 doi           = 10.1561/XXXXXXXXX,
 firstpage     = 1, 
 lastpage      = 18
 }

\usepackage{mwe}

\author[1]{Haikin, Marina}
\author[2]{Gavish, Matan}
\author[3]{Mixon, Dustin G.}
\author[4]{Zamir, Ram}
\affil[1]{Now at Amazon. Previously at Tel Aviv University where this work was performed.}
\affil[2]{Hebrew University of Jerusalem}
\affil[3]{The Ohio State University}
\affil[4]{Tel Aviv University}

\articledatabox{\nowfntstandardcitation}

\begin{document}

\makeabstracttitle

\begin{abstract}
Over-complete systems of vectors, or in short, frames, play the role of analog codes in
many areas of communication and signal processing. To name a few, spreading sequences
for code-division multiple access (CDMA), over-complete representations for multiple-description (MD) source coding, space-time codes, sensing matrices for compressed sensing (CS), and more recently, codes for unreliable distributed computation. 
In this survey paper we 
observe an information-theoretic random-like behavior of frame subsets.
Such sub-frames arise in setups involving erasures (communication), random user activity (multiple access), or sparsity (signal processing), in addition to channel or quantization noise. 
The goodness of a frame as an analog code is a function of the eigenvalues of a sub-frame, averaged over all sub-frames
(e.g., harmonic mean of the eigenvalues relates to least-square estimation error, 
while geometric mean to the Shannon transform, and condition number to the restricted isometry property).

Within the highly symmetric class of Equiangular Tight Frames (ETF), 
as well as other “near ETF” families,
we show a universal behavior 
of the empirical eigenvalue distribution (ESD) of a randomly-selected sub-frame:
$(i)$ the ESD is asymptotically indistinguishable from Wachter’s MANOVA distribution; 
and $(ii)$ it exhibits a convergence rate to this limit that is 
indistinguishable from that of a matrix sequence drawn from MANOVA (Jacobi) ensembles of corresponding dimensions.
Some of these results follow
from careful statistical analysis of empirical evidence, and some are proved analytically using random matrix theory arguments of independent interest. The goodness measures of the MANOVA limit distribution are better, in a concrete formal sense, than those of the Marchenko–Pastur distribution at the same aspect ratio, implying that deterministic analog codes are better than random (i.i.d.) analog codes. We further give evidence that the ETF (and near ETF) family is in fact superior to any other frame family in terms of its typical sub-frame goodness.
%
\end{abstract}


\chapter{Introduction}
\label{section-Introduction}
%
%
%
A frame is an ``over-complete basis'', i.e., a system of vectors that spans the space with
more vectors than the space dimension (real or complex). 
Let us denote the space dimension by $m$, and the number of vectors by $n$, where $n > m$.
The $m \times n$ frame matrix 
\begin{equation}
    \label{frame}
    F = [\bff_1 \cdots  \bff_n]
\end{equation}
is generated by stacking the frame vectors $\bff_1, \ldots, \bff_n$ as columns,
where we restrict attention to unit-norm vectors $\|\bff_i\|=1$.
The relative position of frame vectors is determined by their pairwise cross-correlation matrix
%
\begin{equation}
    \label{intro-gram-matrix}
    F^\dagger F = 
    \Bigl\{ \langle \bff_i, \bff_j \rangle \Bigr\} ,
\end{equation}
%
called also Gram or covariance matrix, which is invariant under unitary operation on the frame vectors (e.g., rotation).

This survey proposes an information-theoretic view on the design and analysis of frames with favorable performance.
We think of a frame as an ``analog code'', which can add redundancy
\cite{wolf1983redundancy,goyal2001quantized,strohmer2003grassmannian,love2003grassmannian},
remove redundancy 
\cite{calderbank2010construction},
or multiplex information directly in the signal space 
\cite{rupf1994optimum,massey1993welch,xia2005achieving}.
Multiplication by the frame matrix can expand the dimension $m:n$ hence add redundancy, 
or reduce the dimension $n:m$ hence compress
(with $\by = F^\dagger \bx$ for the former and $\bx = F \by$ for the latter).
The aspect ratio $n/m$ is often called the ``frame redundancy''.

Although information theory tells us that reliable data transmission (adding redundancy) and compression (removing redundancy) 
can be achieved by {\em digital} codes,
real-world physical-layer communication systems combine {\em analog} modulation techniques
that can be described in terms of frames\footnote{
For example, coded-modulation can be thought of as concatenation of an outer digital code with an inner analog code.
} 
\cite{marshall1984coding,calderbank2010construction,bolcskei1998frame,zaidel2018sparse,goyal1998multiple,boufounos2008causal}.
%
%
We are specifically interested in some old and new applications of frames that
involve a combination of {\em random activity} and {\em noise}.
Performance in these applications is a function of the eigenvalues of (the Gram of) 
a randomly selected $k$-subset $F_S$ of the $n$ frame vectors, 
\begin{equation}
    \label{frame_subset}
    F_S = [\bff_{i_1} \cdots \bff_{i_k}] 
\end{equation}
where $S = \{i_1,\ldots,i_k\} \subset [n]$, $|S|=k$.
(We shall use $[n]$ to denote $\{1,\ldots,n\}$.) 

For example,
in {\em non-orthogonal code-division multiple access} (NOMA-CDMA),
\cite{verdu-shamai1999spectral,tulino2004random,stoica2019massively},
$n$ users are allocated with $m$-length spreading sequences using the frame $F$,
but only $k$ out of the $n$ users are active at any given moment,
and performance is measured by the {\em Shannon capacity} of the
vector Gaussian channel associated with the $m \times k$ sub-matrix $F_S$
(averaged over the subset $S$ of active users).
In transform-based {\em multiple-description} (MD) source coding,
\cite{goyal1998multiple,ostergaard2009multiple},
an $m$-dimensional vector source is expanded into $n$ packets using the frame $F$, only $k$ packets are received, and performance is measured by the {\em remote rate-distortion function} associated with $F_S$
(averaged over the subset $S$ of received packets).
In coded distributed computation (CDC),
\cite{lee2018speeding,li2020coded,fahim2019numerically},
the user (master) node expands $m$ sub-computation tasks into $n$ redundant tasks using the frame $F$, and sends them to $n$ {\em noisy} computation nodes 
(where the noise is due to finite precision computation); 
only $k$ nodes return their answers on time ($n-k$ are stragglers),
and 
the sub-matrix $F_S$ determines
the final average precision (or noise amplification) after the user node decodes 
the desired computation value.    
Other ``$(n,m,k)$ setups'' are listed in Table~\ref{table-nmk-applications}.


%
	\begin{table}[htbp]
		\scriptsize
		\centering
		\caption{$(n,m,k)$ setups featuring analog frame codes} 
		\label{table-nmk-applications}
		\renewcommand{\arraystretch}{0.80}%
		\begin{tabularx}{\textwidth}{XXXXlX}
			~ \\\hline  ~\\
			Application & $n$ & $m$ & $k$ & $m \gtrless k $  & References \\
			~ \\\hline  ~\\
			\midrule
			Source \newline with erasures & block-length & bandwidth  & important samples & $m>k$ &  
			\cite{xia2005achieving,haikin2016analog}
			\\
			\midrule
			NOMA-CDMA & users & resources (spread) & active users & $m>k$   &  
			\cite{strohmer2003grassmannian,rupf1994optimum,zaidel2018sparse}
			\\
			\midrule
			Impulsive channel & block-length & bandwidth & non-erased & $m<k$   &  
			\cite{wolf1983redundancy,tulino2007gaussian} 
			\\
			\midrule
			Space-time coding & space \ \ \ \ \ \ \ \ \ \ \  (diversity) 
			& time  & non-erased antennas & $m<k$  &  
			\cite{tarokh1998space,tse2005fundamentals}, Section~\ref{section-Apps-STC} 
			\\
			\midrule
			$\Delta \Sigma$ \newline modulation & over-sampled & original & non-erased & $m<k$   & 
			\cite{candy1992oversampling,goyal1998quantized} 
			\\
			\midrule
			Multiple descriptions & transmitted & original & received & $m<k$  & \cite{goyal1998multiple,ostergaard2009multiple}
			\\
			\midrule
			Wavelets & coefficients & source & significant \newline coefficients & $m>k$  & 
			\cite{kovacevic2008introduction}
			\\
			\midrule
			Compressed sensing & input & output & sparsity & $m>k$ &  
			\cite{donoho2006stable,candes2006near}
			\\
			\midrule
			Coded \newline computation & workers & computations & non-stragglers & $m<k$  & 
			\cite{lee2018speeding,li2020coded,fahim2019numerically}
			\\
			\midrule
			Neural \newline networks & input & output & features & $m>k$ &  
			\cite{bank2020ETF}
			~ \\\hline  ~\\

		\end{tabularx}
	\end{table}

In an ideal {\em noiseless} setup one could choose the frame redundancy $n/m$ equal to the
reciprocal of the activity ratio $k/n$, 
i.e., $m = k$ = the effective number of users/packets/nodes in the examples above. 
This is similar to digital erasure correction using {\em maximum  distance separable} (MDS) codes for a channel with $k$ out of $n$ non-erased symbols 
\cite{blahut1985algebraic}.
However, when {\em noise} is involved  
(channel/quantization/computation noise in the setups above), 
a better trade-off between noise immunity and information rate
is obtained by choosing a lower/higher frame redundancy;
$k < m < n$ in NOMA-CDMA, or $m < k < n$ in MD and CDC.
Thus, $m$ is a design parameter that we can optimize.

Frame design could be viewed as an attempt to find vectors 
$\bff_1, \ldots, \bff_n$ in $\mathbb{R}^m$ or $\mathbb{C}^m$, $n>m$, 
that are somehow 
``as orthogonal to each other as possible'', 
either
in pairs ($k=2$) or in larger $k$-subsets
\cite{waldron2018introduction,kovacevic2008introduction}.
As we shall see in Chapter~\ref{section-Intro-performance-measures},
the sub-frame performance criteria 
mentioned above (capacity, rate-distortion function, noise amplification),
denoted in general as $\specstat(F_S)$,
depend on the spread of the eigenvalues of the Hessian $F_S F_S^\dagger$ or Gram $F_S^\dagger F_S$ matrices\footnote{The nonzero eigenvalues of both matrices are the same.} of the sub-frame $F_S$.
More mutual orthogonality amounts to a more compact eigenvalue spectrum, 
and ideal performance occurs when the spectrum shrinks to a delta function,
or equivalently, the sub-frame is orthogonal.
The redundant nature of the frame, however, implies that most of its 
subsets are {\em not} orthogonal. 
Our target is therefore to find a frame whose {\em average} performance over all $k$-subsets
\begin{equation}
\label{subset_average_performance_measure}
\bar{\specstat}(F,k)  
=
 \frac{1}{{n \choose k}} \sum_{\substack{S \subset [n] \\ |S|=k} }
\specstat(F_S)    
\end{equation}
is ``good'';
or in other words,
%
a frame whose {\em typical} subset has a {\em compact} eigenvalue spectrum.\footnote{
Simple (though only partial) measures for
spectrum compactness are the variance and kurtosis; see Section~\ref{section-ETF-super--ESV-ESK}.} 

%
%

We borrow from information theory the probabilistic view of a communication channel, 
and the notions of typicality and typical-case (rather than worst-case) goodness
\cite{CoverBook}.
The information-theoretic viewpoint leads us to look for 
frames with a ``typically compact'' subset spectrum
for a given $(n,m,k)$ triplet,
and for frame {\em families} with the
best attainable {\em asymptotic} goodness in the limit as $n$ goes to infinity
for fixed (asymptotic) redundancy ratios $n/m$ and $k/m$. 
This fresh look on frames turns out to be fruitful, and 
opens many interesting questions 
at the intersection of signal processing, random matrix theory, geometry, harmonic analysis and information theory.

%


Sampling theory suggests the {\em low-pass frame} (LPF), the frame analog of band-limited interpolation,  as a practical candidate for signal expansion.
This turns out to be a far-from-optimal choice, as we shall see, due to large noise amplification for a typical subset (which corresponds to noisy reconstruction from a non-uniform sampling pattern
\cite{seidner2000noise,mashiach2013noise,krieger2014multi,venkataramani2000perfect}).
%

Information theory suggests {\em random} (i.i.d.) frames as natural candidates for good analog codes.
To study the spectrum of these objects,
%
{\em Random Matrix Theory} (RMT) 
offers a helpful matrix version of the law of large numbers:
the eigenvalue distribution of a typical random matrix tends to {\em concentrate} towards a fixed distribution in the limit of large dimensions \cite{ZeitouniBook}, \cite{feier2012methods}.
Indeed, if we choose the elements of the frame matrix $F$ as i.i.d. Gaussian variables,
then the subset Gram matrix $F_S^\dagger F_S$ is drawn from a Wishart ensemble,
and its spectrum converges almost surely in distribution to the Marchenko--Pastur (MP) distribution with parameter $\beta = k/m$ \cite{marcenko1967distribution}.
We can thus compute the capacity / rate-distortion function / noise amplification 
(\ref{subset_average_performance_measure})
associated with the Marchenko--Pastur distribution, and obtain some achievable asymptotic performance $\specstat({\rm MP},\beta)$
for the problems described above.

Is the Marchenko--Pastur distribution - corresponding to random i.i.d. frames - the ``most compact'' 
subset spectrum we can hope for?
One of the key results of this survey is that better {\em deterministic} frames do exist.
In fact, a certain class of highly symmetric 
frames 
obeys asymptotic concentration of the spectrum of a randomly-selected subset to a universal limiting distribution -- similarly to the case of a completely random (i.i.d.) matrix.
Crucially, 
this limiting distribution
is {\em more compact} than the MP distribution.


%

%

{\em Equiangular tight frames} (ETF)
are in a sense the most geometrically symmetric family of frames
\cite{waldron2018introduction,casazza2012finite,fickus2011constructing}.
They have numerous applications in communications and signal analysis,  \cite{love2003grassmannian},
and their study brings together geometry, combinatorics, probability, and harmonic analysis. 
Interestingly, as we shall see in Chapter~\ref{section-frame-theory},
one construction of ETFs corresponds to signal expansion with an {\em irregular} Fourier transform, \cite{xia2005achieving}, as opposed to the low-pass frame (LPF)
mentioned above. 

%
A series of recent papers 
\cite{haikin2016analog,haikin2017random,haikin2018frame,MarinaThesis,magsino2020kesten,haikin2021moments},
demonstrated that ETFs, as well as other deterministic tight frames that we term ``near ETFs'', exhibit an RMT-like behavior familiar from Multivariate ANalysis Of VAriance (MANOVA) \cite{muirhead2009aspects,wachter1980limiting,erdHos2013local}.
We shall call this phenomenon the {\em ETF-MANOVA relation}.

\begin{figure*}[htbp]  
		\centering
		\includegraphics[width=1.5in]{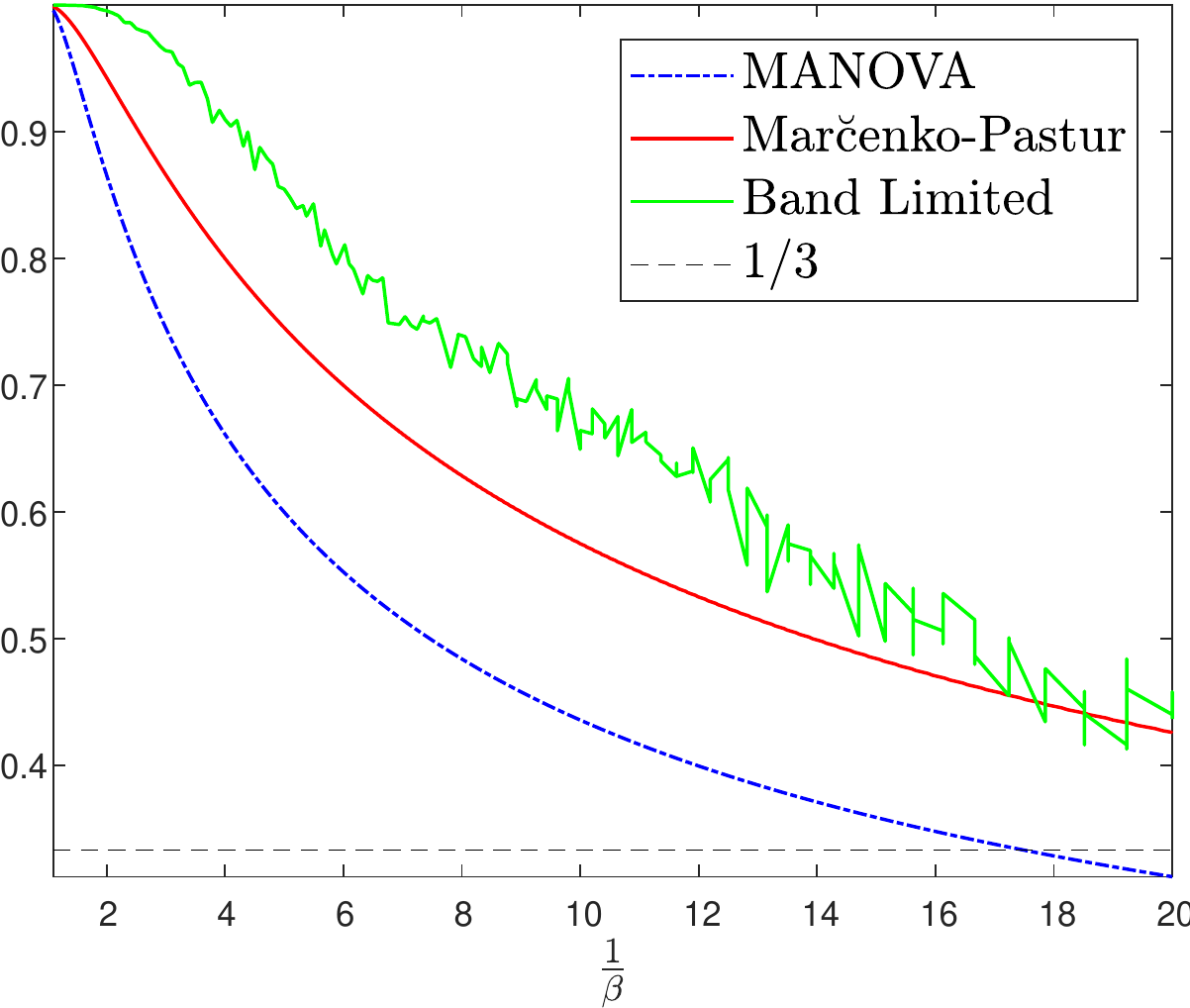}
		\includegraphics[width=1.54in]{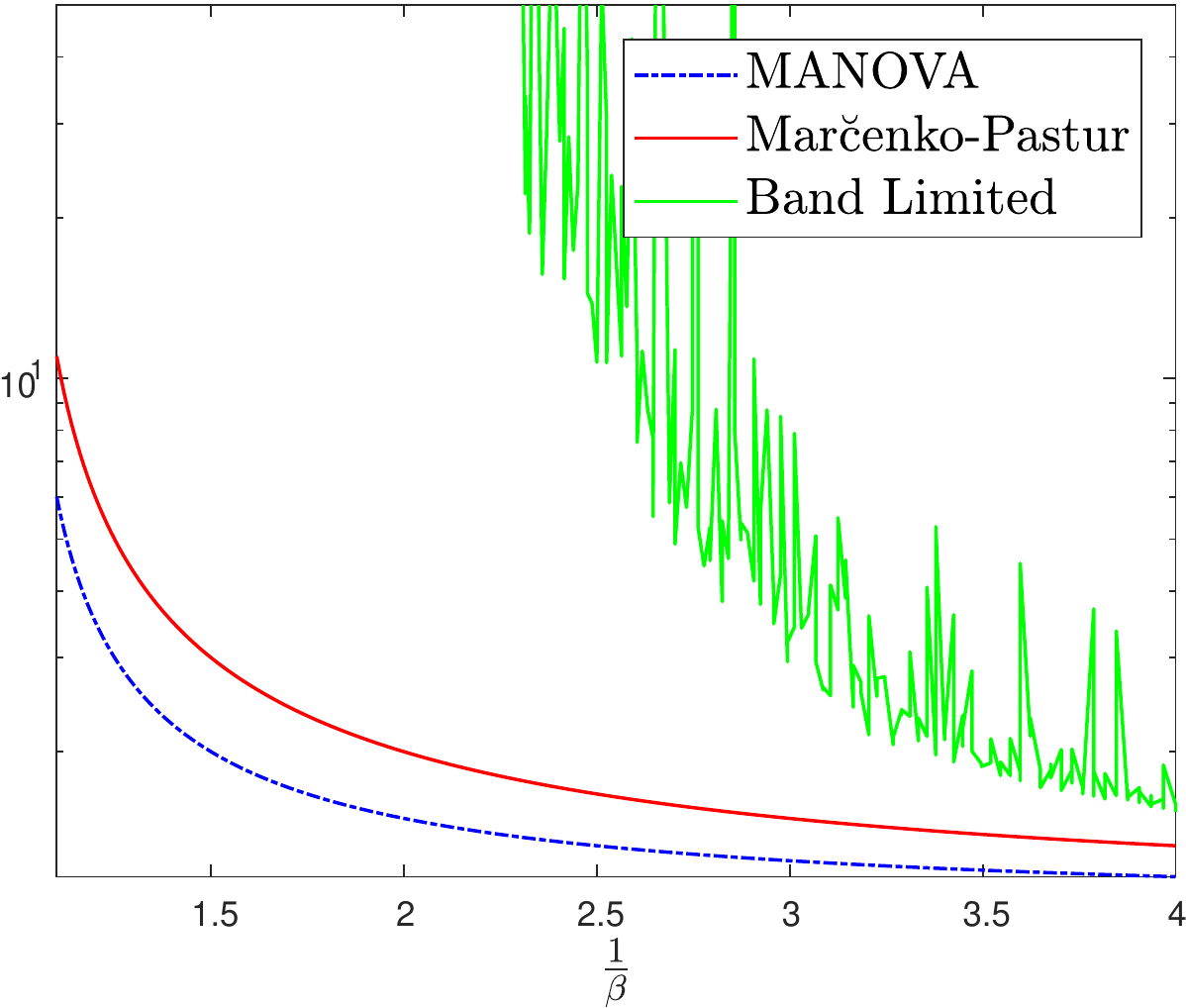}
		\includegraphics[width=1.5in]{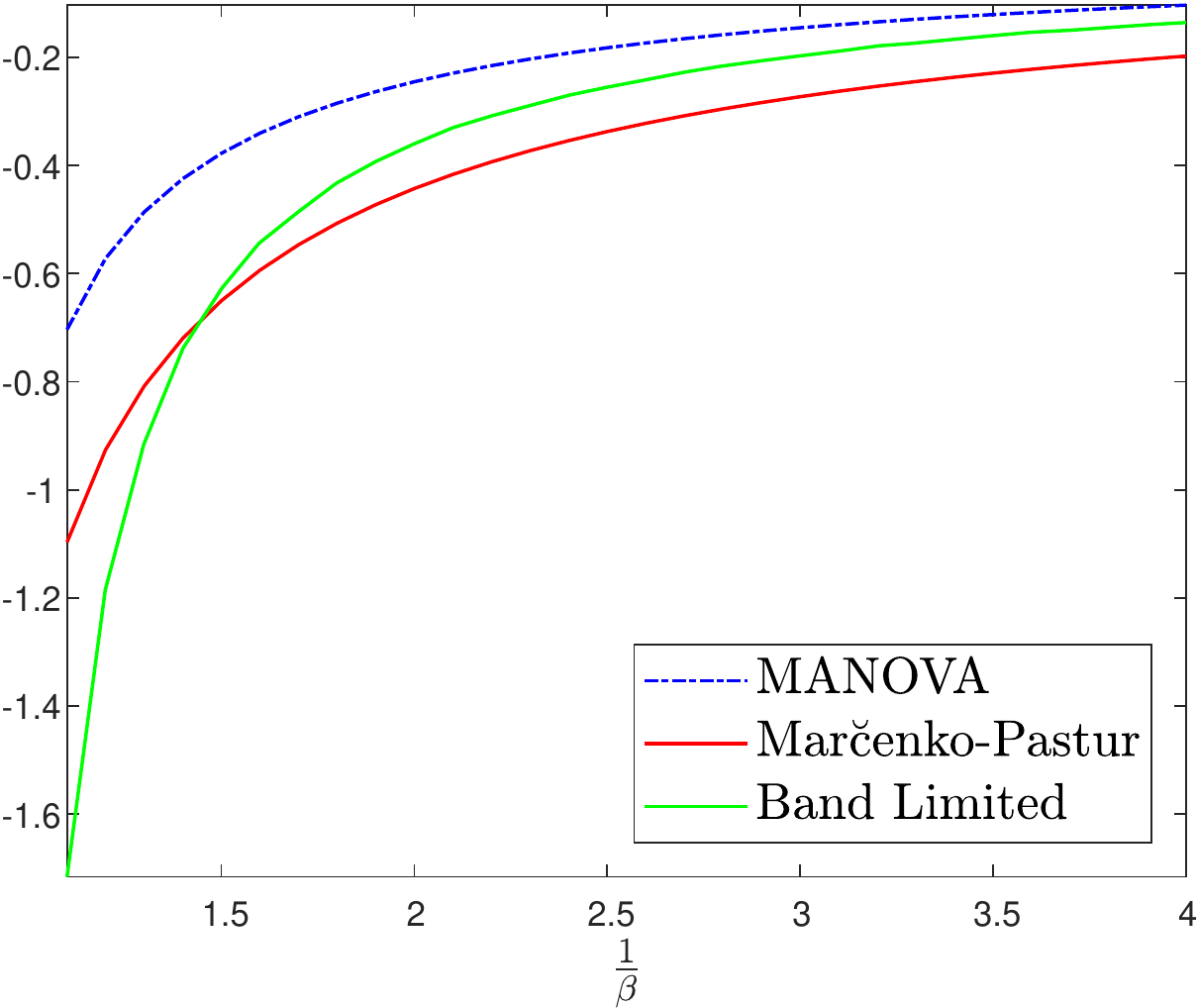}
		\caption{Comparison of the asymptotic performance measures
		$\specstat({\rm MANOVA},\beta,\gamma)$, 
		$\specstat(\mbox{Marchenko-Pastur},\beta)$ and 
		$\specstat(\mbox{Band Limited},\beta,\gamma)$, 
		as a function of $\beta$, for $\gamma = 1/2$
		(corresponding to sub-frame aspect ratio $\beta$ from a redundancy-2 frame, in the families of ETF, random (i.i.d.), and LPF, respectively). 
		Each figure corresponds to a different sub-frame performance measure $\specstat$:
		(Left) restricted isometry property (RIP)
		associated with compressed sensing (lower is better);
		(Middle) noise amplification associated with MD and CDC (lower is better); (Right) Shannon transform associated with the capacity of NOMA-CDMA (higher is better).
		The $1/3$ line in the left figure is the sharp RIP bound for sparse signal and low rank matrix recovery \cite{CAI201374}.
		}
		\label{fig-goodness-comparison-Manova-MP-LPF}
		\end{figure*}

Specifically, the work in \cite{haikin2017random} showed empirically 
that for any frame within this class of frames,
the eigenvalue distribution of a randomly-selected subset
appears to be indistinguishable from that of a random matrix taken from the MANOVA (Jacobi) ensemble. 
The work in
\cite{haikin2018frame,MarinaThesis,magsino2020kesten,haikin2021moments}
further partially proved analytically\footnote{
\label{footnote-partial-proof}
The proof is complete for the case $\gamma = 1/2$ \cite{magsino2020kesten}.
For a general $0<\gamma<1$, 
\cite{haikin2021moments} establishes 
a recursive formula in $r$ for the asymptotic mean $r$th moment of a randomly-selected ETF subset, for $r=1,2,\ldots$.
Using a symbolic computer program, we were able to verify that this formula coincides with the first 10 MANOVA moments  
(above which the complexity explodes).
A proof of the identity for a general $r \in \mathbb{N}$  
remains a fascinating open problem. 
} 
that as $n \rightarrow \infty$, 
for aspect ratios $m/n \rightarrow \gamma$ and $k/m \rightarrow \beta$,
this eigenvalue distribution converges in distribution almost surely to Wachter's limiting MANOVA distribution parameterized by $\gamma$ and $\beta$ \cite{wachter1980limiting}. 
The concluded asymptotic performance  (\ref{subset_average_performance_measure}) of the ETF family,
\begin{equation}
\label{intro-etf=manova}
  \bar{\specstat}({\rm ETF},\gamma,\beta) = \specstat({\rm MANOVA},\gamma,\beta)  ,
\end{equation}
%
is strictly better than $\specstat({\rm MP},\beta)$, the asymptotic performance of random (i.i.d.) frames, for various performance measures $\specstat$.
Figure~\ref{fig-goodness-comparison-Manova-MP-LPF} shows that the gain of MANOVA over MP is $\sim 2$ dB in noise amplification, and $\sim 0.35$ bit in capacity.
We conjecture that in terms of these performance measures, 
the MANOVA distribution is, in fact, the 
{\em most compact} typical sub-frame spectrum achievable by {\em any} unit-norm frame.

In this survey paper we propose a common framework for these topics, located at the intersection of information theory and neighboring fields.
Chapter~\ref{section-Intro-motivations-IT} addresses primarily the information theory audience,
and motivates a passage from digital codes to low-pass interpolation and analog frame codes,
through a side-information source coding problem.
Chapter~\ref{section-Intro-performance-measures} formalizes the notion of a performance measure that depends on the eigenvalue spectrum of a sub-frame; e.g.,
noise amplification 
amounts to the harmonic mean of the spectrum, while capacity (Shannon transform) amounts to the geometric mean of the spectrum.
Chapter~\ref{section-frame-theory}
gives background from frame theory
(in particular, earlier results on the spectral properties of sub-frames  motivated by compressed sensing), 
while Chapter~\ref{section-RMT} gives the relevant background on random matrix theory.

The two highlights of this survey are $(i)$ the {\em ETF-MANOVA relation}, connecting frame theory with random matrix theory, 
and $(ii)$ the (still mostly open) possibility of {\em ETF superiority}.
The first highlight is divided between
two sections:
Chapter~\ref{section-RMT-meets-frames}
describes the 
empirical results of \cite{haikin2017random}
regarding the universal behavior of sub-frames of ETFs and ``near ETFs''; 
and Chapter~\ref{section-ETF-MANOVA-moments} develops analytically the convergence to the MANOVA limit distribution based on the moment method (see footnote~\ref{footnote-partial-proof} above) 
\cite{haikin2018frame,MarinaThesis,magsino2020kesten}. 
To support the ETF superiority claim,  we examine numerically in Chapter~\ref{section-Apps} some of the applications listed in Table~\ref{table-nmk-applications};
and we prove analytically in Chapter~\ref{section-ETF-super} the {\em erasure Welch bound}, \cite{haikin2018frame},
which implies that tight frames have the smallest sub-frame spectral variance among all unit-norm frames, and that ETFs have the smallest sub-frame spectral kurtosis among all 
unit-norm tight frames.
In between these two highlights,
Chapter~\ref{section-inequalities} 
proves some sub-frame performance inequalities (in the flavor of the information-theoretic inequalities of \cite{dembo1991information}), 
which explain the role of the sub-frame aspect ratio $k/m$ as a design parameter. 
%
Finally, 
Chapter~\ref{section-conclusion} concludes and lists interesting open questions and conjectures that arise in this area.


\section{Notation}
A finite sequence of integers $\{1,\ldots,n\}$ is denoted as $[n]$.
Bold letters $\bff, \bx$ etc. denote column vectors. 
The $n \times n$ identity matrix is denoted by $I_n$.
Dagger $(\cdot)^\dagger$ denotes transpose or conjugate (Hermitian) transpose,  
according to the context. 
The set of all $k$-subsets of the set $[n]$ is denoted $\binom{[n]}{k}$,
or $\{ S \subset [n]: \; |S|=k \}$,
or simply $\{ S: \; |S|=k \}$ when the context is clear.
%
$\mathbb{E}$ denotes expectation.
Throughout we try to keep the following glossary:
%

\vspace{10mm}

\begin{tabular}{|l|l|}
     \hline 
    $m$ & vector space dimension \\
    $n$ & frame size \\
    $k$ & sub-frame size \\
    $p$ & selection probability \\
    $\gamma$ & frame aspect ratio $m/n$ \\
    $\beta$ & sub-frame aspect ratio $k/m$ \\    
    $F$ & $m \times n$ frame matrix \\
    $S$ & $k$-subset of $[n]$ \\
    $F_S$ & $m \times k$ sub-frame matrix \\
    $\specstat$ & performance measure \\
        \hline
\end{tabular}


\chapter{An information-theoretic toy example} 
\label{section-Intro-motivations-IT}
%
%
To set the stage for the information theory audience, 
we begin with an ``$(n,m,k)$'' source coding problem 
that originally motivated this study
\cite{martinian2008source,haikin2016analog}.


\section{Source coding with erasures}
\label{section-distortion-side-information}
\begin{figure*}[htbp]
	\centering
	\hspace{-1.5em}
	\def\svgscale{3}
	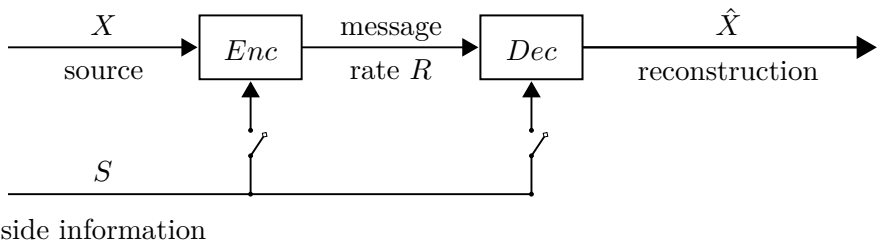
	\caption{Source coding with distortion side information. Our problem of interest is when only the first switch in connected.
}
	\label{fig-source-with-erasures}
\end{figure*}

Consider the setup shown in Figure~\ref{fig-source-with-erasures}, 
of a source with ``distortion side information'' at the encoder \cite{martinian2008source}.
Specifically, 
the encoder of a (white) Gaussian source $X \sim N(0,\sigma_x^2)$ has access to a statistically independent 
Bernoulli$(p)$ variable $S$ that reveals whether the source
sample $X$ is ``important'' (to be reconstructed under a squared-error distortion) or not.
The distortion measure is thus given by
\begin{equation}
\label{distortion-side-information}
d(x,\hx,s) = \left\{ 
\begin{array}{lr}
(\hx -x)^2, & \mbox{if $s=1$ \ (``important'')}
\\
0, & \mbox{if $s=0$ \ (``don't care''),}
\end{array}
\right.
\end{equation}
and the mean-squared error (MSE), i.e., the average distortion per important sample in encoding the block $X_1,\ldots, X_n$, is given by\footnote{For simplicity, the normalization is by the {\em expected} number of important samples: $np = \mathbb{E}\{ \sum_{t=1}^{n} 1_{\{S_t=1\}} \}$.}
\begin{equation}
\label{distortion}
D = \frac{1}{n p} \sum_{t=1}^{n} 1_{\{S_t=1\}} \cdot \mathbb{E}\{(\hX_t-X_t)^2\}  .  
\end{equation}
%
\\
If also the decoder had access to the side information,
then the best achievable rate-distortion trade off is given by the conditional 
rate-distortion function of $X$ given $S$ 
\cite{CoverBook,berger1971rate}:
\begin{equation}
\label{RDF}
R_{X|S}(D) 
= \min_{\hX} I(X;\hX|S)
= \frac{p}{2} \cdot \log_2 \left( \frac{\sigma_x^2}{D}\right) 
\end{equation}
bits per source sample,
where a possible minimizing $p(\hx|x,s)$ in (\ref{RDF}) is the optimal quadratic-Gaussian test-channel $p^*(\hx|x)$ for $s=1$, 
and a degenerate channel (say, $\hX \equiv 0$) for $s=0$.   
This simply amounts to ignoring the non-important samples, and communicating the $k \approx n p$ out of $n$ important samples at a rate of $(1/2) \log_2( \sigma_x^2/D )$ bits per sample.
Interestingly, 
like in some other favorite side-information problems in information theory, \cite{CoverBook,cover-chiang2002duality}, 
there is no loss for having access to the side information at just one end of the communication channel:
the same rate-distortion trade off (\ref{RDF}) can be achieved even if $S$ is available only at the encoder.\footnote{
This setup can be seen as a counterpart of the Slepian--Wolf problem,
\cite{slepian-wolf1973,CoverBook}, of ``statistical side information'' at the decoder.
}
A formal proof is obtained by computing 
the rate-distortion function of the equivalent source pair $(X,S)$, under the distortion measure (\ref{distortion-side-information}):
\begin{equation}
\label{RDF2}
R_{X,S}(D) = \min_{\hX} I(X,S;\hX) 
= \frac{p}{2} \cdot \log_2 \left( \frac{\sigma_x^2}{D}\right) ,
\end{equation}
where the minimizing $p(\hx|x,s)$ now has the optimum marginal $p^*(\hx)$ for {\em both} values of $s$; 
i.e., $p(\hx|x,s=1) = p^*(\hx|x)$ and $p(\hx|x,s=0) = p^*(\hx)$.
It follows from the chain rule that the minimizing $\hX$ in (\ref{RDF2}) has {\em zero} mutual information with the side information $S$,
i.e., $I(S;\hX) = 0$,
which is a necessary feature of a zero-loss solution \cite{martinian2008source}.
%


\subsection{Complexity of the all-digital solution}
\label{section-complexity}
The information-theoretic promise 
comes however at a high price: the complexity of $n$-dimensional vector quantization.
Specifically, to achieve the rate-distortion function (\ref{RDF2}) 
one needs to draw $2^{n R}$ codewords of size $n$ at random, and to encode the source vector $X_1,\ldots, X_n$ into a  jointly typical codeword: close to the important samples and arbitrary at the non-important samples \cite{CoverBook}.
Clearly, the encoding complexity is exponential in $n$.
A simple naive solution would be first to encode the side information, i.e., describe the location of the important samples at a rate of $H_b(p) = p \log (1/p) + (1-p) \log (1/(1-p))$, e.g., 1 bit per sample for $p=1/2$;
and then
to quantize only the important samples using a 
simple {\em scalar} quantizer 
(with just a small loss compared to optimal vector quantization)\footnote{Entropy-coded 
scalar quantization exceeds the rate-distortion function by only $\approx 1/4$ bit per sample at high resolution conditions 
\cite{gersho2012vector,zamir1992universal}.
}.
If however we do not want to pay the rate penalty of $\sim 1$ bit per sample for encoding the side information, 
then we cannot replace the complex $n$-dimensional codebook by simple scalar quantization of the important samples
because their location is a priori unknown. 
%
Can we simplify the solution without sacrificing too much in rate-distortion performance?


%
\begin{figure*}[htbp]
	\centering
	\hspace{-1.5em}
	\def\svgscale{2.1}
	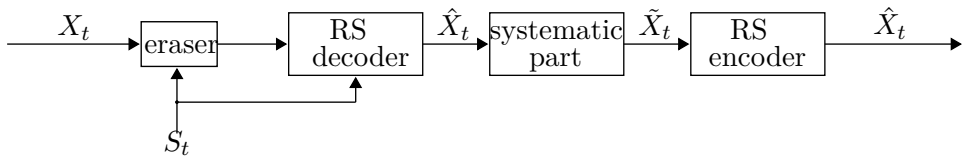
	\caption{Reversed Reed--Solomon coding for a discrete source with erasures known at the encoder.
	Source = $X_1,\ldots,X_n$, reconstruction = $\hX_1,\ldots,\hX_n$, compressed version = $\tX_1,\ldots,\tX_k$.
}
	\label{fig-Reverse-RS}
\end{figure*}
%


\subsection{A simple solution in the discrete case}
Martinian et al. suggested an intriguing simple yet optimal solution based on a {\em reversed Reed-Solomon (RS) code} for a {\em lossless} variant of this problem
\cite{martinian2008source}.
Here, the source $X$ is uniform over a finite alphabet $\cX$, the important samples are encoded under a Hamming distortion measure:
\begin{equation}
\label{distortion-side-information-Hamming}
d(x,\hx,s) = \left\{ 
\begin{array}{lr}
1_{\hx \not= x}, & \mbox{if $s=1$ \ (``important'')}
\\
0, & \mbox{if $s=0$ \ (``don't care''),}
\end{array}
\right.
\end{equation}
and we target a small ($D \rightarrow 0$) average distortion.
The idea of \cite{martinian2008source} 
was to treat the non-important samples as ``erasures'', and use an erasure-correction (e.g., RS) code in {\em reverse}:
the encoder applies RS decoding to ``correct'' 
the erasures to the codeword $\hat{\bx}$ (that matches the source at the non-erased (important) samples $\{ t: s_t=1\}$),
and then it transmits the systematic part of $\hat{\bx}$; 
the decoder uses RS encoding to reconstruct $\hat{\bx}$ from the systematic part.
See Figure~\ref{fig-Reverse-RS}.
Clearly, the decoder has no idea which samples are important and which are not, 
so $I(\bS,\hat{\bX})=0$ like in (\ref{RDF2}).
Moreover, assuming that the number of important samples 
\begin{equation}
\label{k=pn}
k = p n   
\end{equation}
is (roughly) constant,
the maximum-distance separable (MDS) property of RS codes implies that perfect erasure correction is possible using an $(n,k)$ code.\footnote{Implicit in this construction is that the size of the alphabet $\cX$ is a prime power, and meets the Galois field requirements of the RS code.}
Thus, the size of the systematic part is $k$, and the coding rate is $R = (k/n) \log_2 |\cX|$ bits per sample,
as if both the encoder and decoder knew which samples are important and which not.


\section{Band-pass interpolation}
\label{section-band-limited-interpolation}
\begin{figure*}[htbp]
	\centering
	\hspace{-1.5em}
	\def\svgscale{3}
	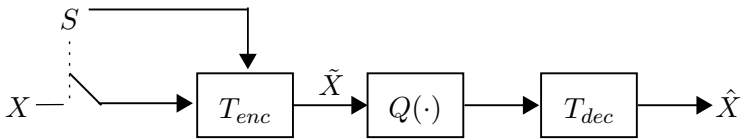
	\caption{Band-limited interpolation coding scheme. 
	Depicted are the source $X_1,\ldots,X_n$; 
	important samples $\{ X_t:  s_t=1 \}$;  
	encoder $m \times k$ transformation 
	$T_{enc} = F_S (F_S^\dagger F_S)^{-1}$;
	quantizer input $\tX_1,\ldots,\tX_m$; 
	quantizer output $Q(\tX_1),\ldots,Q(\tX_m)$; 
	decoder $n \times m$ transformation  $T_{dec} = F^\dagger$; 
	and reconstruction $\hX_1,\ldots,\hX_n$.
	%
	This combination of DFT to handle ``erasures'' and quantization to introduce ``noise''
can be viewed as {\em hybrid analog--digital} coding.
}
	\label{fig-analog-source-coding-scheme}
\end{figure*}

It is tempting to mimic the philosophy of the ``reverse'' RS-code solution above in the quadratic-Gaussian problem of Figure~\ref{fig-source-with-erasures}.
We are inspired by the view of a cyclic code (in particular, an RS code) as consisting of all codewords whose Fourier transform (over the corresponding Galois field) is zero in a certain band of frequencies \cite{blahut1985algebraic,blahut1990digital}. 
Hence, erasure correction can be thought of as ``band-limited interpolation'', 
while the optimum ratio $k/n$ corresponds to the ``Nyquist sampling rate'' \cite{yaroslavsky2015can}.
This interpretation suggests a discrete Fourier transform (DFT)-based 
``analog'' coding scheme for real- or complex-valued vector sources, \cite{haikin2016analog},
as shown in Figure~\ref{fig-analog-source-coding-scheme}.
Each codeword is a ``low-pass'' vector, i.e., an $n$-dimensional vector whose high 
DFT coefficients are zero. 
The encoder looks for a low-pass vector that matches the important samples,
and transmits a quantized version (scalar or vector quantization) of its nonzero DFT coefficients.
The decoder then reconstructs the source by inverse DFT. 
%

%
%

\subsection{Coding and decoding}
To define the scheme precisely,
assume for now that as in (\ref{k=pn}) the number of important samples is constant and equal to $k = p n$.
Namely, the importance side-information law is ``combinatorial'' ($n$ choose $k$) rather than Bernoulli($p$). 
Let $m$, the codewords bandwidth, $k \leq m \leq n$, be a design parameter to be optimized later, and  
let $F$ denote the upper $m \times n$ (lowpass) section from the $n \times n$ DFT matrix: 

%
\begin{equation}
\label{LPF-frame}
F =  \frac{1}{\sqrt{m}}
\left[
\begin{array}{ccccc}
1     & w & w^2 & \ldots & w^{n-1} \\
1     & w^2 & w^4 & \ldots & w^{2(n-1)} \\
\vdots & & & & \\
1     & w^m & w^{2m} & \ldots & w^{m(n-1)} 
\end{array}
\right]
\,,
\end{equation}
where $w = e^{- j 2 \pi /n}$ and $j = \sqrt{-1}$,
and where the $1/\sqrt{m}$ pre-factor normalizes each column to a unit norm.  
The vertical axis corresponds to ``frequency'' and the horizontal axis to ``time''.
The matrix $F$ defines an 
``$m$ lowpass'' codebook,
i.e., vectors $\bx' \in \mathbb{C}^n$ whose higher $n-m$ DFT coefficients are zero;
these vectors live in the row space of $F$,
so $\bx' = F^\dagger \tilde{\bx}$ for some ``frequency vector'' $\tilde{\bx}$ in $\mathbb{C}^m$.
%
The subset of the time instances of important samples, 
$S = \{ t: s_t=1\} \subset [n]$, 
defines the $m \times k$ sub-matrix $F_S$, (\ref{frame_subset}),
and a corresponding $k$ sub-vector $\bx_S$ of the source vector $\bx = (x_1,\ldots, x_n)$.   %
%
%
%
%
%
%
The encoder looks for a codeword $\bx'$ that matches the important samples sub-vector $\bx_S$, and computes its corresponding frequency vector $\tilde{\bx}$:
\begin{equation}
\label{xtilde}
\bx_S = (\bx')_S = F_S^\dagger \tilde{\bx}.
\end{equation}
If $m=k$, then the solution $\tilde{\bx} = (F_S^{\dagger})^{-1}  \bx_S$ for (\ref{xtilde}) 
is unique.\footnote{Any band of rows of the DFT matrix is ``full spark'',
i.e., any square sub-matrix of a DFT band is invertible \cite{fickus2015group,stevenhagen1996chebotarev}.}
For $m > k$ the solution is generally not unique, and for reasons to be clear soon,  
a good estimate for $\tilde{\bx}$ is the least-squares (LS) ``pseudo inverse'' solution 
\begin{equation}
\label{LS}
\tilde{\bx}_{\rm LS} = F_S \cdot (F_S^\dagger \cdot F_S)^{-1} \cdot \bx_S ,
\end{equation}
i.e., the $\tilde{\bx}$ satisfying (\ref{xtilde}) with the minimum squared-norm.\footnote{Note that (\ref{LS})
is the unique solution $\tilde{\bx}$ for (\ref{xtilde}) in the column space of $F_S$ (which is a $k$-dimensional subspace of $\mathbb{C}^m$), and any other solution can be written as 
$\tilde{\bx}_{\rm LS} + \bz$ for some $\bz$ in the orthogonal subspace.} 
After estimating $\tilde{\bx}$, the encoder transmits a quantized version $Q(\tilde{\bx})$ to the decoder, 
which reconstructs the source via inverse DFT:\footnote{An equivalent solution for the case where $n/m$ is an integer,
is to {\em uniformly} re-sample $\bx' = F^\dagger \tilde{\bx}_{\rm LS}$, transmit the $m$ quantized uniform samples, and
reconstruct by $m:n$ band-limited interpolation.}
\begin{equation}
    \label{analog-reconstruction}
    \hat{\bx} = \mbox{IDFT}[Q(\tilde{\bx}) | 0,\ldots,0] =  F^\dagger \cdot Q(\tilde{\bx}) .
\end{equation}
%
Indeed, if we neglect the quantization operation $Q(\cdot)$ in (\ref{analog-reconstruction}), then by (\ref{xtilde}) the important samples are perfectly reconstructed, i.e., $\hx_t = x_t$, for all $t$ in $S$.


\subsection{Rate-distortion performance}
The rate-distortion performance of this analog coding scheme depends on the characteristics of the quantization operation $Q(\cdot)$.
An efficient analytical tool is the additive-noise ``test channel'' model of 
entropy-coded (subtractive) dithered quantization (ECDQ) 
\cite{zamir1992universal,zamir2014lattice}.
For a given pattern $S$ of important samples, the ECDQ coding rate in bits per source sample \cite{haikin2016analog} 
is given by 
\begin{align}
    \label{Ranalog1}
R_{\rm analog}(S,D) 
& = \frac{m}{2n} \cdot \log 
\left( 1 + \frac{\frac{1}{m} \mathbb{E} \{ \|\tilde{\bX} \|^2 \} }{D} \right)
\end{align}
%
where 
the expectation $\mathbb{E} \{\cdot\}$ is with respect to randomness of the source $\bX$.
See Appendix~\ref{section-ECDQ}. 
As explained earlier, LS estimation (\ref{LS}) minimizes the squared norm of $\tilde{\bX}$,
hence it minimizes the coding rate (\ref{Ranalog1}) among all possible solutions for (\ref{xtilde}).
Using (\ref{LS});
recalling the definition of the system parameters: the non-erasure probability $p$, the aspect ratio of the $m \times n$ matrix $F$, and the aspect ratio of the $m \times k$ sub-matrix $F_S$, 
\begin{equation}
     \label{aspect-ratios}
     p = \frac{k}{n} ,  \  \ \
    \gamma = \frac{m}{n}
        \ \ \ \mbox{and} \ \ 
    \beta = \frac{k}{m} \ ;
\end{equation}
%
and assuming a white-Gaussian source $\bX \sim N(0,\sigma_x^2 I)$,
we can rewrite the ECDQ coding rate 
(\ref{Ranalog1}) as
\begin{align}
\label{Ranalog2}
R_{\rm analog}(S,D) =
&   \frac{p}{2 \beta} \cdot \log 
\left( 1 + \frac{\sigma_x^2}{D} \cdot \beta \cdot \specstat_{\rm MSE}(F_S)  \right)
\end{align}
%
where
\begin{equation}
    \label{signal-amplification}
\specstat_{\rm MSE}(F_S) 
\Ddef 
\frac{ \tr{(F_S^\dagger F_S)^{-1}} }{k}    
\end{equation}
is the {\em signal amplification factor} of the sub-matrix $F_S$.\footnote{
The notation uses ``MSE'' because (\ref{signal-amplification}) plays the role of mean-square error (MSE) amplification in other problems; 
see \cite{venkataramani2000perfect,mashiach2010multiple,krieger2014multi,seidner2000noise,haikin2016analog} and Chapter~\ref{section-Apps}.}
Finally, the {\em operational rate-distortion function} of the system
is the average of the ECDQ coding rate 
over all $k$-subsets $S \subset [n]$:
\begin{equation}
        \label{Ranalog-operational}
\bar{R}_{\rm analog}(D) 
=  
 \frac{1}{{n \choose k}} \sum_{S: |S|=k}
 R_{\rm analog}(S,D) 
    \end{equation}
as in (\ref{subset_average_performance_measure}).

In view of (\ref{Ranalog1})-(\ref{Ranalog-operational}), 
the system performance differs from the information-theoretic bound (\ref{RDF2}) in three aspects,
which contribute to the loss of the analog coding scheme with respect to the optimum performance (\ref{RDF}):
(i) the ``$+1$'' term inside the log; 
(ii) the design parameter $\beta \leq 1$ that appears both as a pre-log factor and inside the log; 
and (iii) the signal-amplification factor (\ref{signal-amplification}).
%
%
The ``$+1$'' term is negligible for high signal-to-distortion ratio $\sigma_x^2 \gg D$,
and it can be eliminated (for $\beta = \specstat_{MSE} =1$) if we replace LS estimation by minimum mean-squared error (MMSE) estimation \cite{haikin2016analog}.
As for the role of $\beta$,
since the function $x \log(1 + a/x)$ is monotonically {\em increasing} in $x$, 
choosing $\beta < 1$ can only be justified if 
the effect on reducing the signal amplification factor $\specstat_{\rm MSE}(F_S)$ 
is stronger 
for sufficiently many subsets $S$. 

%

\section{Non-uniform sampling and signal amplification}
The signal amplification factor (\ref{signal-amplification})
plays a key role in understanding frame goodness.
As we shall see, it is greater than or equal to 1 for every subset $S$,
and it tends to increase with the subset size $k$, i.e., with $\beta$ (see Chapter~\ref{section-inequalities}).
The equality $\specstat_{MSE}(F_S) = 1$ holds if and only if the column vectors of $F_S$ are orthogonal.
Indeed, if $l = n/k$ is an integer and the pattern $S$ happens to fall on a uniform grid
(i.e., $t_i = t_1 + (i-1)l$ modulo $n$, for $i=1,\ldots,k$),
then for $\beta=1$, the sub-matrix $F_S$ becomes a $k \times k$ DFT matrix, up to a constant phase shift.
Thus, $F_S$ is unitary 
and $F_S^\dagger F_S$ is an identity matrix,
$\specstat_{MSE}=1$, and the coding rate
$R_{\rm analog}(S = \mbox{uniform}, D)$ coincides with the information-theoretic bound (\ref{RDF2}) at high
signal-to-distortion ratio.
\begin{figure}[htbp] 
	\begin{center}
		\includegraphics[width=2in]{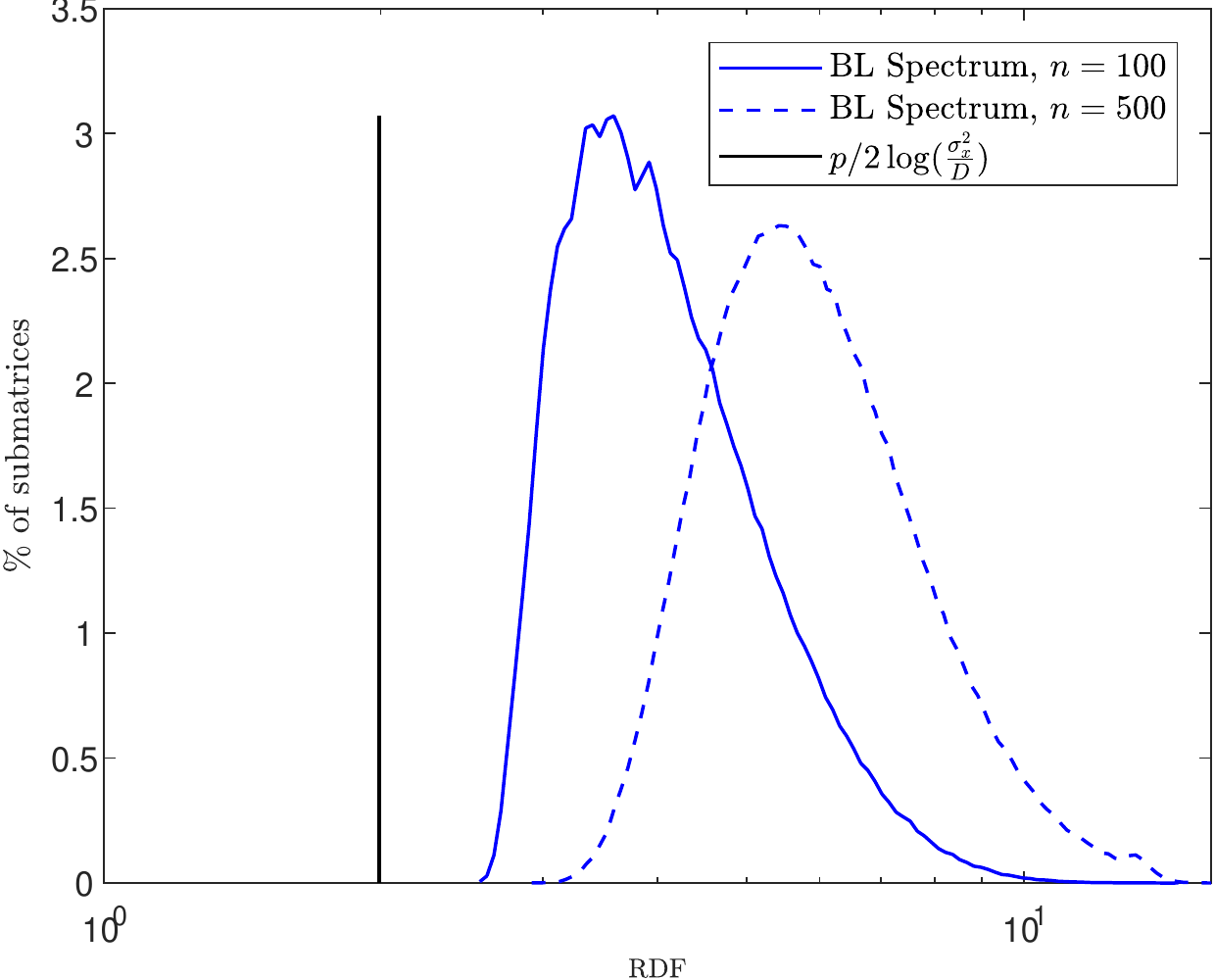}
		\includegraphics[width=2in]{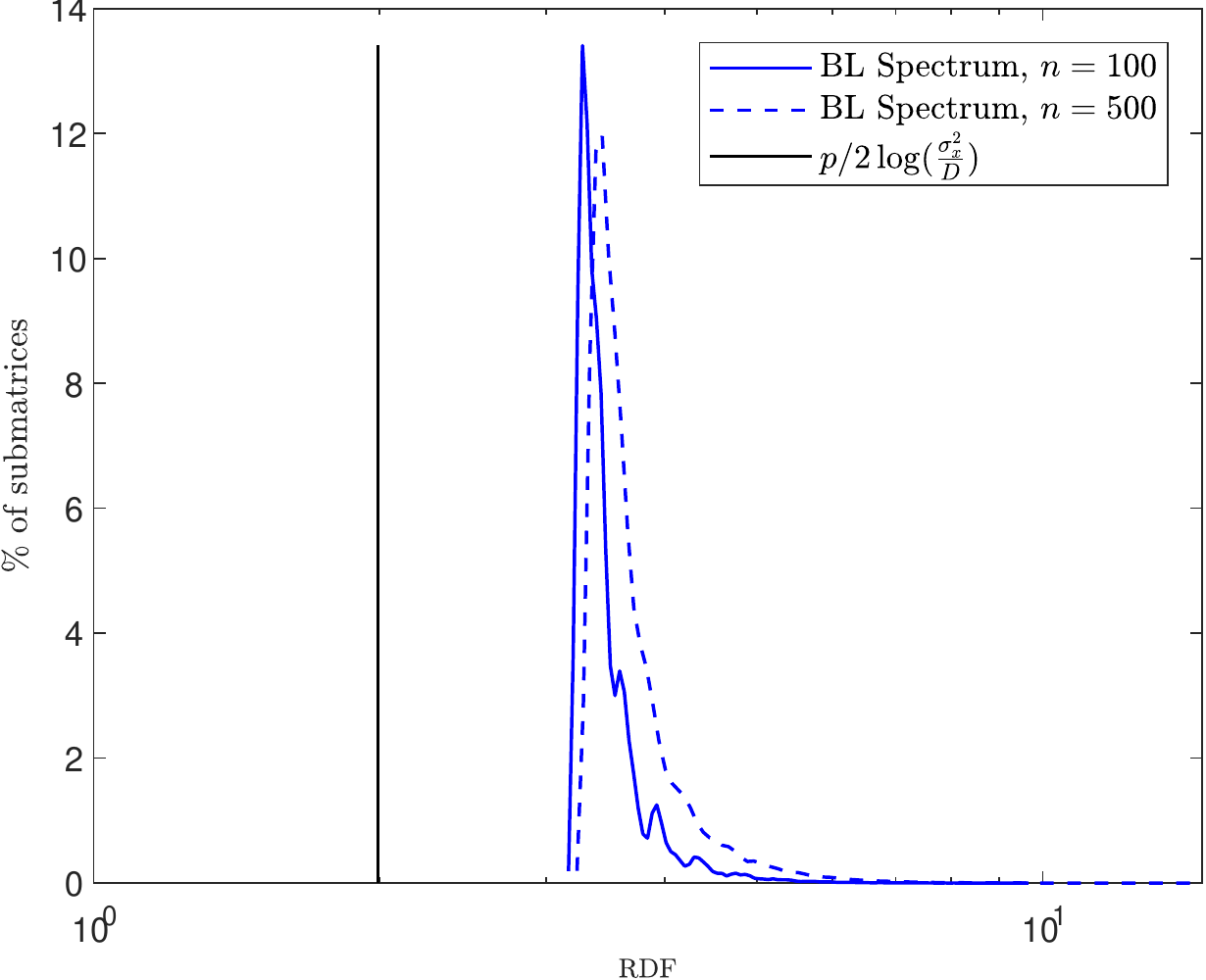}
		\caption{Histogram of the operational RDF $R_{\rm analog}(S,D)$ for random subsets $S$ from a low-pass (BL) frame with two frame sizes ($n=100$ and $500$), 
		for $\sigma_x^2/D = 1000$, $p=0.4$, and:
		(left) $\gamma = 1/2$ and $\beta = 0.8$;
		(right) $\gamma = 2/3$ and $\beta = 0.6$.
		The vertical line shows Shannon's RDF (\ref{RDF}).
		Note that the overall effect of decreasing $\beta$ below one is to reduce the rate,
and that a larger frame size ($n=500$) is worse than a smaller one ($n=100$).
		}
		\label{fig-analog-loss}
	\end{center}
\end{figure}

A typical pattern $S$ of a Bernoulli process 
is, however, far from being uniform,
and so 
the signal amplification factor (\ref{signal-amplification})
is typically strictly greater than one.
This phenomena is known in the literature as {\em noise amplification} in signal recovery from
noisy non-uniform samples \cite{seidner2000noise,mashiach2013noise,krieger2014multi,venkataramani2000perfect}.
It implies that the operational rate-distortion function (\ref{Ranalog-operational}) 
fails to achieve the information-theoretic 
bound (\ref{RDF2}), as was already observed (for $\beta=1$)
in \cite{martinian2008source} and \cite{mashiach2013sampling}.
Figure~\ref{fig-analog-loss} shows a histogram of 
$R_{\rm analog}(S = \mbox{random}, D)$
for two values of $\beta$.
In both cases, the average loss compared to (\ref{RDF2}) is still
much higher than the $\sim 1$ bit penalty of the naive 
side information coding scheme (Section~\ref{section-complexity}).


\section{From interpolation to frame codes}
Fortunately, the analog ``band-limited interpolation'' coding approach can be saved by 
replacing the low-pass DFT matrix (\ref{LPF-frame}) by frames which are more robust to the erasure pattern.
Analog codes follow a long line of works in the communications literature, 
that used linear transform-based codes
for noise and erasures, 
under names like ``DFT codes'' \cite{wolf1983redundancy,rath2004frame},
``coding over the reals'' \cite{marshall1984coding}, 
``analog BCH/RS codes'' \cite{tomlinson2017analogue,raviv2020gradient}, 
and in a broad sense also in {\em sigma-delta modulation} \cite{chou2015noise}, 
{\em code-division multiple access} \cite{rupf1994optimum},
{\em multiple-description coding} \cite{goyal1998multiple,ostergaard2009multiple},
{\em space-time coding} \cite{tarokh1998space},
{\em array imaging} 
\cite{krieger2014multi,taghavi2018high},
and {\em coded distributed computation} \cite{lee2018speeding}.
In the context of the current work, the low-pass DFT matrix $F$ 
is a special case of a {\em harmonic frame},
i.e., a subset of the rows of the DFT matrix \cite{waldron2018introduction,thill2017low}. 
This viewpoint opens our discussion to general frames in the form of (\ref{frame}),
and the rich frame theory.
Interestingly, 
in a sharp contrast to the consecutive band of frequencies in the low-pass frame (\ref{LPF-frame}), 
the performance of analog coding is optimized by 
an ``irregular'' (random-like) selection of the DFT rows,\footnote{
Low-pass spectrum can fit (or ``explain'') well arbitrary sampling values over a {\em uniform} time grid.
Irregular spectrum provides a better (lower squared-norm) fit when the sampling times lie on a {\em non}-uniform grid 
\cite{marshall1984coding,venkataramani2000perfect,feng1996spectrum}. 
} 
which eventually will lead us to {\em equiangular tight frames} \cite{xia2005achieving}.

%
%
%




\chapter{Sub-frame performance measures}
\label{section-Intro-performance-measures}
%
The previous chapter provided an information-theoretic motivation to assess a frame by averaging a specific performance measure (noise amplification) over sub-frames of a fixed size.
In this section we 
show that this, as well as other relevant frame performance measures \cite{tulino2004random},
%
%
%
%
are functions of the {\em eigenvalue spectrum} of sub-frames.
The sub-frame-average viewpoint (which seems to be new) will motivate our search for ``good'' frames.
We start yet with another well known frame performance measure, the mean-squared pairwise cross correlation,
and the notions of the Welch bound and a {\em tight frame}.


\section{The Welch bound}
The development of code-division multiple access (CDMA) triggered the introduction of 
the Welch bound (WB) \cite{welch1974lower} on the average squared-correlation between all distinct pairs
of an over-complete set of unit-norm vectors $\{ \bff_1, \ldots, \bff_n \}$ in $\mathbb{R}^m$ or $\mathbb{C}^m$:
%
\begin{equation}
    \label{WB}
    \frac{1}{(n-1)n}  \sum_{i \not= j} |\langle \bff_i, \bff_j \rangle|^2 
    \geq 
    \frac{n-m}{(n-1)m} .
\end{equation}
The Welch bound is achieved with equality if and only if 
the frame $F = [\bff_1  \cdots  \bff_n ]$ is {\em tight},
i.e., $F \cdot F^\dagger = (n/m) \cdot I_m$
\cite{kovacevic2008introduction,xia2005achieving,rupf1994optimum}.
For example, the low-pass frame (\ref{LPF-frame}) and the repetition frame 
$F = [I_m  \cdots  I_m]$ (or any union of orthonormal bases) are tight. 
The weakness of the Welch bound (\ref{WB}), 
however, is that it measures the frame as a {\em whole}.
While ``the whole is greater than the sum of its parts'',
in our problems of interest every (or almost every) individual part must be good,
so tightness is not sufficient. 
%


\section{Noise amplification}
%
%
%
%
Let us return to the erasure source coding problem of Section~\ref{section-distortion-side-information}. 
At high signal-to-distortion ratio,
each term in the operational rate-distortion function (\ref{Ranalog-operational}) becomes
$(m / 2n) \cdot \log ( \specstat_{\rm MSE}(F_S) )$,
plus some constant (independent of $F$); 
this gives rise to a subset-average performance measure of the form (\ref{subset_average_performance_measure}): 
\begin{equation}
    \label{AHMR}
    L_{\rm MSE}(F,k) = 
    \frac{1}{{n \choose k}} \sum_{S: |S|=k} \log( \specstat_{\rm MSE}(F_S) ),
    \end{equation}
to which we refer as the ``MSE loss'' of the frame $F$.
Here,
$\specstat_{\rm MSE}(F_S)$ is the ``noise amplification'' factor (\ref{signal-amplification}),
and the sum is over all 
$k$-subsets $S$ of $[n]$, 
i.e., over all $m \times k$ sub-matrices $F_S$ of $F$.

Note that for $k=2$, 
the MSE loss (\ref{AHMR}) can be lower bounded using the  Welch bound 
(\ref{WB}): 
$L_{\rm MSE}(F,2)$ is the average of  
$\log( 1 / (1 - |\langle \bff_i, \bff_j \rangle|^2) )$  
over all distinct pairs of frame vectors,
which, by Jensen's inequality and the Welch bound, 
is greater than or equal to 
$\log( (n-1)m / (m-1)n )$   
(which is roughly zero for large $m$ and $n$),
with equality if and only if the frame is tight and all the absolute pairwise cross-correlations are equal,
i.e., if and only if $F$ is an equi-angular tight frame.

The MSE performance measure (\ref{AHMR}) implies a frame optimization criterion
for any triplet $n \geq m \geq k$: 
\begin{equation}
    \label{AHMR*}
    L_{\rm MSE}^*(n,m,k) = \min_F  L_{\rm MSE}(F,k) ,
    \end{equation}
where the minimization 
is taken over all $m \times n$ unit-norm frames (\ref{frame}).\footnote{
The spark of $F$ must be larger than $k$, otherwise the sum in (\ref{AHMR}) is infinite
\cite{fickus2015group}.
}
%
%
%
Once the minimal loss function (\ref{AHMR*}) 
is known, we can go back
and optimize the free bandwidth parameter $m$ 
in the operational rate distortion function 
(\ref{Ranalog-operational}), 
for fixed $n$, $k$ and signal-to-distortion ratio $\sigma_x^2/D$. 

We note that other problems of interest give rise to the complementary case where $n \geq k \geq m$ (i.e., $\beta = k/m$ is now greater than $1$),
in which case the MSE performance measure (\ref{signal-amplification}) is computed with respect to
the sub-frame Hessian $F_S F_S^\dagger$ rather than Gram matrix $F_S^\dagger F_S$:
\[
\specstat_{\rm MSE}(F_S) 
= \frac{\tr{(F_S F_S^\dagger)^{-1}}}{k}  .   
\]




\subsection{Asymptotic goodness}
Information theory 
tends to look on the asymptotic performance of large block codes, at some fixed redundancy. It is therefore tempting to explore the large dimensional behavior of (\ref{AHMR*}),
when the aspect ratios $\beta = k/m$ and $\gamma = m/n$ in (\ref{aspect-ratios}) are held fixed:
%
\begin{align}
 L_{\rm MSE}^* (\gamma, \beta) 
\label{AHMR-asymptotic}
 \Ddef
\liminf_{n \rightarrow \infty}
 L_{\rm MSE}^*
 \Bigl( n, m = \lfloor n \gamma \rfloor, k = \lfloor n \gamma \beta \rfloor \Bigr)  ,    
    \end{align}
and to find a frame family with aspect ratio going to $\gamma$ 
that asymptotically achieves it.
%
Alternatively, we can use Bernoulli($p$) (rather than combinatorial)
selection of the pattern $S$ 
(as in the original source coding problem of Section~\ref{section-distortion-side-information}),
where the subset size $k$ becomes a random variable with expectation $p n$.
We expect that  
rare events where $k$ significantly deviates from $pn$ are asymptotically negligible, so
\begin{equation}
    \label{AHMR-asymptotic-Bernoulli}
    L_{\rm MSE}^*(\gamma,p) = L_{\rm MSE}^*(\gamma,\beta= p / \gamma) .
\end{equation}
Assuming this asymptotic MSE goodness measure 
is not zero, it reflects a fundamental loss of
analog coding with respect to the information theoretic bound (\ref{RDF}).
%

%

\section{Spread of eigenvalues}
The performance measures
(\ref{signal-amplification}) and (\ref{AHMR}) are functions of the $k \times k$ sub-frame Gram matrix $F_S^\dagger F_S$.
It is useful and insightful to
express them in terms of the eigenvalues $\lambda_1,\ldots,\lambda_k$ of 
$F_S^\dagger F_S$ (i.e., the squared nonzero singular values of the sub-matrix $F_S$).
Recall that the trace of a matrix is equal to the sum of its eigenvalues, so
$\lambda_1 + \ldots + \lambda_k = 
\tr{(F_S^\dagger F_S)} = k$,
where the last equality follows because $F_S$ is composed of $k$ unit-norm vectors.
Also, the eigenvalues of an inverse matrix are the inverse eigenvalues. 
Thus, the signal amplification factor (\ref{signal-amplification}) of a specific sub-matrix 
can be written as
\begin{align}
\specstat_{\rm MSE}(F_S) 
& =
(1/k) \left[ \lambda_1^{-1} + \ldots + \lambda_k^{-1} \right] 
\label{signal-amplification-eigenvalues}
=\frac{ \Bigl[ \frac{\lambda_1 + \ldots + \lambda_k}{k} \Bigr] }{\left[\frac{1/\lambda_1 + \ldots + 1/\lambda_k}{k}\right]^{-1}} , 
    \end{align}
%
i.e., as the 
{\em arithmetic-to-harmonic means ratio} (AHMR) of the eigenvalues of the Gram sub-matrix $F_S^\dagger F_S$.
By the {\em means inequality}, the AHMR is greater than or equal to one, with
equality if and only if all the eigenvalues are equal (and in our case, if all the eigenvalues are equal to one). 
Thus, $\specstat_{\rm MSE}(F_S)$ is greater than or equal to one, with equality if and only if $F_S^\dagger F_S$ is an identity matrix, i.e.,
the column vectors of $F_S$ are orthogonal.

\begin{figure}[htbp]
	\centering
		\includegraphics[width=2.5in]{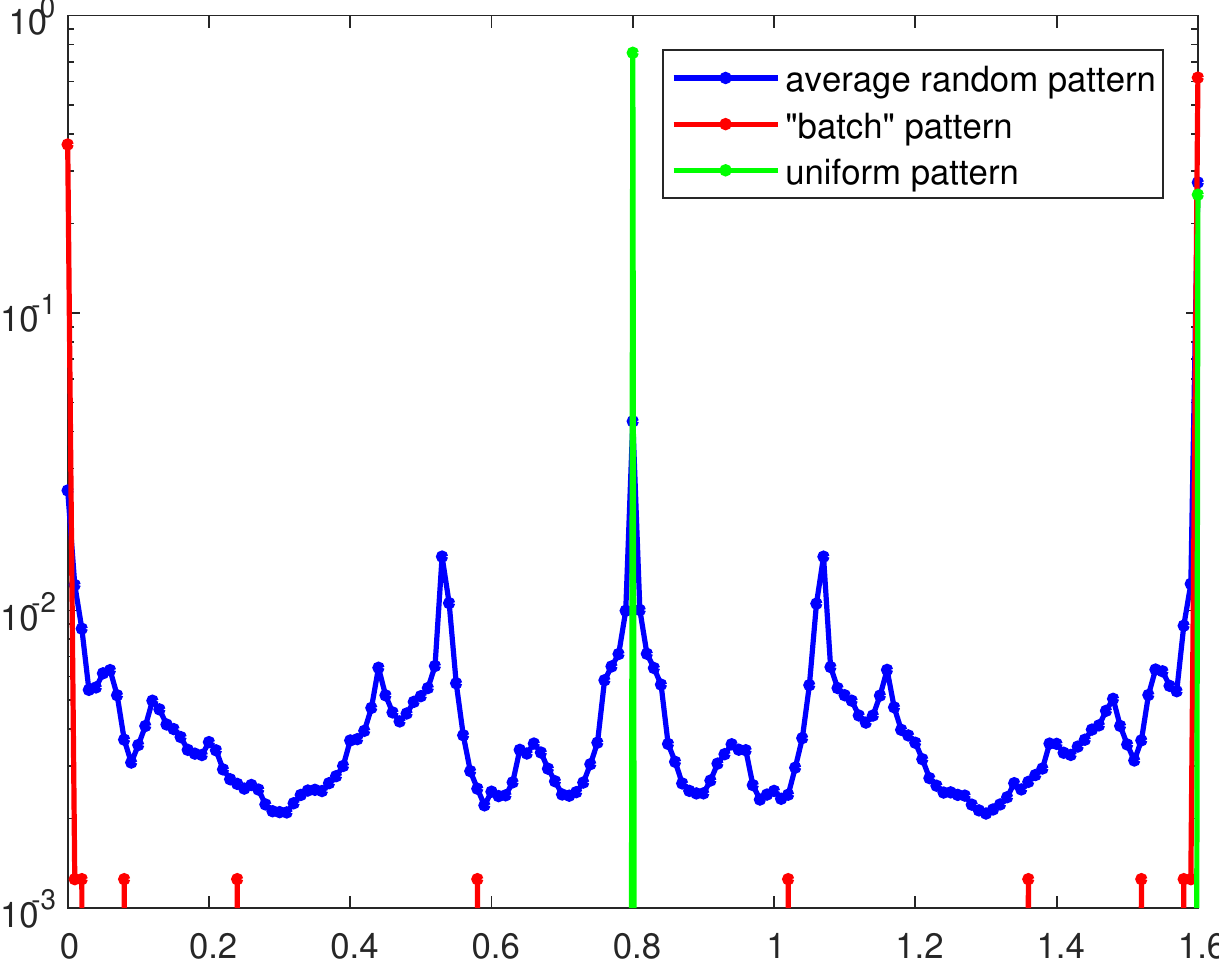}
\caption{Eigenvalues of subsets of the LPF frame (\ref{LPF-frame}),
	with 
	$p=1/2$, $\gamma=0.625$, $\beta=0.8$ ($n=1600$, $m=1000$, $k=800$), 
	for several possible patterns $S$: uniform, random, batch.
}
\label{fig-eigenvalue-spread-LPF}
\end{figure}
%

Clearly, since $n > m$, not all subsets of $\{\bff_1,\ldots,\bff_n\}$ can be orthogonal,
so $\specstat_{\rm MSE}(F_S)$ must be greater than one for some patterns $S$.
%
Figure~\ref{fig-eigenvalue-spread-LPF} illustrates the spread of the eigenvalues $\lambda_1,\ldots,\lambda_k$ of
$F_S  F_S^\dagger$ around their mean (i.e., $1$) for the low-pass frame (\ref{LPF-frame}), 
for three possible patterns: uniform sampling, random-like sampling, and consecutive samples (batch).
We observe that for this frame the eigenvalue spread is very sensitive to the pattern $S$.
Thus, a ``good'' frame should be robust in the sense that the eigenvalue spread of
(most of) its subsets is small.


\section{Shannon transform}
%
The Shannon transform of the sub-matrix $F_S$ is the capacity 
of a multiple-input multiple-output (MIMO) AWGN channel with a transfer matrix $F_S$
\cite{tulino2004random,tse2005fundamentals,verdu-shamai1999spectral,zaidel2001multicell}.
%
At high signal-to-noise ratio, 
and assuming $n \geq k \geq m$,
the Shannon transform gives rise to a ``log-det'' 
{\em geometric-to-arithmetic means ratio} goodness measure of the $m \times k$ sub-frame $F_S$: 
%
\begin{subequations}
\label{shannon-transform-00}
\begin{align}
\label{shannon-transform-0}
\specstat_{\rm Shannon}(F_S) 
&=  \frac{1}{m} \cdot \log \left( \det [\beta^{-1} \cdot F_S  F_S^\dagger] \right)
\\
\label{shannon-transform}
&= 
\log \left(
\frac{ \sqrt[m]{\lambda_1 \cdots \lambda_m}}{\frac{\lambda_1 + \ldots + \lambda_m}{m}} 
\right) .    
    \end{align}
\end{subequations}
%
%
Similarly to (\ref{signal-amplification-eigenvalues}), the argument of the logarithm is smaller than or equal to one, with equality if and only if the eigenvalues of $F_S F_S^\dagger$ are identical.
Thus, $\specstat_{\rm Shannon}$ is another measure for the spread of the eigenvalues of $F_S$.
%
The corresponding average sub-frame performance measure (\ref{subset_average_performance_measure}) is the 
``Shannon goodness'':
%
\begin{equation}
    \label{GAMR}
    L_{\rm Shannon}(F,k) = 
    \frac{1}{{n \choose k}} \sum_{S: |S|=k}  \specstat_{\rm Shannon}(F_S) 
\end{equation}
where the sum is over all $k$-subsets of $[n]$.
As above, we define the associated frame optimization criterion 
\begin{equation}
    \label{GAMR*}
    L_{\rm Shannon}^*(n,m,k) = 
    \max_F  L_{\rm Shannon}(F,k) ,
\end{equation}
%
as well as the asymptotic counterparts $L_{\rm Shannon}^*(\gamma,\beta)$ and $L_{\rm Shannon}^*(\gamma,p)$ 
(for Bernoulli($p$) selection) as in (\ref{AHMR*})-(\ref{AHMR-asymptotic-Bernoulli}).


\section{Statistical RIP and condition number}
We now turn to performance measures which look only on the {\em edges} of the sub-frame spectrum.
%
In compressed sensing, 
the Restricted Isometry Property (RIP)
is a well-known sufficient condition for sparse recovery \cite{candes2006near}.
A ``sensing'' frame $F$ satisfies a $\delta$-RIP of order $k$ if the spectral radius
of $F_S^\dagger F_S - I_k$, i.e.,
\begin{equation}
    \label{Intro-RIP}
 \specstat_{RIP}(F_S) 
 = \max \{\lambda_{\rm max}(F_S)-1, 1 - \lambda_{\rm min}(F_S) \} ,  
\end{equation}
is uniformly bounded by $\delta$ for all $k$-subsets $S$ of $[n]$, 
where $\lambda_{\rm max}$ and $\lambda_{\rm min}$ are the maximum and minimum eigenvalues of the Gram matrix $F_S^\dagger F_S$.
For  
$\delta < 1/2$, this condition guarantees perfect recovery via $\ell_1$-minimization in sensing any $k/2$-sparse vector, 
while $\delta < 1/3$ allows one to sense any $k$-sparse vector 
\cite{candes2006near,candes2008restricted,foucart2009sparsest,CAI201374}.
%

The RIP is, however, dominated by the worst subset, which is too restrictive in view of the ``typical'' approach of information theory.
The {\em statistical RIP} allows some fraction of the subsets to exceed the uniform bound \cite{calderbank2010construction}. 
The fraction of ``good'' subsets can be written as an average of the indicator of the event $\{ \specstat_{RIP}(F_S) \leq \delta \}$ over the subset $S$,
and we may wish to maximize it over all $m \times n$ frames $F$:
\begin{align}
\label{Intro-StRIP}
L_{StRIP}^* (n,m,k,\delta) 
= \max_F  \frac{1}{{n \choose k}} \sum_{ S: \; |S|=k }  1_{\{ \specstat_{RIP}(F_S) \leq \delta \}}.
\end{align}

To complete the picture, we shall also define the closely related notion of sub-frame condition number
\begin{equation}
    \label{condition_number_performance_measure}
    \Psi_{\rm cond}(F_S) 
    = \frac{\lambda_{\rm max}(F_S)}{\lambda_{\rm min}(F_S)}
\end{equation}
and the associated sub-frame average (\ref{subset_average_performance_measure}) (or statistical $\delta$-average as in (\ref{Intro-StRIP})),
as well as the corresponding optimum ``$(n,m,k)$ frame'' as in (\ref{AHMR*}) and (\ref{GAMR*}).
%
%
%


\section{Sub-frame rectangularity}
In Chapter~\ref{section-inequalities} we show that the sub-frame performance measures above 
(in particular, $L_{\rm MSE}(n,m,k)$ and $L_{\rm Shannon}(n,m,k)$)
exhibit a natural monotonic behavior as a function of the sub-frame aspect ratio $k/m$.
The more rectangular is the sub-frame, the better is the average goodness of the frame.
See also Figure~\ref{fig-analog-loss}.







\chapter{Frame theory}
\label{section-frame-theory}
In this section we formalize the discussion of frames and introduce a few important examples.
Frame theory is the study of overcomplete systems in a Hilbert space.
In particular, a sequence $\{\bff_i\}_{i\in I}$ of vectors in a Hilbert space $H$ is said to be a {\em frame} if there exist constants $a,b>0$ such that
\begin{equation}
    \label{frame_bounds}
a\|\bx\|^2
\leq\sum_{i\in I}|\langle \bx,\bff_i\rangle|^2
\leq b\|\bx\|^2    
\end{equation}
for every $\bx\in H$.
If $a=b$, then $\{\bff_i\}_{i\in I}$ is known as a {\em tight frame}, in which case the frame enjoys a painless reconstruction formula~\cite{daubechies2009painless}:
\[
\bx=\frac{1}{a}\sum_{i\in I}\langle \bx,\bff_i\rangle \bff_i.
\]
As an example, one may verify that the \textit{low-pass frame} (\ref{LPF-frame}) obtained from the first $m$ rows of the $n\times n$ discrete Fourier transform matrix is a tight frame for $\mathbb{C}^m$.
In general, one may select any $m$ rows from any $n\times n$ unitary matrix to produce a tight frame of $n$ vectors in $\mathbb{C}^m$.
As another example, one may combine $r$ orthonormal bases for $\mathbb{C}^m$ to produce a tight frame consisting of $n=rm$ vectors.
For instance, the \textit{repetition frame} consists of $r$ copies of the same basis, while the \textit{spike-and-sine frame} combines the identity basis with the Fourier basis.
Note that for {\em unit-norm tight} frames, the constants in (\ref{frame_bounds}) satisfy $a = b = n/m$.

Frames were introduced in 1952 by Duffin and Schaeffer~\cite{duffin1952class} in the context of nonharmonic Fourier series.
More recently, frames have been used as models for data processing, as information about $\bx\in H$ can be recovered from the frame coefficients $\{\langle \bx,\bff_i\rangle\}_{i\in I}$.
Most of the work in this direction has focused on the case in which $H$ has finite dimension, in which case we put $I=[n]$, and we abuse notation by letting $F$ denote both the sequence $\{\bff_i\}_{i\in [n]}$ and the matrix whose $i$th column is $\bff_i$.
Some frames are particularly well suited for information recovery after corruption in the coefficient domain.
For example, of all frames with unit-norm elements, tight frames are the most robust to a single erasure under additive white gaussian noise~\cite{goyal2001quantized}.

Intuitively, a frame is more robust to data corruption when no two frame elements capture similar information.
Suppose each frame element has unit norm, and for the sake of illustration, consider $\bx$ drawn from a gaussian distribution with identity covariance.
Then each frame coefficient $\langle \bx,\bff_i\rangle$ has standard gaussian distribution, and the covariance between $\langle \bx,\bff_i\rangle$ and $\langle \bx,\bff_j\rangle$ is $\langle\bff_i,\bff_j\rangle$.
As such, if we want to minimize the common information between all pairs of frame elements, we are inclined to minimize the {\em coherence} of $F$, defined as
\[
\nu(F)
\Ddef
\max_{\substack{i,j\in [n]\\i\neq j}}|\langle \bff_i,\bff_j\rangle|.
\]
In fact, of all unit norm tight frames, the ones that minimize coherence are most robust to two erasures under additive white gaussian noise~\cite{holmes2004optimal}.

An alternative interpretation is available in the context of coding theory.
Recall that given a compact metric space $X$ with metric $d$, an {\em optimal code} in $(X,d)$ of size $n$ is a subset $C\subseteq X$ for which the minimum distance
\[
\delta(C)
\Ddef 
\min_{\substack{x,y\in C\\x\neq y}}d(x,y)
\]
is maximized.
(Here, we ask $X$ to be compact so that an optimal code necessarily exists.)
There has been extraordinary success in studying optimal codes in Hamming space and the sphere.
In this survey paper, we are concerned with another metric space.
Specifically, for each $\mathbb{F}\in\{\mathbb{R},\mathbb{C}\}$, let $\mathbb{FP}^{m-1}$ denote {\em projective space}, i.e., the set of $1$-dimensional subspaces of $\mathbb{F}^m$.
One may define the \textit{chordal distance} over projective space by
\[
d(S,T)
\Ddef 
\|P_S-P_T\|_F,
\qquad
S,T\in\mathbb{FP}^{m-1}.
\]
Here, $P_S$ denotes the $m\times m$ orthogonal projection matrix onto $S$, while $\|\cdot\|_F$ denotes the Frobenius matrix norm.
(Notably, antipodal points on the sphere are as far apart as possible in the spherical metric space, but they determine identical points in projective space; despite such differences, both spaces have been treated simultaneously to some extent under the common umbrella of $2$-point homogeneous spaces~\cite{levenshtein1992designs,levenshtein1998universal,cohn2016optimal}.) 
For unit vectors ${\bf x}$ and ${\bf y}$, it is straightforward to verify that 
\[
d(\operatorname{span}\{{\bf x}\},\operatorname{span}\{{\bf y}\})
=\sqrt{2(1-|\langle {\bf x},{\bf y}\rangle|^2)}.
\]
It follows that optimal codes in projective space correspond to ensembles of vectors that minimize coherence.

In the following subsection, we discuss a particularly nice family of optimal projective codes known as \textit{equiangular tight frames}.
Later, we briefly survey other known optimal projective codes before discussing frames with well-conditioned subensembles.


\section{Equiangular tight frames}
This subsection is concerned with a particularly special type of frame.
We start with a definition:


%
\begin{definition}
We say $\{\bff_i\}_{i\in[n]}$ in $\mathbb{C}^m$ is an {\em equiangular tight frame (ETF)} if there exist constants $\nu, r \geq 0$ such that
\begin{subequations}
  \label{ETF_definition}
\begin{equation}
\label{ETF_definition_1}
|\langle \bff_i,\bff_j\rangle|
=\left\{\begin{array}{cl}1&\text{if }i=j\\ \nu &\text{if }i\neq j,\end{array}\right.    
\end{equation}
and
\begin{equation}
\sum_{i\in[n]}\bff_i\bff_i^*
= r I_m.    
\end{equation}
\end{subequations}
\end{definition}
Notice that if $F$ is an $m \times n$ ETF, then 
$\nu$ and $r$ are equal to its coherence $\nu(F)$ and redundancy $n/m$, respectively. 
As an example, one may verify that
\[
F
\Ddef
\left[
\begin{array}{ccc}
1&-\frac{1}{2}&-\frac{1}{2}\\
0&\frac{\sqrt{3}}{2}&-\frac{\sqrt{3}}{2}
\end{array}
\right]
\]
is an ETF with $\nu = \frac{1}{2}$ and $r = \frac{3}{2}$.
This example is affectionately known as the \textit{Mercedes-Benz frame} due to its resemblance to the luxury vehicle logo:
\begin{center}
\begin{tikzpicture}
\coordinate (0) at (0,0);
\coordinate (1) at ({2*cos(0*360/3)},{2*sin(0*360/3)});
\coordinate (2) at ({2*cos(1*360/3)},{2*sin(1*360/3)});
\coordinate (3) at ({2*cos(2*360/3)},{2*sin(2*360/3)});
\filldraw[black] (0) circle (2pt);
\draw [line width=1.5pt,-stealth] (0)--(1);
\draw [line width=1.5pt,-stealth] (0)--(2);
\draw [line width=1.5pt,-stealth] (0)--(3);
\end{tikzpicture}
\end{center}
Equiangular tight frames are important in part because they correspond to a certain type of optimal projective code.
The following result (a strengthened form of (\ref{WB})) makes this explicit:
\begin{prop}[The ``maximum'' Welch bound~\cite{welch1974lower,strohmer2003grassmannian}]
Suppose $F=\{\bff_i\}_{i\in[n]}$ with $\bff_i\in\mathbb{C}^m$ satisfying $\|\bff_i\|=1$ for each $i\in[n]$.
Then
\begin{equation}
    \label{WB_MAX}
\nu(F)
\geq \sqrt{\tfrac{n-m}{m(n-1)}},    
\end{equation}
with equality if and only if $F$ is an equiangular tight frame.
In particular, the constant $\nu$ in (\ref{ETF_definition_1}) is equal to the lower bound in (\ref{WB_MAX}).
\end{prop}
%

Not only are ETFs optimal projective codes, they simultaneously minimize an infinite family of potential energies, specifically, those of the form
\[
\sum_{\substack{i,j\in[n]\\i\neq j}}g(1-|\langle \bff_i,\bff_j\rangle|^2),
\]
where $g\colon(0,1]\to\mathbb{R}$ is any function satisfying $(-1)^kg^{(k)}(x)\geq0$ for every $x\in(0,1]$ and $k\in\mathbb{N}$; see~\cite{cohn2007universally} for details.
For example, one may take $g(x)=x^{-c}$ or $g(x)=e^{-cx}$ for any $c>0$.
We note that it is generally difficult to identify configurations that minimize a given potential energy, and there is a substantial literature dedicated to this line of inquiry, e.g.,~\cite{smale1998mathematical,benedetto2003finite,bilyk2019optimal,glazyrin2019moments,glazyrin2020repeated,chen2020universal,xu2021minimizers}.
Along these lines, ETFs can be equivalently characterized as equiangular minimizers of the \textit{frame potential}~\cite{benedetto2003finite}, which appears on the left-hand side of~\eqref{WB}.
For all of these reasons, 
we should think of ETFs as extremely special objects, and indeed, they appear to be quite rare.
ETFs do not exist for most pairs $(m,n)$, and most of the known ETFs arise from combinatorial designs.
In what follows, we briefly describe the most fundamental examples; see~\cite{colbourn2006handbook} for surveys on the relevant combinatorial designs, and see~\cite{fickus2015tables} for a survey on ETFs.

\subsection{Strongly regular graphs}

Suppose $F=\{\bff_i\}_{i\in[n]}$ is an ETF with $\bff_i\in\mathbb{R}^m$ for every $i\in[n]$.
Then the Gram matrix of $F$ takes the form
\begin{equation}
    \label{Seidel_matrix}
F^\dagger F = I_n+\nu(F)\cdot S,    
\end{equation}
where $S$ is the so-called \textit{Seidel adjacency matrix} of a regular $2$-graph~\cite{seidel2014geometry}; these correspond to a certain class of \textit{strongly regular graphs}~\cite{waldron2009construction}.
To illustrate, we consider the cyclic graph on $5$ vertices:
\begin{center}
\begin{tikzpicture}
\coordinate (0) at ({2*cos(0*360/5)},{2*sin(0*360/5)});
\coordinate (1) at ({2*cos(1*360/5)},{2*sin(1*360/5)});
\coordinate (2) at ({2*cos(2*360/5)},{2*sin(2*360/5)});
\coordinate (3) at ({2*cos(3*360/5)},{2*sin(3*360/5)});
\coordinate (4) at ({2*cos(4*360/5)},{2*sin(4*360/5)});
\draw[-] (0)--(1)--(2)--(3)--(4)--(0);
\filldraw[black] (0) circle (2pt);
\filldraw[black] (1) circle (2pt);
\filldraw[black] (2) circle (2pt);
\filldraw[black] (3) circle (2pt);
\filldraw[black] (4) circle (2pt);
\end{tikzpicture}
\end{center}
This graph is strongly regular since every vertex has the same degree (i.e., $2$), adjacent vertices all have the same number of common neighbors (i.e., $0$), and non-adjacent vertices have the same number of common neighbors (i.e., $1$).
Let $V$ denote the vertex set of this graph, and consider the $6\times6$ Seidel adjacency matrix $S$ with rows and columns indexed by $V\cup\{\infty\}$, where $S_{ii}=0$, $S_{ij}=-1$ if $i,j\in V$ and $i\leftrightarrow j$, and otherwise $S_{ij}=1$:
\begin{equation}
    \label{Seidel_matrix_pentagon}
S = 
\left[\begin{array}{c|ccccc}
0&+&+&+&+&+\\\hline
+&0&-&+&+&-\\
+&-&0&-&+&+\\
+&+&-&0&-&+\\
+&+&+&-&0&-\\
+&-&+&+&-&0
\end{array}\right].
\end{equation}
One may verify that $G:=I_6+\frac{1}{\sqrt{5}}S$ has eigenvalues $0$ and $2$, both with multiplicity $3$, meaning $G$ is the Gram matrix of an ETF of $6$ vectors in $\mathbb{R}^3$.
In fact, the vectors in this ETF can be obtained by grouping the $12$ vertices of the icosahedron into $6$ antipodal pairs and selecting a representative point from each pair.
In other words, the $6$ lines spanned by the icosahedron's vertices form an optimal projective code.
While the real ETFs correspond to a special class of strongly regular graphs, the more general complex ETFs do not enjoy a perfect correspondence with combinatorial designs.
Regardless, there exist a few important combinatorial constructions in the complex case, as we review below. 

\subsection{Difference sets}
\label{section-difference-set}

Next, we consider \textit{harmonic frames}, which are constructed by sampling rows of the character table of a finite abelian group $G$.
Specifically, we take $n=|G|$ and select a subset of characters $D\subseteq\hat{G}$ so that for each $g\in G$, we have $\bff_g\in\mathbb{C}^D$ defined by $\bff_g(\chi)=\chi(g)$ for every $\chi\in D$.
Such frames $\{\bff_g\}_{g\in G}$ are necessarily tight, and they form ETFs (after scaling by $|D|^{-1/2}$) precisely when $D$ is a \textit{difference set}, meaning there exists $\lambda\in\mathbb{N}$ such that every non-identity member of $\hat{G}$ can be expressed as a difference of two members of $D$ in exactly $\lambda$ ways~\cite{strohmer2003grassmannian,xia2005achieving,ding2007generic}.

For an illustrative example, take $G$ to be the abelian group $\mathbb{Z}/7\mathbb{Z}$.
We claim that $\{1,2,4\}$ is a difference set in $G$.
This can be easily seen from the difference table:
\[
\begin{array}{c|ccc}
-&1&2&4\\\hline
1&0&6&4\\
2&1&0&5\\
4&3&2&0
\end{array}
\]
For example, $1-2=-1\equiv6\bmod 7$, and so $6$ appears in the difference table at entry $(1,2)$.
Since the off-diagonal entries of the difference table evenly cover the nonzero members of $\mathbb{Z}/7\mathbb{Z}$ (specifically, with $\lambda=1$), it follows that $\{1,2,4\}$ is a difference set.
Next, we consider the characters of $G$.
Let $\chi\in\hat{G}$ denote the function $\chi\colon G\to\mathbb{C}\setminus\{0\}$ defined by $\chi(g):=e^{2\pi j g / 7}$, 
where $j = \sqrt{-1}$.
Then $\chi$ is a homomorphism of $G$ and its entrywise powers $\{\chi^k\}_{k=0}^6$ form the character group $\hat{G}\cong G$.
It follows that $D \Ddef \{\chi^1,\chi^2,\chi^4\}$ is a difference set for $\hat{G}$.
Denoting $\omega \Ddef e^{2\pi j / 7}$, then the above procedure gives
\[
F 
\Ddef  
\frac{1}{\sqrt{3}}\left[\begin{array}{ccccccc}
\omega^0&\omega^1&\omega^2&\omega^3&\omega^4&\omega^5&\omega^6\\
\omega^0&\omega^2&\omega^4&\omega^6&\omega^1&\omega^3&\omega^5\\
\omega^0&\omega^4&\omega^1&\omega^5&\omega^2&\omega^6&\omega^3
\end{array}\right].
\]
Equivalently, we select the rows of the $7\times7$ discrete Fourier transform matrix that are indexed by $\{1,2,4\}$.
This approach allows one to design ETFs by selecting rows from other character tables such as the appropriate $2^k\times2^k$ Hadamard matrix.
Observe that these ETFs have the property that every entry has the same modulus, thereby making them desirable for use in CDMA~\cite{tropp2003cdma}.
However, since a size-$n$ group admits a size-$m$ difference set only if $n-1$ divides $m(m-1)$ 
(specifically, $m(m-1) = \lambda (n-1)$), this construction is somewhat restrictive. 

\subsection{Block designs}

A third combinatorial construction of ETFs arises from block designs.
Let $B$ denote any $(v,b,r,k,\lambda=1)$-\textit{balanced incomplete block design (BIBD)}, i.e., $B$ is a collection of $b$ size-$k$ subsets of $[v]$ (called \textit{blocks}) such that every point resides in exactly $r$ blocks and every pair of points resides in exactly $\lambda=1$ block.
For each $x\in [v]$, let $B_x$ denote the size-$r$ set of blocks that contain $x$, let $H_x$ denote any complex Hadamard matrix of order $r+1$ with rows indexed by $B_x\cup\{\infty\}$, and for each $z\in[r+1]$, consider the vector $\bff_{(x,z)}\in\mathbb{C}^B$ defined by
\[
\bff_{(x,z)}(y)
\Ddef
\left\{\begin{array}{cl} H_x(y,z)&\text{if }y\in B_x\\0&\text{otherwise.}\end{array}\right.
\]
Then $\{\bff_{(x,z)}\}_{x\in[v],z\in[r+1]}$ forms an ETF (after scaling by $r^{-1/2}$)~\cite{fickus2012steiner}.
This construction is known as the \textit{Steiner equiangular tight frame}.

As an example, take $v:=3$ and $B:=\{\{1,2\},\{1,3\},\{2,3\}\}$.
Then $B$ is a BIBD with $b=3$, $k=2$, $r=2$, and $\lambda=1$.
Notice that $B_1=\{\{1,2\},\{1,3\}\}$, $B_2=\{\{1,2\},\{2,3\}\}$, and $B_2=\{\{1,3\},\{2,3\}\}$.
Put $\omega \Ddef e^{2\pi j / 3}$ and take
\[
H 
\Ddef 
\left[\begin{array}{lll}
1^{\phantom{2}}&1&1\\
1&\omega&\omega^2\\
1&\omega^2&\omega
\end{array}\right].
\]
For each $x\in\{1,2,3\}$, we take $H_x$ to be a copy of $H$ in which the first row is indexed by $\infty$ and the other two rows are indexed by $B_x$.
Then the above procedure gives
\begin{equation}
\label{eq.3x9 example}
F
\Ddef 
\frac{1}{\sqrt{2}}\left[\begin{array}{lllllllll}
1^{\phantom{2}}&\omega&\omega^2&1^{\phantom{2}}&\omega^2&\omega&0&0&0\\
1&\omega^2&\omega&0&0&0&1^{\phantom{2}}&\omega&\omega^2\\
0&0&0&1&\omega&\omega^2&1&\omega^2&\omega
\end{array}\right].
\end{equation}

One may verify that $(v,k,\lambda)$ determines the full list of parameters $(v,b,r,k,\lambda)$ of a BIBD by virtue of the following identities:
\[
bk=vr,
\qquad
\lambda(v-1)=r(k-1),
\]
which in turn determines the size of the corresponding Steiner ETF; specifically, there are $n=(r+1)v$ vectors in $m=b$ dimensions.
For every fixed $k_0\geq2$, there is an infinite family of BIBDs with $k=k_0$ and $\lambda=1$.
This produces an infinite sequence of Steiner ETFs with $m\to\infty$ and $\gamma\to\frac{1}{k_0}$.
Observe that the first three columns in \eqref{eq.3x9 example} are linearly dependent.
In fact, for every Steiner ETF, the first $O(\sqrt{m})$ columns are linearly dependent, and as we will soon see, this feature bears negative consequences in compressed sensing in an adversarial setting, albeit negligible in an average-case setting. 

\subsection{Zauner's conjecture}

The longest standing open problem concerning ETFs was posed in 1999 by Zauner in his PhD thesis~\cite{zauner2011quantum}.
This conjecture predicts the existence of ETFs of $m^2$ vectors in $\mathbb{C}^m$ for every dimension $m$; notably, $\gamma\to0$ as $m\to\infty$.
Furthermore, these ETFs appear to be constructed from orbits of the Heisenberg--Weyl group.
To be explicit, let $\{{\bf e}_k\}_{k=0}^{m-1}$ denote the identity basis of $\mathbb{C}^m$ with zero-based indices, and consider the translation and modulation operators defined by
\[
T{\bf e}_k={\bf e}_{(k+1) \bmod m},
\qquad
M{\bf e}_k=e^{2\pi ik/m}\cdot {\bf e}_k,
\]
and extended linearly.
Zauner predicts that for each dimension $m\in\mathbb{N}$, there exists a vector $\bff\in\mathbb{C}^m$ such that the time--frequency shifts $\{T^aM^b\bff\}_{a,b=0}^{m-1}$ form an ETF.
Such ETFs play a significant role in the quantum information theory literature, where they are known as symmetric, informationally complete, positive operator--valued measures (SIC-POVMs).
Accordingly, there has been a flurry of work to prove Zauner's conjecture.
Today, we know that for every $m\leq 151$, there exists $\bff\in\mathbb{C}^m$ such that $\{T^aM^b\bff\}_{a,b=0}^{m-1}$ is within machine precision of satisfying the ETF conditions~\cite{fuchs2017sic}.
Furthermore, for a fraction of these dimensions, an exact construction is known~\cite{appleby2018constructing}.
It appears that a general construction will depend on a strengthened version of the Stark conjectures from algebraic number theory~\cite{kopp2019sic}, which explains why it has been so difficult to prove Zauner's conjecture.


\section{Other optimal projective codes}
\begin{figure*}
\begin{center}
\includegraphics[width=\textwidth]{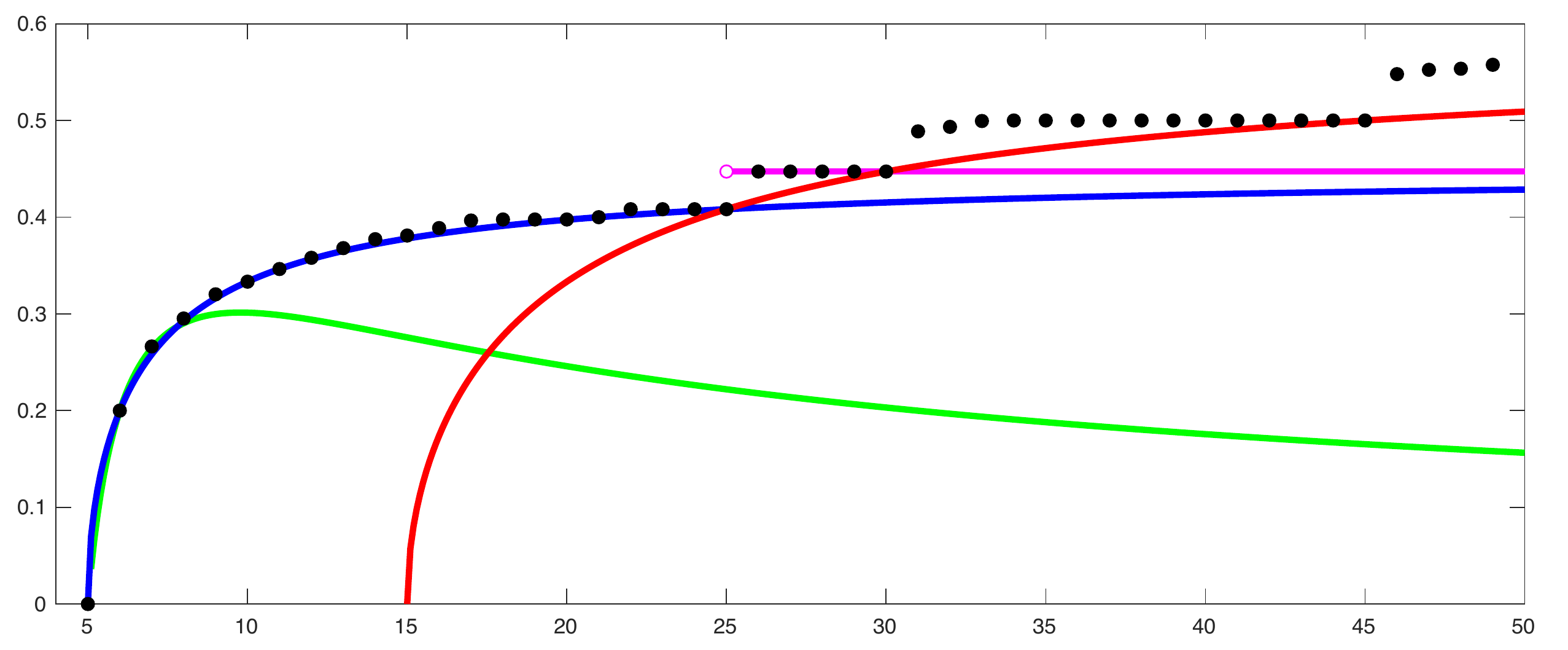}
\end{center}
\caption{From~\cite{jasper2019game}.
Various bounds on coherence of $n\in\{5,\ldots,49\}$ unit vectors in $\mathbb{C}^5$.
The Bukh--Cox bound is in green, the ``maximum'' Welch bound (\ref{WB_MAX}) in blue, the orthoplex bound in pink, and the Levenshtein bound in red.
Points in black denote the coherence of the best known code of $n$ points in $\mathbb{CP}^4$, maintained in~\cite{jasper2019game}.
}
\label{fig.bounds}
\end{figure*}


While most work on projective codes has focused on ETFs, other types of optimal projective codes have been discovered by achieving equality in alternative bounds; see Figure~\ref{fig.bounds} for an illustration.
The most successful construction of this sort arises from mutually unbiased bases.
Here, we say that $m\times m$ unitary matrices $U_1,\ldots,U_k$ are mutually unbiased if for every $i,j\in[k]$ with $i\neq j$, it holds that every entry of $U_i^\dagger U_j$ has modulus $m^{-1/2}$.
Given $k=m+1$ mutually unbiased bases, it holds that the $m(m+1)$ columns of $[U_1\cdots U_{m+1}]$ form an optimal projective code.
In fact, they achieve equality in both the orthoplex bound~\cite{conway1996packing} and the Levenshtein bound~\cite{levenshtein1982bounds,haas2017levenstein}.
Such projective codes are known to exist whenever $m$ is a power of a prime~\cite{wootters1989optimal}.
Beyond mutually unbiased bases, there are few known constructions that achieve equality in the Levenshtein bound~\cite{haas2017levenstein,mikhail2020infinite}.
By comparison, the orthoplex bound is much easier to work with: 
Given a projective code of size at least $m^2+2$ that achieves equality in the orthoplex bound, removing any vector produces another code that achieves equality in the orthoplex bound.
Going the other direction, one may achieve equality in the orthoplex bound by appending the identity basis to an appropriately incoherent harmonic frame with at least $m^2-m+1$ frame elements~\cite{bodmann2016achieving}.
Recently, Bukh and Cox~\cite{bukh2020nearly} introduced a new bound on projective codes of size $m+O(\sqrt{m})$.
Notably, equality in this bound occurs precisely when there exists $k\in\mathbb{N}$ such that $n$ is a multiple of $k^2$, $m=n-k$, and there exists an ETF of $k^2$ vectors in $\mathbb{C}^k$.
(Mind the relationship to Zauner's conjecture.)
Interestingly, the Welch, Levenshtein, and Bukh--Cox bounds can all be proved using a common \textit{Delsarte linear programming} framework~\cite{haas2017levenstein,magsino2019delsarte}. 

For $(m,n)$ such that $2\leq m\leq 7$ and $m+2\leq n\leq 49$, the best known codes of $n$ points in $\mathbb{CP}^{m-1}$ are maintained in~\cite{jasper2019game}.
For most choices of $(m,n)$, it appears to be impossible to achieve equality in the above general bounds, and so other techniques are required to perform the optimization.
In the special case where $m=2$, we note that $\mathbb{CP}^1$ is isometrically isomorphic to the sphere $S^2$, and so our problem reduces to the famous Tammes problem.
Here, optimal codes are known for $n\leq 14$ and $n=24$~\cite{musin2015tammes}.
In the absence of a tight lower bound, the techniques here are rather different: one enumerates a large number of candidate solutions (think ``critical points'') before pruning the list until identifying the complete set of solutions.
The most recent results of this form have relied on computer assistance.
To date, this accounts for all known optimal codes for $\mathbb{CP}^{m-1}$.
In the case of $\mathbb{RP}^{m-1}$, additional codes have been proved optimal using a Tammes-inspired approach~\cite{benedetto2006geometric,mixon2019optimal} as well as techniques from computational algebraic geometry~\cite{fickus2018packings,mixon2021globally}, but the latter techniques do not transfer easily to the complex case.

\section{Frames with well-conditioned subensembles}
\label{section-earlier}

Compressed sensing spurred substantial interest in a particular matrix design problem:
Given integers $k<n$ and $\delta\in(0,1/2)$, find the smallest $m\in\mathbb{N}$ for which there exists $F=\{\bff_i\}_{i\in[n]}$ with $\bff_i\in\mathbb{C}^m$ 
such that for every $k$-subset $S \in \binom{[n]}{k}$, the Gram matrix $G_S$ of $\{\bff_i\}_{i\in S}$  satisfies the matrix inequality
\[
(1-\delta)I
\preceq G_S
\preceq (1+\delta)I.
\]
(Note that this is slightly stronger than asking every $G_S$ to be well conditioned, since the eigenvalues must reside in the fixed interval $[1-\delta,1+\delta]$.)
This is 
the {\em restricted isometry property}
(compare with (\ref{Intro-RIP}) with $G_S = F_S^\dagger F_S$),
and any ensemble with the property can be used as columns of a sensing matrix in compressed sensing.
Of course, we must have $m\geq k$, since otherwise $G_S$ would be rank deficient.
Also, the interesting regime is $m\leq n$ since one may otherwise design $F$ to consist of orthonormal vectors.
If the entries of each vector in $F$ are independent with distribution 
$N(0,1/m)$, then $F$ exhibits the restricted isometry property with high probability in the regime $m\asymp (k/\delta^2)\log(n/k)$~\cite{foucartmathematical}.

Different applications of compressed sensing lead to different design constraints, and this has led researchers to develop restricted isometries with additional structural properties, such as partial Fourier, partial circulant, or Gabor structure~\cite{krahmer2014suprema,haviv2017restricted}.
All of these constructions exhibit the restricted isometry property in the regime $m\asymp k\operatorname{poly}(1/\delta,\log n)$, and they do so with the help of the probabilistic method; note that for a fixed $\delta$, this is identical to the scaling of the gaussian construction up to log factors in $n$.
Every known restricted isometry with this scaling is a random matrix that exhibits some small failure probability.
This motivated Terry Tao to pose the derandomization problem of finding explicit restricted isometries.

This derandomization problem is an instance of what Avi Wigderson calls a \textit{hay in a haystack problem}~\cite{wigderson2019mathematics}, in which random objects satisfy a given property with high probability, while explicit constructions consistently produce objects that do not exhibit that property; other examples include Ramsey graphs (see~\cite{grolmusz2000low} and references therein) and the Gilbert--Varshamov bound for linear codes~\cite{varshamov1957estimate}. 
To make our derandomization problem well defined, one might leverage ideas from computational complexity, which in turn forces us to pose things asymptotically.
Given an infinite $\mathcal{M}\subseteq\mathbb{N}$, we say a sequence $\{F_m\}_{m\in \mathcal{M}}$ with each $F_m$ being a tuple of $n(m)$ vectors in $\mathbb{C}^m$ is {\em explicit} if there exists an algorithm that on input $m\in \mathcal{M}$ takes $\operatorname{poly}(m,n(m))$ time to output $F_m$.
This model of explicitness precludes certain unwanted solutions to the derandomization problem.
As an example of an unwanted solution, one could consider all $2^{mn}$ possible $n$-tuples of vectors in $\mathbb{C}^m$ with entries in $\{\pm1/\sqrt{m}\}$, and for each of these, one could further compute the spectrum of the Gram matrix of the vectors indexed by $S$ for each $S \in \binom{[n]}{k}$.
This deterministic process allows one to identify a restricted isometry with $m\asymp (k/\delta^2)\log(m/k)$, but the algorithm is too slow to be considered explicit.

Having made precise the notion of explicitness, we may now enunciate the explicit restricted isometry problem:
Find the largest $k=k(m)$ for which there exists an explicit $(k,\delta)$-restricted isometry of $n=(1+\Omega(1))\cdot m$ vectors in $\mathbb{C}^m$ for some fixed $\delta\in(0,1/2)$,
and an infinite sequence $m$ in $\mathbb{N}$.
For $k=1$, we may simply select $F$ to consist of unit-norm vectors, since each $S \in\binom{[n]}{1}$ takes the form $S=\{i\}$, in which case $G_S=\|\bff_i\|^2=1$.
Next, for $k=2$, we have $S=\{i,j\}$ with $i\neq j$.
If we assume that $F$ consists of unit-norm vectors, then we have
\[
G_S
=\left[\begin{array}{cc}
1&\langle \bff_i,\bff_j\rangle\\
\langle\bff_j,\bff_i\rangle&1
\end{array}\right].
\]
Considering $\tr{G_S} = 2$ and $\operatorname{det}(G_S) = 1-|\langle\bff_i,\bff_j\rangle|^2$, it follows that the eigenvalues of $G_S$ are $1\pm|\langle\bff_i,\bff_j\rangle|$.
As such, finding a $(2,\delta)$-restricted isometry with unit-norm vectors and minimal $\delta$ is equivalent to minimizing coherence $\nu(F)$, i.e., finding an optimal projective code.
Intuitively, optimal projective codes are good candidates for restricted isometries for even larger values of $k$.

This is made rigorous with the Gershgorin circle theorem~\cite{varga2004gervsgorin}, which implies that the eigenvalues of $G_K$ must reside between $1-(k-1)\nu(F)$ and $1+(k-1)\nu(F)$.
Pleasingly, a smaller coherence implies that the spectrum resides in a smaller interval.
In particular, $F$ is necessarily a $(k,\delta)$-restricted isometry whenever $\nu(F)\leq\delta/(k-1)$.
However, combining the ``maximum'' Welch bound (\ref{WB_MAX}) and the constraint $n=(1+\Omega(1))\cdot m$ gives
\[
\tfrac{\delta}{k-1}
\geq\nu(F)
\geq\sqrt{\tfrac{n-m}{m(n-1)}}
\geq\Omega(\tfrac{1}{\sqrt{m}}),
\]
and so $k\lesssim \delta\sqrt{m}$.
This is a far cry from the linear scaling of $k$ that is obtained from probabilistic constructions.
Interestingly, certain optimal projective codes reveal that Gershgorin isn't to blame for this performance gap, as it sometimes gives the correct scaling.
For example, equation~\eqref{eq.3x9 example} determines $9$ vectors in $\mathbb{C}^3$ that achieve equality in the Welch bound, meaning $\nu(F)=1/2$.
According to Gershgorin, every subensemble of size $k<1/\nu(F)+1=3$ is linearly independent, and conversely, the first $3$ columns vectors are linearly dependent.
In general, Steiner ETFs all have the property that the Gershgorin analysis is sharp, and so one must consider higher-order statistics in order to break the square-root bottleneck $k\lesssim \delta\sqrt{m}$.

To this end, Bourgain et al.~\cite{bourgain2011explicit} introduced the $(k,\theta)$-{\em flat restricted isometry property}, which states that
\[
\bigg|\sum_{i\in I}\sum_{j\in J} \langle \bff_i,\bff_j \rangle\bigg|
\leq \theta\sqrt{|I||J|}
\]
for every disjoint $I,J\subseteq[n]$ with $|I|\leq k$ and $|J|\leq k$.
A technical argument involving convexity and dyadic decomposition~\cite{bourgain2011explicit,bourgain2011breaking,mixon2015explicit} gives that every $(k,\theta)$-flat restricted isometry of unit-norm vectors is also a $(2k,150\theta\log k)$-restricted isometry.
With this higher-order statistic, Bourgain et al.\ leverage ideas from additive combinatorics to show that a certain ensemble with $n\asymp m^{1+\epsilon}$ is a $(k,1/3)$-restricted isometry with $k\asymp m^{1/2+\epsilon}$ for $\epsilon\approx 3.6\cdot10^{-15}$.
Alternatively, this higher-order statistic can be used to show that the Paley equiangular tight frame of $2m$ vectors in $\mathbb{C}^m$ is a $(k,1/3)$-restricted isometry with $k\asymp m^{1/2+\epsilon}$ for larger choices of $\epsilon>0$, conditioned on a conjecture in number theory concerning cancellations in the Legendre symbol~\cite{bandeira2017conditional}.
Despite a decade of work in this area, these are the only results that break the square-root bottleneck in the explicit restricted isometry problem, leaving a substantial gap compared to random constructions.

Interestingly, given an equiangular tight frame $F$ for $\mathbb{C}^m$ and $k\gg m^{1/2}$, the proportion of $S \in \binom{[n]}{k}$ for which either $G_S \prec (1-\delta)I$ or $G_S \succ(1+\delta)I$ is quite small.
That is, if there are any subsets precluding an equiangular tight frame from being a restricted isometry, they are relatively few.
This was first observed by Tropp~\cite{tropp2008conditioning}, who demonstrated that equiangular tight frames (and similar ensembles) satisfy $(1-\delta)I\preceq G_S \preceq (1+\delta)I$ for all but a fraction $1/\operatorname{poly}(k)$ of $S \in\binom{[n]}{k}$ whenever $k\leq cm/\log m$ for an appropriately small choice of $c>0$.
In fact, this phenomenon allows one to obtain strong compressed sensing guarantees for random sparse signals.
This behavior motivated Calderbank et al.~\cite{calderbank2010construction} to introduce a notion of \textit{statistical restricted isometry} and construct several examples.
(See (\ref{Intro-StRIP}).)
In pursuit of a more detailed description of this phenomenon, Gurevich and Hadani~\cite{gurevich2008statistical} demonstrated that for appropriately incoherent ensembles, the spectrum of $G_S$ satisfies Wigner's semicircle law when $S \sim\mathsf{Unif}\binom{[n]}{k}$ with $k=o(m)$.
This can be viewed as an edge case of the
``ETF-MANOVA relation'',
\cite{haikin2017random,haikin2018frame},
that the spectrum of $G_S$ empirically satisfies a MANOVA law when $k\sim cm$ for some $c\in(0,1)$.
This observation also enjoys precursors in the context of random Parseval frames with various distributions~\cite{tulino2010capacity,farrell2011limiting,anderson2014asymptotically}.
More on that in Chapters~\ref{section-RMT-meets-frames} and~\ref{section-ETF-MANOVA-moments}.

\chapter{Random matrix theory} 
\label{section-RMT}
The main focus of this survey paper is a relation between subsets of deterministic frames and random matrix ensembles
\cite{haikin2017random,haikin2018frame}.
%
Random matrix theory (RMT) studies properties, and in particular spectral properties, of matrix-valued random variables. 
%
The 
empirical spectral distribution (ESD) of a symmetric (or Hermitian in the complex case) $n \times n$ matrix $G_n$, 
with eigenvalues $\lambda_1 \leq \ldots \leq \lambda_n$ (which are real-valued due to symmetry), 
is defined as 
\begin{equation}
    \label{ESD}
    \mu_n(x) = \mu_{G_n}(x) =
    \frac{1}{n} \sum_{i=1}^n \mathbf{1}_{\{ \lambda_i \leq x \} }
\end{equation}
where $\mathbf{1}_{E}$ is the indicator of the event $E$.
The cumulative distribution (\ref{ESD}) can be thought of as the integral of the ``empirical spectral density function'', 
%
\begin{equation}
    \label{ESD-density}
    f_n(x) 
    = \frac{d \mu_n(x)}{dx}
    = \frac{1}{n} \sum_{i=1}^n \delta(x - \lambda_i) 
\end{equation}
where $\delta(x)$ is Dirac's delta function.
%

Early results in RMT typically looked at a sequence of random matrices, say $X_n$, and asserted that the ESD of
(a normalized version and/or the Gram matrix of) $X_n$ converges in distribution to some specified limiting spectral distribution $\mu$. 
The theory came to the world in the 1950's with Wigner's {\em semi-circle law}, for the limiting ESD of
a symmetric matrix whose diagonal and upper triangular entries are i.i.d. random variables
\cite{feier2012methods}.

Another famous example, closer to the topic of our survey paper, is the Marchenko--Pastur law that states the following. 
Let $X_n$ be a sequence of random matrices, such that $X_n$ is $m_n$-by-$n$, and such that the entries of $X_n$ are all i.i.d of zero mean and unit variance.  
Assume that $m_n/n \to \beta$ as $n\to\infty$, with $\beta \geq 1$. 
In 1967, Marchenko and Pastur proved that 
the empirical distribution $\mu_{n}(x)$ 
of the eigenvalues
of the ``Wishart'' random matrix $X_n^\dagger X_n /m_n$ converges in distribution to the distribution $\mu_{\beta}(x)$
with density
\begin{equation}
    \label{MP_density}
f_{\beta}(x) = \frac{\sqrt{(\lambda_+ - x) (x - \lambda_-)}}
{2\pi \beta x }\cdot \mathbf{1}_{[\lambda_- \leq x \leq \lambda_+]}(x)    
\end{equation}
where the support edge points are 
$\lambda_\pm = (1\pm\sqrt{\beta})^2$.

A similar yet slightly less known result arises in multivariate statistics. 
Let $A$ and $B$ be $k$-by-$(n-m)$ and $k$-by-$m$ matrices, respectively, whose entries are Gaussian i.i.d. Consider the random matrix
\begin{equation}
    \label{MANOVA_ensemble}
\frac{n}{m}(AA^\dagger+BB^\dagger)^{-\frac{1}{2}}BB^\dagger (AA^\dagger+BB^\dagger)^{-\frac{1}{2}}    
\end{equation}
and denote its distribution by ${\rm MANOVA}(n,m,k)$ 
\cite{muirhead2009aspects}.
Wachter \cite{wachter1980limiting}  discovered that, as $k/m\to\beta\leq 1$ and $m/n\to\gamma$, the empirical distribution of the eigenvalues of a random matrix drawn from ${\rm MANOVA}(n,m,k)$ converges in distribution to the distribution with density 
\begin{align}
  \nonumber
f_{\beta,\gamma}(x) 
& = 
\frac
{\sqrt{(\lambda_+ - x) (x - \lambda_-)}}
{2\beta\pi x (1-\gamma x)}\cdot
\mathbf{1}_{[\lambda_- \leq x \leq \lambda_+]}(x)
\\
      \label{MANOVA_density}
& +
\left(
1 + \frac{1}{\beta} - \frac{1}{\beta \gamma} \right)^{+} \cdot
\delta \left ( x -\frac{1}{\gamma} \right) 
\end{align}
where 
$\lambda_\pm = [\sqrt{\beta(1-\gamma)} \pm \sqrt{1-\beta \gamma}]^2$,
which is known as {\em Wachter's asymptotic MANOVA distribution}.\footnote{The $x$-axis in (\ref{MANOVA_density}) is scaled by $\gamma$ compared to \cite{dubbs2015infinite}, due to the factor $n/m$ in (\ref{MANOVA_ensemble}) which does not exists in the ensemble definition in \cite{dubbs2015infinite}.}
Recently, \cite{erdHos2013local} 
proved convergence to this limiting spectral distribution even when the Gaussian assumption is removed, namely, when the entries of the matrices above are simply i.i.d, in analogy with the Wigner and the Marchenko--Pastur laws.

The MANOVA distribution (\ref{MANOVA_density}) reduces to the Marchenko--Pastur distribution (\ref{MP_density})
in the limit as $\gamma \rightarrow 0$.
A second interesting case is $\gamma = 1/2$, where the MANOVA density (\ref{MANOVA_density}) is symmetric around its center $(\lambda_+ + \lambda_-)/2$, and its centeralized version is known as the Kesten--McKay distribution
\cite{mckay1981expected,dubbs2015infinite}.

%
%
More recent results in RMT established connections between the MANOVA distribution and the spectrum of matrices with a reduced (non-i.i.d.) randomness. 
One example is a fixed sub-rectangle of a random matrix,
taken from a Haar distribution over all 
orthogonal/unitary matrices \cite{edelman2008beta}.
Another example is a random sub-rectangle (a random set of rows and columns) of the discrete Fourier transform matrix \cite{farrell2011limiting}.\footnote{Though the author of \cite{farrell2011limiting} did not identify the limiting law as the Wachter distribution.}
The MANOVA $(\beta, \gamma)$ parameters are determined by the fraction of rows/columns in the sub-rectangle.
We shall return to these examples, which are close in spirit to the ETF-MANOVA relation, in Chapter~\ref{section-RMT-meets-frames}.


%

\section{Methods of proof in RMT}
%
\label{section-RMT-methods}
In his original work, Wigner proved convergence of the moments of the ESD 
\cite{ZeitouniBook,feier2012methods}.
The $r$-th moment of a (cumulative) distribution $\mu(x)$
(with a density $f(x) = d\mu(x)/dx$) 
is defined as 
\begin{align}
\label{Md-mu}
M_r(\mu) & = 
\int x^r 
d \mu(x)
= \int x^r f(x) dx
\end{align}
for $r = 1,2,\ldots$.
The $r$-th moment 
of a symmetric/Hermitian matrix $G_n$ with an ESD $\mu_n$ (\ref{ESD}) 
is defined similarly as 
\begin{align}
\label{Md}
M_r(\mu_n) & = 
\int x^r 
d \mu_{n}(x)
 = \frac{1}{n} \sum_{i=1}^{n} \lambda_i^r 
 = \frac{1}{n} \tr{G_n^r}      
\end{align}
%
where the second equality follows from (\ref{ESD})--(\ref{ESD-density}), and the last equality follows because $(i)$ the eigenvalues of the power of a matrix are powers of the eigenvalues;
and $(ii)$ the trace of a matrix is equal to the sum of its eigenvalues.

The {\em moments method} aims to show that there exists some law $\mu$, such that $M_r(\mu_n)$ 
(which is a random variable for a random $G_n$) converges almost surely as $n$ goes to infinity to the moment $M_r(\mu)$
for all $r$, 
and to conclude that $\mu_n$ converges in distribution to $\mu$.
Typically, the analysis is split into two steps: 
$(i)$ convergence of the {\em mean} moment to the limiting moment, i.e., 
\begin{equation}
    \label{moment-method-mean}
    \mathbb{E} \{ M_r(\mu_n) \} \rightarrow M_r(\mu) 
\end{equation}
%
and $(ii)$ concentration, 
i.e, that the {\em variance} of $M_r(\mu_n)$ vanishes sufficiently fast in $n$,
say, 
that
\begin{equation}
    \label{moment-method-variance}
\sum_{n=1}^\infty  {\rm var} \{ M_r(\mu_n) \}  < \infty .
    \end{equation}
%
The two conditions (\ref{moment-method-mean}) and (\ref{moment-method-variance}) imply 
the almost sure convergence of $M_r(\mu_n)$ to $M_r(\mu)$; 
which, if held for all moments $r$, implies the desired convergence in distribution of the ESD to the limiting law $\mu$ 
\cite[exercise 2.4.6]{tao12}.

In Chapter~\ref{section-ETF-MANOVA-moments} we use the moment method to prove the asymptotic ETF-MANOVA relation.
The question of whether other proof strategies in random matrix theory, such as
the Stielties transform and free probability 
\cite{ZeitouniBook,feier2012methods}, may be useful in proving the ETF-MANOVA relation, remains an interesting open problem.

%
%
%
%
%



\chapter{Empirical ETF-MANOVA relation}
\label{section-RMT-meets-frames}
%
 In this and the next chapter, we shall see 
%
evidence for an intriguing RMT phenomenon that arises in a class of deterministic frames, 
which we call the {\em ETF-MANOVA relation}.  

Equiangular tight frames (ETFs) are highly symmetric objects constructed to minimize a {\em local} condition, namely, the pairwise absolute correlation (\ref{ETF_definition}). It stands to reason that such a highly symmetric structure will exhibit emergent {\em global} structure, namely, that some form of emergent structure can be found in the group correlations 
in subsets of vectors from an ETF. A natural place to look for such structure is the spectrum of the covariance matrix, 
which
determines, up to rotation, the global point arrangement in Euclidean space \cite{torgerson1952multidimensional}.

Is it possible that 
deterministic frames (like
ETFs) would demonstrate some regularity in the spectrum of covariance of subsets selected at random from the frame? 
Hints that this might be the case, and specifically that the MANOVA distribution may be a universal limit for the covariance spectrum, have appeared in the literature since 1980's in the case of tight frames of random design. Wachter's original paper \cite{wachter1980limiting} has noted that his limiting law arises as the limiting empirical spectral distribution of covariance matrix of Haar frames, namely, frames formed by taking sub-rectangles of Haar-distributed orthogonal matrices. This was formally proved by Edelman \cite{edelman2008beta} and Farrell \cite{farrell2011limiting}. In the same paper, Farrell showed that Wachter's distribution arises also as the limiting empirical spectral distribution of subsets taken from random Fourier frames, namely, frames obtained by taking a random subset of rows from the DFT matrix.

%


Another hint, 
pointing to the statistical properties of ETF subsets, appeared in the work in \cite{haikin2016analog}, which studied the toy example of Chapter~\ref{section-Intro-motivations-IT}.
This work was inspired by ideas on robust interpolation in the analog coding literature \cite{marshall1984coding}.
It observed that if a deterministic frame is generated by selecting rows from the DFT matrix corresponding to a {\em difference set} 
(which is a specific way to construct ETFs - see Section~\ref{section-difference-set}),
then the spectra of random subsets exhibit a typical behavior, which is favorable in terms of its noise amplification.  
%
%

The ETF-MANOVA relation was revealed and made explicit in \cite{haikin2017random}.
This work observed, through systematic analysis of massive empirical evidence, that the MANOVA distribution is in fact the universal limit - not only for Haar and random Fourier frames, but for a diverse family of 
ETFs, and even specific tight frames called ``near ETFs''.  
%
%
The work in \cite{haikin2017random} also observed that
this ``rabbit hole'' seems to go much deeper than just universality of the {\em limiting} covariance spectrum of frame subsets. It conclusively demonstrated that even the rate of convergence itself - both of the entire covariance spectrum and of numerous interesting functionals of the covariance spectrum
as described in Chapter~\ref{section-Intro-performance-measures} -  
is universal within the families under investigation: 
ETFs, and near ETFs (deterministic frames),
and Haar and random Fourier (random frames).
 
 

%
%
%

We shall next describe the empirical method and findings of \cite{haikin2017random}.



\section{Distance measures}
Let us consider two observables, one based on a measure of distance between empirical spectral distributions, and one based on the difference between specific test functions applied to the empirical spectral distributions.
%
Specifically, let 
\begin{equation}
    \label{KS}
\norm{\mu_1 - \mu_2}_{KS} 
\triangleq  
\max_x | \mu_1(x) - \mu_2(x) |
\end{equation}
%
denote the Kolmogorov--Smirnoff (KS) distance between two cumulative distribution functions (CDFs).
For an $m \times k$ sub-frame $F_S$ with a Gram matrix $G_S = F_S^\dagger F_S$,
taken from an $m \times n$ frame $F$,
define 
\begin{equation}
    \label{Delta_KS_frame_manova}
 \Delta_{KS}(G_S) 
 \triangleq
 \norm{\mu_{G_S}(x) - \mu^{\rm MANOVA}(x) }_{KS}     
\end{equation}
where 
$\mu_{G_S}$ is the ESD (\ref{ESD}) of $G_S$,
and $\mu^{\rm MANOVA}(x)$ is the MANOVA distribution with a density $f_{\beta,\gamma}(x)$  
(\ref{MANOVA_density}), where $\gamma = m/n$ and $\beta = k/m$.
%
This is one of many possible ways to measure the distance between 
the empirical spectral distribution of a given frame-subset covariance matrix $G_S$ and the limiting empirical spectral law.
Second, for a functional $\specstat$ that can be applied to an ESD, 
i.e., to the vector of eigenvalues of a Gram matrix, we let
\begin{equation}
    \label{Delta_Psi_frame_manova}
\Delta_\Psi(G_S) 
\triangleq
\Big|
\Psi(\mu_{G_S}) - \Psi(\mu^{\rm MANOVA}) 
\Big |
\end{equation}
where the functional of a general distribution $\mu$ (or a density $f = d\mu / dx$) is defined in a similar way,
i.e.,
\begin{equation}
    \label{functional_of_distribution}
    \Psi(\mu) = \int \Psi(x) d \mu(x)
    = \int \Psi(x) f(x) dx ;
\end{equation}
%
e.g., for the MSE goodness (\ref{signal-amplification-eigenvalues}) it is 
$\Psi_{\rm MSE}(\mu) = \int x^{-1} d \mu(x)$,
and for the (high-SNR) Shannon transform (\ref{shannon-transform}) it is $\Psi_{\rm Shannon}(\mu) = \int \log(x) d \mu(x)$.
(Note that $\int d \mu(x) = \int f(x) dx = 1$.)
See Chapter~\ref{section-Intro-performance-measures} for various functionals $\Psi$ under study.
Unlike the KS distance, which measures total distance between entire spectral distributions, the value $\Delta_{\Psi}$ measures  the difference as seen through the specific prism of $\Psi$.
%


\section{Frames under study}
 Table \ref{FramesTable} shows the frames studied in \cite{haikin2017random}.
 They include several frames of both random and deterministic design. 
 All frames studied were tight; some were tight but not equiangular, while some were ETFs. Both real and complex frames were studied. 
	\begin{table*}[htbp]
		\scriptsize
		\centering
		\caption{Frames under study} 
		\label{FramesTable}
		\renewcommand{\arraystretch}{0.85}%
		\begin{tabularx}{\textwidth}{XXcXXlX}
			\toprule
			Label & Name & $\mathbb{R}$ or $\mathbb{C}$ & Natural \newline $\gamma$ & Tight \newline frame &
			Equiangular & References \\
			\midrule
			& & & & &  & \\
			{\bf Deterministic \newline frames} & & & & &  & \\
			DSS & Difference-set \newline spectrum & $\mathbb{C}$  & & Yes & Yes & \cite{xia2005achieving}\\
			GF & Grassmannian frame & $\mathbb{C}$ & $1/2$& Yes & Yes &  \cite[Corollary 2.6b]{strohmer2003grassmannian}
			\\
			RealPF & Real Paley's construction & $\R$ & $1/2$ & Yes & Yes & \cite[Corollary 2.6a]{strohmer2003grassmannian}
			\\
			ComplexPF & Complex Paley's construction & $\mathbb{C}$ & $1/2$ & Yes & Yes & 
			\cite{paley1933orthogonal}
			\\
			Alltop & Quadratic Phase Chirp & $\mathbb{C}$ & $1/L$ & Yes & No &  
			\cite[eq. S4]{monajemi2013deterministic} 
			\\
			SS & Spikes and Sines & $\mathbb{C}$ & $1/2$ & Yes & No & \cite{Elad2010} \\
			SH & Spikes and Hadamard & $\R$  & $1/2$ & Yes & No & \cite{Elad2010}
			~ \\\hline  ~\\
			{\bf Random frames} & & & & &  & \\
			HAAR 
			& Unitary Haar frame & $\mathbb{C}$ & & Yes & No & \cite{farrell2011limiting,edelman2008beta} \\
			RealHAAR 
			& Orthogonal Haar frame & $\R$ & & Yes & No & \cite{edelman2008beta} \\
			RandDFT  
			& Random Fourier transform & $\mathbb{C}$ & & Yes & No & \cite{farrell2011limiting} \\
			RandDCT  
			& Random Cosine transform & $\R$ & & Yes & No &  \\
			
			\bottomrule
		\end{tabularx}
		\label{frames:tab}
	\end{table*}


\section{Empirical observation method}
For a specific frame family with a fixed aspect ratio $\gamma=m/n$, 
and a fixed sub-frame aspect ratio $\beta=k/m$, and
for each frame size $n$ under study, we have generated an $m \times n$ frame matrix of a given size and sampled frame subsets uniformly at random over $S \subset [n], |S|=k$. We then computed the values $\Delta_{KS}$, as well as  $\Delta_\Psi$ for all functionals $\Psi$ under study. This was repeated for increasing frame sizes available in the frame family. 
Separately, for each triplet $(n,m,k)$, we have performed independent draws
from the ${\rm MANOVA}(n,m,k)$ ensembles \eqref{MANOVA_ensemble} and calculated the analogous quantities.
To inspect the possibility of convergence to a limit of these quantities as the frame size grows, and the convergence rate, we have plotted the value of $\Delta_{KS}$, as well as  $\Delta_\Psi$  over the frame size. To observe an exponential decrease of the form $Cn^{-b}$ for some exponent $b$, simple linear regression of $\log(\Delta)$ against $\log(n)$ (with an intercept term) was performed to extract the value $b$ and its standard error. To observe an exponential decrease of the form $Cn^{-b}\log^{-a}(n)$ for some exponents $a$ and $b$, simple linear regression of $\log(\Delta)$ against $\log(n)$ and $\log(\log(n))$ (with an intercept) was performed to extract the values $a$ and $b$ and their standard errors.  
%


\section{Findings}
\label{section-RMT-ETF-findings}

	\begin{figure}[htbp]
		\centering
		\includegraphics[width=3in]
		{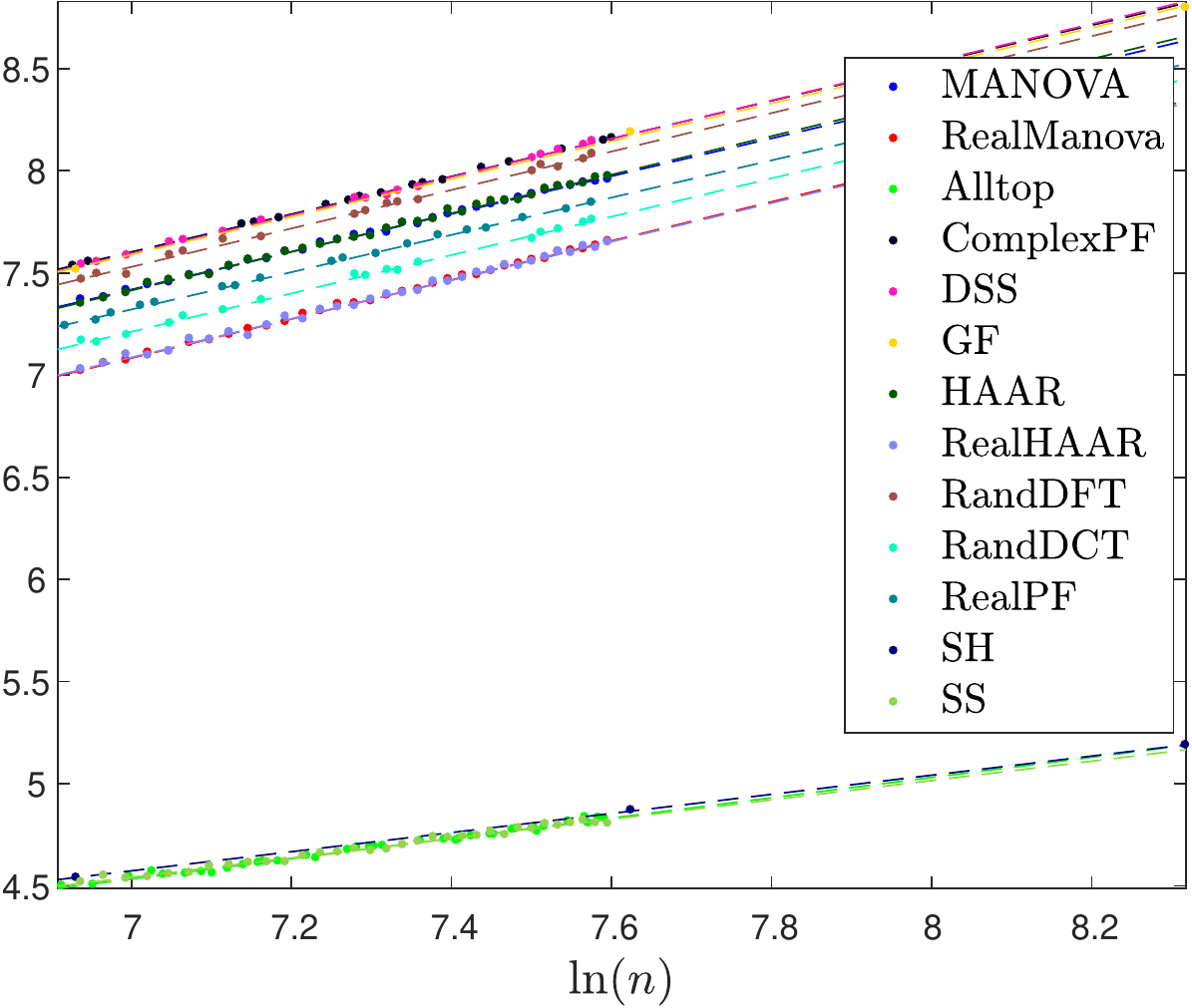}
		\caption{
		KS-distance (\ref{Delta_KS_frame_manova}) for $\gamma=0.5$ and $\beta=0.8$. Plot shows $-\frac{1}{2}\ln {\rm Var}_{K}(\Delta_{KS}(G_S^{(n)}))$ over $\ln(n)$. }
		\label{fig:KS1}
	\end{figure}


	\begin{figure}[htbp]
		\centering
		\includegraphics[width=3in]{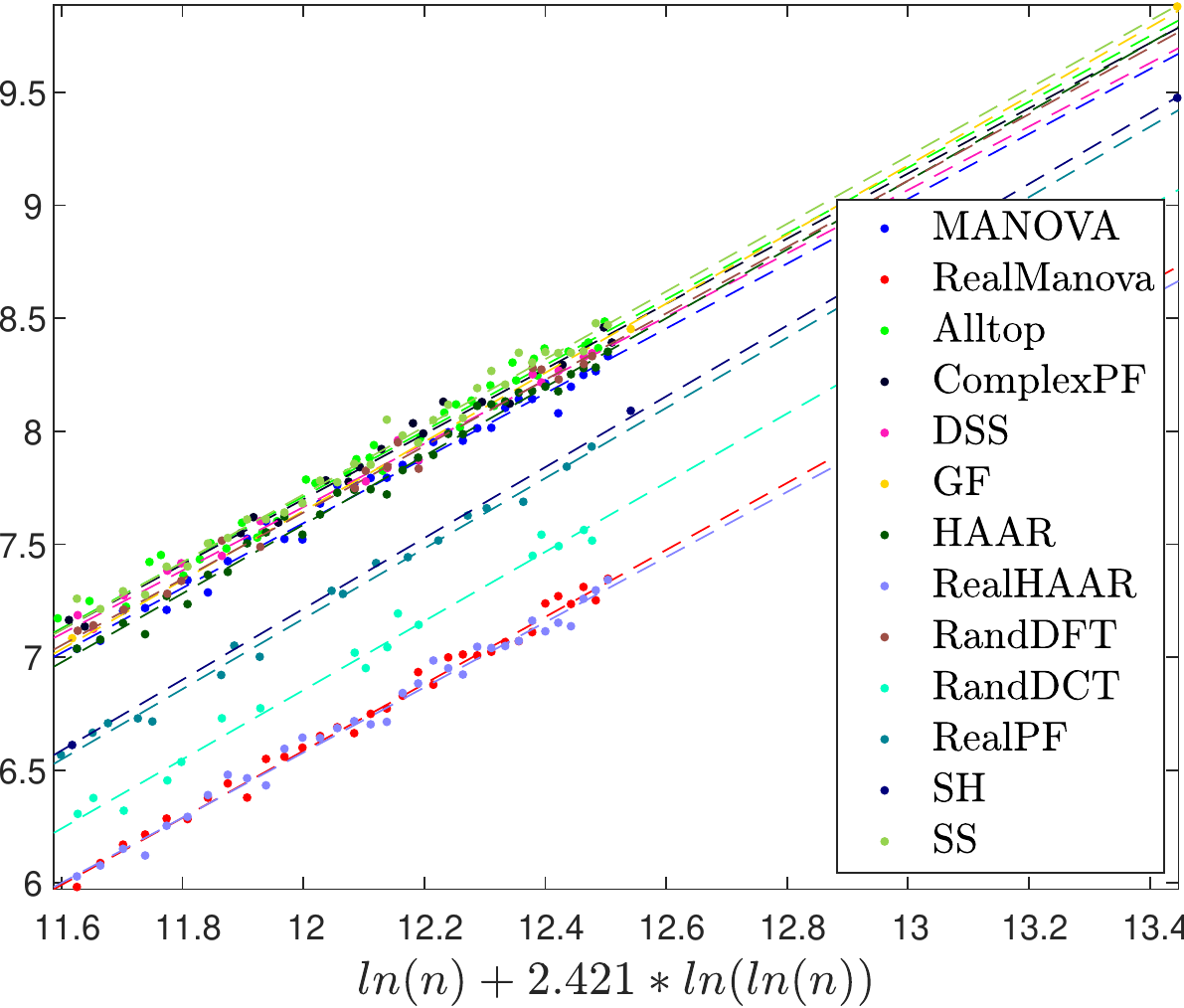}
		\caption{
		Functional distance (\ref{Delta_Psi_frame_manova}) for $\specstat_{\rm MSE}$, $\gamma=0.5$ and $\beta=0.8$. Plot shows $-\ln\E_{K}(\Delta_\specstat(G_S^{(n)})^2)$.
		}
		\label{fig:f_AC}
	\end{figure}

	
Figures~\ref{fig:KS1} and~\ref{fig:f_AC}, taken from \cite{haikin2017random}, show examples of these tests for both distance measures.  
The results demonstrate clearly that 
$\Delta_{KS}$
and $\Delta_{\Psi}$ 
decrease to zero. 
The fluctuations, namely the rate of convergence, 
show an excellent fit to the exponential rates. 
%
Moreover, for ETFs, we observe that the exponent $b$ measured in ETFs is statistically indistinguishable from the exponent measured for the MANOVA distribution, in the sense that the null hypothesis that they are identical cannot be rejected.
For the deterministic near-ETF cases (SS, SH and Alltop),
the slope in Figure~\ref{fig:KS1} (KS distance) is slightly smaller,
while the slope in Figure~\ref{fig:f_AC} (functional distance) is statistically indistinguishable from that of the MANOVA ensemble.
These empirical evidence lead us to the striking conclusion that not only is the MANOVA limit universal, but also the {\em convergence rate} to the MANOVA limit appears to be universal within the ETF family, and almost universal in the larger class of near ETFs. 







\chapter{Moments of an ETF subset}
%
\label{section-ETF-MANOVA-moments}
%
%
%
The empirical ETF-MANOVA relation 
of Chapter~\ref{section-RMT-meets-frames}
leaves several far-reaching conjectures for further theoretical study and explanation. 
%
In the spirit of information theory, 
we turn to focus our attention on  the {\em typical asymptotic} behavior of an ETF subset.

In this chapter we demonstrate how the classical moments method of random matrix theory,
\cite{ZeitouniBook,feier2012methods}, 
described in Chapter~\ref{section-RMT}, can be used to
prove analytically the convergence of the 
empirical spectral distribution 
(\ref{ESD})
of a randomly-selected subset of an ETF to 
Wachter's asymptotic MANOVA distribution (\ref{MANOVA_density}).
The proof is complete for $\gamma=1/2$, \cite{magsino2020kesten}, and is still partial for a general $0<\gamma<1$, \cite{haikin2018frame,MarinaThesis,haikin2021moments}.
Similar tools were used in \cite{babadi2011spectral} to establish convergence of the spectra of subsets of binary code-based frames to the Marchenko-Pastur distribution. 


%
For any $m \times n$ (deterministic) unit-norm frame $F$, 
consider the $m \times m$ random matrix
\begin{equation}
    \label{subframe_hessian}
F P F^\dagger,     
\end{equation}
where 
the randomness comes from the ``selection matrix'' $P$.
That is, 
\begin{equation}
    \label{selection_matrix}
P = {\rm diag}[P_1, \ldots, P_n]    
\end{equation}
is a diagonal matrix with $\{0,1\}$ elements on its diagonal, 
where 
$P_i = 1$ if $i$ is in the (random) subset $S$, and $P_i=0$ otherwise.
%
Specifically, we shall assume that the $P_i$'s are generated Bernoulli($p$),
\begin{equation}
    \label{bernoulli-p-selection}
    P_i = \left\{
\begin{array}{ll}
1,     &  \mbox{with probability $p$,} \\
0,     &  \mbox{with probability $1-p$;}
\end{array}
\right.    
\end{equation}
in other words, each frame vector is selected independently with probability $p$.
Compared with the  combinatorial ($n$ choose $k$) selection of the subset $S$ assumed in the empirical results of Chapter~\ref{section-RMT-meets-frames},
Bernoulli selection $(i)$ fits the information-theoretic i.i.d. model of Chapter~\ref{section-Intro-motivations-IT}, $(ii)$ is easier for analysis, 
and $(iii)$ does not seem to affect the asymptotic results. 
%
%

Since in (\ref{subframe_hessian}) all columns $i \not\in S$ 
of $F$ are zeroed, we have 
\[
F P F^\dagger = F_S F_S^\dagger
\]
which is the Hessian of the selected sub-frame $F_S$ (\ref{frame_subset}).
By properties of the trace operator, 
the trace-power of the Gram matrix is equal to the trace-power of the Hessian:  
\[
\tr{ (F_S^\dagger F_S)^r } = \tr{ (F_S F_S^\dagger)^r } ,
\]
for any power $r = 1,2, \ldots$. 
Hence, in light of the moment method (\ref{Md}), 
we define the expected $r$th moment of a randomly-selected subset from a frame $F$, 
as  
\begin{equation} 
\label{moment_r}
m_r 
\triangleq 
\frac{1}{n}\mathbb{E}\left[\mbox{trace} \left((FPF^\dagger)^r\right)\right] 
\end{equation}
%
where the expectation $\mathbb{E}[\cdot]$ is over the diagonal (Bernoulli $p$) elements of $P$.
\footnote{For mathematical convenience, 
we normalize by the frame size $n$, rather than the frame dimension 
$m = \gamma n$, or the sub-frame (expected) size 
$k = |S| = p n$.} 

\section{Low-order moments: exact analysis} 
%
\label{section-ETF-MANOVA-moments-d1to4}
For a small moment order $r$, we can compute the exact (non-asymptotic)
mean (\ref{moment_r}) and variance of the sub-frame $r$th moment, for any ETF,
as a function of $m$ and $n$ (for which an $m \times n$ ETF indeed exists).
Furthermore, we can compare the mean to the $r$th moment of Wachter's MANOVA distribution 
\cite[table 4]{dubbs2015infinite}:\footnote{In \cite{dubbs2015infinite} the normalization outside the moment (\ref{moment_r}) is $pn$ instead of $n$, while the $x$-axis is normalized by $\gamma$, so overall their moment is $\gamma^r/p$ times the moment in (\ref{MANOVA-moment}).}
%
%
\begin{align}
   \nonumber
     m^{\rm MANOVA}_r & (\gamma,p) = p(x+1)^{r-1}  \\
      \label{MANOVA-moment}
     & - (1-p)^2\sum_{j=0}^{r-2}(1-p)^j(x+1)^{r-2-j}N_{j+1} \left(x \frac{p}{1-p} \right)
\end{align}
for $0 \leq \gamma, p \leq 1$, and $r = 1,2,\ldots$,
where $N_j(q) = \sum_{i=1}^{j}N(j,i)q^i$ is the Narayana polynomial,
$N(j,i)=\frac{1}{j}\binom{j}{i}\binom{j}{i-1}$ is the Narayana number,
and
\begin{equation}
    \label{x_definition}
 x \triangleq 
\frac{n}{m} -1  
= 
 \frac{1}{\gamma}-1   
\end{equation}
is the ``net redundancy''.
%
%
%
%
%
For $r = 1,\ldots,4$, 
the MANOVA moments (\ref{MANOVA-moment}) become
\begin{subequations}
\label{Manova_moments}
\begin{align} 
\label{Manova_moments1}
m^{\rm MANOVA}_1(\gamma,p) &= p\\
\label{Manova_moments2}
m^{\rm MANOVA}_2(\gamma,p) &= p+p^2x\\
\label{Manova_moments3}
m^{\rm MANOVA}_3(\gamma,p) &= p+p^23x+p^3(x^2-x)\\
m^{\rm MANOVA}_4(\gamma,p) &= p+p^26x+p^3(6x^2-4x)+p^4(x^3-3x^2+x) .
\label{Manova_moments4}
\end{align}
\end{subequations}
Let us define:
\begin{equation} 
\label{delta EWBd}
\Delta(\gamma,p,r,n)\triangleq
\begin{cases}
0,& r=2,3  \\
p^2(1-p)^2\frac{x^2}{n-1},& r=4.  \\
\end{cases}
\end{equation}


\vspace{5mm}

\begin{theorem}[Expected moments \cite{haikin2018frame,MarinaThesis}] 
\label{th_m_1_4}
If $F$ is any $m$-by-$n$ unit-norm (real or complex) ETF (\ref{ETF_definition}), 
then the expected sub-frame $r$th moment (\ref{moment_r}) is given by
	\begin{equation}
	\label{moment1-4_th}
	m_r^{\rm ETF}  = m^{\rm MANOVA}_r(\gamma,p) +\Delta(\gamma,p,r,n)
		\end{equation}
for $r=1,2,3$ and 4. 
\end{theorem}


%
Note that the second term in (\ref{moment1-4_th}) is either zero or it  vanishes asymptotically as $n$ goes to infinity.
%
Therefore, the asymptotic $r$th moment of an ETF sequence with aspect ratio $\gamma$ is equal to $m^{\rm MANOVA}_r(\gamma,p)$, for $r=1,2,3$ and $4$,
in line with the empirical results of \cite{haikin2017random} (Chapter~\ref{section-RMT-meets-frames}). 


\begin{proof}
For any frame $F$, the trace-power in (\ref{moment_r}) can be written as 
$\tr{ (P F^\dagger F)^r }$,
where $P$ is given in (\ref{selection_matrix}), 
and $F^\dagger F$ is the Gram matrix of the frame. 
%
%
Now, recall from linear algebra that the trace of a matrix product is given by sums of cyclic chains of products of  elements.
In particular, if $A^{(1)},\ldots,A^{(r)}$ are $n \times n$ matrices with elements 
$a^{(k)}_{ij}$, then
\begin{align}
\label{trace_matrix_product}
\tr{ A^{(1)} \cdots A^{(r)} }=\sum_{i_1=1}^n \ldots \sum_{i_r=1}^n
a^{(1)}_{i_1 i_2} a^{(2)}_{i_2 i_3} \cdots a^{(r)}_{i_r i_1} .
\end{align}
Thus, the mean $r$th moment (\ref{moment_r})
involves sums of length-$r$ chains of products of cross correlations $\langle \bff_i, \bff_j \rangle$, and powers of $p$ depending on how many distinct pairs 
$i \not= j$ appear in the chain. 
%

In the case where $F$ is an ETF, the Gram matrix (\ref{intro-gram-matrix}) can be written in the form 
\begin{equation}
    \label{Gram-Seidel}
F^\dagger F = \{ \langle \bff_i, \bff_j \rangle \} = I_n + \mu(F) S    
\end{equation}
as in (\ref{Seidel_matrix}),
where $S$ is either the Seidel adjacency matrix  (\ref{Seidel_matrix_pentagon}) in the real ETF case
(with $\pm 1$ on the off diagonal elements), 
or a ``generalized adjacency matrix'' (with phase values on the off diagonal elements) in the complex ETF case.
The first moment ($r=1$) 
is thus given by 
\begin{align}
m_1^{\rm ETF}
& = 
\frac{1}{n}\mathbb{E}\left[ \sum_{i=1}^{n}\langle \bff_i, \bff_i \rangle P_i\right]
=\frac{1}{n}\mathbb{E}\left[ \sum_{i=1}^{n}P_i\right]
\label{m1}
=\frac{1}{n}\sum_{i=1}^{n}p = p    
\end{align}
%
%
where we used the unit-norm condition 
$\langle \bff_i, \bff_i \rangle = 1$, 
and $\mathbb{E}\left[ P_i\right] = p$.
The second moment ($r=2$) is given by
%
\begin{subequations}
\label{m2}
\begin{align}
\label{m2aa}    
    m_2^{\rm ETF} 
    &= \frac{1}{n}\mathbb{E}\left[ \sum_{i,j = 1}^{n}|\langle \bff_i, \bff_j \rangle|^2 P_i P_j\right]
    \\ \label{m2a}
        &= p+p^2\frac{1}{n}\sum_{i\neq j}^{n}|\langle \bff_i, \bff_j \rangle|^2
    \\ 
    \label{m2b}
    &= p+p^2x
    \end{align}  
\end{subequations}
where in (\ref{m2a}) we used the unit-norm condition for $i=j$, and $E[P_i P_j] = p^2$ for $i\not=j$;
and in (\ref{m2b}) we used the fact that the (mean-squared) Welch bound \eqref{WB} amounts to 
\begin{equation} 
\label{WB_x}
\begin{split}
\frac{1}{n}\sum_{i\neq j}^{n}|\langle \bff_i, \bff_j \rangle|^2\ge 
\frac{1}{\gamma}-1
=
x
\end{split}
\end{equation}
with equality for a tight frame. 
The full computation in \cite{haikin2018frame,MarinaThesis} shows that tightness is sufficient also for the third moment $m_3$ in (\ref{moment1-4_th}),
while the ETF condition (\ref{ETF_definition}) is {\em necessary} for the fourth moment $m_4$ in (\ref{moment1-4_th}).  
See Chapter~\ref{section-ETF-super}.
\end{proof}


The {\em empirical spectral variance} (ESV) is defined as the variance of the ESD, given by the second moment of the ESD minus the squared first moment of the ESD.
This should not be confused with the {\em variance of the moment} as a random variable, which is considered in the next theorem.



\vspace{5mm}

\begin{theorem}[Moment variance]
\label{th_var_1_2}
The variance of the $r$th moment 
\begin{equation}
    \label{moment_variance}
V_r \triangleq {\rm Var} \left[ \frac{1}{n} \tr{(F P F^\dagger)^r} \right]    
\end{equation}
%
is given for an ETF for $r = 1$ and 2 by
	\begin{align}
	\label{var1}
	V_1^{\rm ETF}  & =  \frac{p - p^2}{n}
	\\
	\label{var2}
	V_2^{\rm ETF} & = \frac{1}{n} \left[ 
	p + t_2 p^2 + t_3 p^3 + t_4 p^4 \right]
	\end{align}
	where $t_2 = -1 + 4x + 2x^2/(n-1)$, $t_3 = 4x [ -1 + x(n-2)/(n-1) ]$, and $t_4 = x^2 (6-4n)/(n-1)$, and where $x$ is the net redundancy (\ref{x_definition}).
	Equality (\ref{var1}) holds for any unit-norm frame $F$, 
	while (\ref{var2}) holds only if $F$ is an ETF.
\end{theorem}


\begin{proof}
    See Appendix~\ref{section-variance-proof}.
    \end{proof}


Note that both variances in Theorem~\ref{th_var_1_2} vanish as $O(1/n)$;
and that they are equal if $\gamma=1$ ($x=0$), in which case the sub-frame spectrum is supported only on 0 and 1.
Returning to the empirical spectral variance mentioned before the theorem,
since both $m_1$ and $m_2$ are asymptotically deterministic, it follows that the ESV of an ETF is given asymptotically by $m_2 - m_1^2 = p + p^2 x - p^2$.
See Chapter~\ref{section-ETF-super}.


%
%


\section{General moments: asymptotic analysis} 
\label{section-ETF-MANOVA-moments-NewNew}
%
%
The exact expression for the expected $r$th moment (\ref{moment_r}) of an ETF subset 
becomes complicated for $r>4$, 
and it contains more terms (beyond (\ref{delta EWBd})) that vanish as the frame size $n$ goes to infinity.
These vanishing terms might depend also on the specific ETF construction, and not only on $n$ and the ratios $\gamma$ and $p$. 
We next focus our attention on the non-vanishing terms,
in order to apply the asymptotic moment method (\ref{moment-method-mean})-(\ref{moment-method-variance}) of RMT.

Since an $m \times n$ ETF does not exist for every $m,n$ pair, 
we look on a subsequence of dimensions 
$n_1, n_2, \ldots$,
and a corresponding sequence of $m_i$'s for which $(i)$ an $m_i \times n_i$ ETF exists, and $(ii)$ 
the aspect ratios $m_i/n_i$ converge to some fixed $0<\gamma<1$.
For such a sequence $F_1, F_2, \ldots$ of ETFs, the selection probability $0 \leq p \leq 1$ induces a corresponding sequence of expected $r$th moment $m_r^{(n_1)}, m_r^{(n_2)}, \ldots$
according to (\ref{moment_r}).
Let us define the asymptotic $r$th moment
\begin{equation}
     \label{ETF-d-moment-asymptotic}
     m_r^{(\infty)}({\rm ETF},\gamma,p) 
     \triangleq
     \lim_{i \to \infty} 
     m_r^{(n_i)} 
\end{equation}
provided that this limit exists and is unique (independent of the specific ETF sequence).

\subsection{Recursive computation of moments}
We next define a recursive procedure for computing
$m_r^{(\infty)}({\rm ETF}, \gamma, p)$,
for $r = 1, 2, \ldots$.
We start by defining a real-valued sequence $A_1, A_2, A_3, \ldots$,
by its two initial values 
\begin{equation}
    \label{A-sequence-initial}
A_1 = \frac{2-\gamma^{-1}}{2 \sqrt{\gamma^{-1}-1}} \  , \ \ \ \
A_2 = 1 , 
\end{equation}
%
and the recursive relation: 
\begin{equation}
    \label{A-sequence}
    A_{s+1} = \sum_{i=1}^{s} A_i A_{s+1-i}  \ ,  \ \ s = 2, 3, \ldots,  
\end{equation}
for $0 < \gamma < 1$.
Note that for $\gamma = 0.5$ 
all the odd elements (including $A_1$) are zero.

Second, given a moment order $r$, define $\Pi(r,t)$
as the set of partitions of $[r]$ into $t$ sets, for $1 \leq t \leq r$.\footnote{In view of the trace-product formula (\ref{trace_matrix_product}), $\Pi(r,t)$ amounts to all cases with $t$ distinct indices in the expression for the $r$th moment (\ref{moment_r}). 
}
Note that the size of $\Pi(r,t)$ is exponential in $r$, for $1 < t < r$.
The extreme cases
$\Pi(r,1)$ and $\Pi(r,r)$ contain only one partition:
the whole set $(1 2 \ldots r)$, and singletons (1,2,\ldots,r), respectively.
As examples for the general case,
$\Pi(4,2)$ contains seven partitions (four with one singleton and one triplet, and three with two doubles):
(1,234), (2,134), (3,124), (4,123), (12,34), (13,24), (14,23);
while
$\Pi(5,3)$ contains twenty five partitions 
(ten with two singletons and one triplet, fifteen with two doubles and one singleton).
%
%
%
A partition is called a ``crossing partition'' if for some two sets $\cA$ and $\cB$ in $\pi$,
and four elements $a_1, a_2 \in \cA$ and $b_1, b_2 \in \cB$,
we have $a_1 < b_1 < a_2 < b_2$.
In the first example above (13,24) is a crossing partition,
while in the second example above 
(13,24,5), (13,25,4), 
(14,25,3), (14,2,35), 
(24,1,35)
are crossing partitions
\cite{nica2006lectures}.
%
%

\begin{figure}[htbp]
	\centering
		\includegraphics[width=4in]{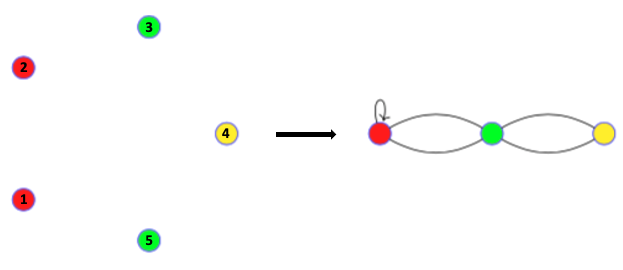}
    \caption{$G(\pi)$ of the partition $\pi = (12,35,4)$}
	\label{graph}
\end{figure}

For each non-crossing partition $\pi \in \Pi(r,t)$,
define an associated graph $G(\pi)$ with $r$ edges and $t$ vertices, where each pair of consecutive elements in $[r]$ are connected with an edge
in a cyclic manner,
i.e., $1 \rightarrow 2$, $2 \rightarrow 3$, \ldots, $r \rightarrow 1$, 
and each subset of $\pi$ is a vertex.
For example, the graph of $\pi = (12,34)$ has two vertices (12) and (34), 
with two edges going between them, and two loops = an edge from each vertex to itself.   
The non-crossing property implies that 
the graph $G(\pi)$ has the shape of a ``cactus'', \cite{magsino2020kesten}, 
hence it can be decomposed into cycles,
and we denote by ${\rm cyc}(\pi) = \{ s \}$ the set of lengths $s$ of cycles in $G(\pi)$.
Note that the sum of $s \in {\rm cyc}(\pi)$
is equal to the number of edges $r$.
Figure \ref{graph} shows the graph of the (non-crossing) partition $\pi = (12,35,4)$ in $\Pi(5,3)$, 
which contains two cycles of length 2 and one cycle of length 1 (loop),
hence $2+2+1 = 5 = r$.

Now, given both $0 < \gamma < 1$ and $r$, 
the ``value'' of a cycle of length $s$ is defined as $A_s$  (\ref{A-sequence}).
The ``value'' of a non-crossing partition $\pi$ is defined as the product of the values of its cycles, 
while the ``value'' of a crossing partition is defined as zero:
\begin{equation}
    \label{V*_pi}
V^*(\pi,\gamma) = 
\left\{
\begin{array}{ll}
\prod_{s \in {\rm cyc}(\pi)} A_s ,   &  \mbox{$\pi$ = non-crossing,}
\\
0 ,     &  \mbox{$\pi$ = crossing,}
\end{array}
\right.
\end{equation}
where $\prod$ denotes product.
%
For example, 
the value of the partition in Figure~\ref{graph} is 
$V^*((12,35,4),\gamma) = A_1 A_2^2$.
In the extreme case of $t=1$,  
there is a unique $\pi = (12 \ldots r)$, and $G(\pi)$ contains one vertex with $r$ loops, 
so $V^*(\pi,\gamma) = A_1^r$.
In the other extreme of $t=r$, 
there is a unique $\pi = (1,2, \ldots, r)$, and $G(\pi)$ contains one cycle between $r$ vertices, 
so $V^*(\pi,\gamma) = A_r$.

Finally, for a given probability $0 < p < 1$, we use the formula (\ref{V*_pi}) for the value of partitions to define the centralized quantity:
\begin{equation}
    \label{ETF-d-moment-centralized-formula}
    m^{\rm central}_r(\gamma,p) = \sum_{t=1}^r \Bigl( \sum_{\pi \in \Pi(r,t)} V^*(\pi,\gamma)  \Bigr) p^t ,
\end{equation}
from which we compute the decentralized quantity:
\begin{align}
    \label{ETF-d-moment-non-centralized-formula}
    m^*_r(\gamma,p) &= 
    \left(\frac{\gamma^{-1}}{2} \right)^r \cdot p   
    +
\sum_{j = 1}^r {r \choose j} \cdot K_{r,j} 
    \end{align}
where
\begin{equation}
      \label{ETF-d-moment-non-centralized-formula2}
K_{r,j} = 
(\gamma^{-1}-1)^{j/2}  \cdot \left( \frac{\gamma^{-1}}{2} \right)^{r-j} 
\cdot m^{\rm central}_j(\gamma,p)  .      
\end{equation}

The relation between the decentralized quantity 
(\ref{ETF-d-moment-non-centralized-formula}), 
which is the output of the recursive computation procedure above, and the $r$th moment of a random subset of an ETF, is stated in the next theorem.

\vspace{5mm}

\begin{theorem}[Recursive formula for moments \cite{magsino2020kesten,haikin2021moments}]
\label{Thm-Marina_New}
%
Let $m_r^{(n_i)} = (1/n_i) E \left[ \tr{  (F_i P_i F_i^\dagger )^r  }\right]$ denote the $r$th moment (\ref{moment_r}) of a random subset of the frame $F_i$,
where 
$F_i$, $i=1, 2, \ldots$ is any sequence of $m_i \times n_i$ ETFs of increasing dimensions with a limiting aspect ratio 
$m_i / n_i \rightarrow \gamma$ as $i \rightarrow \infty$.
Then, 
the sequence $m_r^{(n_1)}, m_r^{(n_2)}, \ldots$
has a limit $m_r^{(\infty)}({\rm ETF}, \gamma, p)$
(see (\ref{ETF-d-moment-asymptotic})),
and this limit is equal to the output (\ref{ETF-d-moment-non-centralized-formula}) of the recursive computation procedure above: 
\begin{equation}
\label{ETF-moments-in-theorem}
m_r^{(\infty)}({\rm ETF}, \gamma, p)
=
m^*_r(\gamma,p) . 
\end{equation}
%
%
\end{theorem}

The proof of Theorem~\ref{Thm-Marina_New} for real-valued redundancy-2 ($\gamma = 1/2$) ETFs appears in \cite{magsino2020kesten}.
The extension 
to general ($0<\gamma<1$, complex) ETFs 
was established during the writing of this survey paper, and
it appears in \cite{haikin2021moments}.
%

\subsection{The redundancy-$2$ case}
\label{section-ETF-moments-redundancy2case}
%
The work in \cite{magsino2020kesten} established a closed-form expression for $m^*_r(1/2, p)$, the $\gamma = 1/2$ case of the right hand side of (\ref{ETF-moments-in-theorem}), using special properties that hold in this case:
(i) the relation $n_i = 2 m_i$ holds {\em exactly} for each $i$ for an infinite sequence of Paley ETFs;
(ii) the Gram-Seidel adjacency matrix relation (\ref{Seidel_matrix}),  (\ref{Gram-Seidel}) is particularly simple;
(iii) the MANOVA$(1/2,\beta)$ distribution is symmetric around its mean;
(iv) its centralized version is equal to the Kesten-McKay distribution \cite{dubbs2015infinite};
and (v)
the $A_s$'s in (\ref{A-sequence}) are zero for odd $s$, and 
\[
A_s = C_{s/2 - 1}
\]
for even $s$,
where $C_i = \frac{1}{i+1} {2i \choose i}$, $i=1,2,\ldots$ 
are the Catalan numbers. 
Luckily,
the inner sum in (\ref{ETF-d-moment-centralized-formula})
(sum of products of Catalan numbers over all partitions in $\Pi(r,t)$), 
has an explicit closed form.
Moreover, after the outer summation,
(\ref{ETF-d-moment-centralized-formula}) coincides with the Kesten-McKay moments \cite{mckay1981expected}. 
This result is summarized in the following theorem.

\vspace{5mm}

\begin{theorem}
\label{Thm-PaleyKM-mean}
{\bf (Paley-Kesten-McKay relation \cite{magsino2020kesten})}
%
For $\gamma = 1/2$ we have
\begin{subequations}
\begin{equation}
m^{\rm central}_r(1/2,p) = m^{\rm KM}_r(p) 
\end{equation}
and therefore
\begin{equation}
m^*_r(1/2, p) = m^{\rm MANOVA}_r(1/2, p)    
\end{equation}
\end{subequations}
%
where 
$m^{\rm central}_r(\gamma,p)$ 
and $m^*_r(\gamma, p)$
are the centralized and non-centralized quantities in (\ref{ETF-d-moment-centralized-formula}) and (\ref{ETF-d-moment-non-centralized-formula}), respectively,
and where
$m^{\rm KM}_r(p)$ 
and $m^{\rm MANOVA}_r(\gamma, p)$
are the $r$th moments of the
Kesten-McKay distribution \cite{mckay1981expected}, \cite[table 4]{dubbs2015infinite}
and the MANOVA distribution (\ref{MANOVA-moment}), respectively.
%
\end{theorem}

Theorems~\ref{Thm-Marina_New} and~\ref{Thm-PaleyKM-mean} imply that for a sequence of redundancy-2 ($\gamma=1/2$) ETFs of growing dimensions, the mean $r$th moment converges to the $r$th MANOVA moment:
\begin{equation}
m_r^{(\infty)}({\rm ETF}, 1/2, p) 
= m^*_r(1/2, p) 
= m^{\rm MANOVA}_r(1/2, p) .
\end{equation}
%

The status of the analysis for a general $0<\gamma<1$ is somewhat subtle.
While the recursive formula (\ref{ETF-moments-in-theorem}) for the asymptotic mean $r$th moment is fully proved, 
\cite{haikin2021moments},
we are still missing a closed-form expression for $m^*_r(\gamma,p)$ 
(the right hand side of (\ref{ETF-moments-in-theorem})).
Thus, the identity 
\begin{equation}
    \label{MarinaNew=Manova?}
 m^*_r(\gamma,p)  \stackrel{?}{=}  m^{\rm MANOVA}_r(\gamma,p)   
\end{equation}
remains as a conjecture, which, together with the identity (\ref{ETF-moments-in-theorem}), would imply 
the desired result: 
\begin{equation}
    \label{MarinaNew=Manova?B}
    m_r^{(\infty)}({\rm ETF}, \gamma, p)   \stackrel{?}{=}  m^{\rm MANOVA}_r(\gamma,p) .   
\end{equation}
We used a symbolic computer program to verify that (\ref{MarinaNew=Manova?}) indeed holds (for all $0<\gamma<1$) 
for a finite sequence of moments $r= 1,2,\ldots, r_m$.\footnote{
The computation complexity is high due to the exponential growth 
of the set $\Pi(r,t)$ with $r$. 
So far we were able to verify (\ref{MarinaNew=Manova?}) for $r = 1,\ldots, 10$, and we intend to extend it further
using parallel (``cloud'') computation.
} 
%


%
%
%


\section{Concentration} 
\label{section-ETF-MANOVA-moments-concentration}
%
%
The moment method (Section~\ref{section-RMT-methods}) requires to establish two 
facts in order to conclude the desired convergence of the ESD of sub-frames
to the limiting MANOVA law:
($i$) that the {\em mean} moments  
converge to the corresponding moments of the MANOVA law (\ref{moment-method-mean}); and
($ii$) that the moments {\em concentrate} 
(\ref{moment-method-variance}). 
While the first fact was partially established for ETFs in Section~\ref{section-ETF-moments-redundancy2case} (the proof is complete for $\gamma=1/2$, and ``almost'' complete for a general $0<\gamma<1$),
the second fact holds for general {\em tight} frames (not necessarily ETFs),
provided that their subset mean centralized moments converge.
Specifically,
the analysis of Paley (redundancy-2) ETFs in \cite[Thm.~1]{magsino2020kesten} 
directly gives 
the following result.



\begin{theorem}[ESD concentration]
\label{thm-concentration}
Fix a bounded probability distribution $\mu$ on the real line, parameters $\gamma,p\in(0,\tfrac{1}{2})$, a lacunary sequence $L\subset\mathbb{N}$, and a sequence $\{F^{(n)}\}_{n\in L}$ of $
[\gamma n] \times n$ unit norm tight frames. 
Suppose that when drawing random subsets $S_n\subseteq[n]$ with $\operatorname{Bernoulli}(p)$ elements, the random subensembles $H_n:=F_{S_n}^{(n)}$ exhibit convergence in expectation of the centralized $r$th moments:
\[
\mathbb{E}[\tfrac{1}{pn}\operatorname{tr}[(\tfrac{1}{p}(H_n^\dagger H_n-I))^r]\to\int_\mathbb{R} x^r d\mu(x)
\]
for each $r \in \mathbb{N}$. 
Then, the empirical spectral distribution of $\frac{1}{p}(H_n^\dagger H_n-I)$ converges almost surely to $\mu$.
\end{theorem}

\vspace{10mm}

The concentration result of Theorem~\ref{thm-concentration}, 
together with a confirmation of the currently open conjecture (\ref{MarinaNew=Manova?B}),
would establish 
an analytic proof for the asymptotic behavior predicted by the empirical observations of Chapter~\ref{section-RMT-meets-frames}.
Namely, the spectrum of a typical subset from a general ETF converges almost surely to a MANOVA spectrum:
%
\begin{equation}
    \label{ETF-MANOVA-conjecture}
\mu^{(n)}_{G_S} \rightarrow \mu^{\rm MANOVA}_{\gamma,\beta}
\end{equation}
as $n$ goes to infinity, 
where $G_S = F_S F_S^\dagger$ is the Gram matrix of a sub-frame $F_S$ selected with probability $p$ (\ref{bernoulli-p-selection}); \
$\mu^{(n)}_{G_S}$
is the sub-frame ESD
(\ref{ESD}); \ 
$\gamma$ and $\beta$ are the asymptotic aspect ratios (\ref{aspect-ratios}) of the sequence of frames $F^{(n)}$ 
and sub-frames $F_{S_n}^{(n)}$, respectively; and $\mu^{\rm MANOVA}_{\gamma,\beta}$ is the MANOVA distribution whose density is given in (\ref{MANOVA_density}).
%



\chapter{Sub-frame inequalities} 
\label{section-inequalities}
%
In the applications discussed in Chapters~\ref{section-Introduction} and~\ref{section-Intro-motivations-IT},
the frame dimension $m$ 
is a design parameter that has an optimal value.
On the one hand, the more {\em rectangular} is the subset
($m > k$ in source coding with erasures, $m < k$ in vector
Gaussian channels), the 
more compact is its typical eigenvalue spread.
On the other hand, the deviation from a {\em square} subset corresponds to wasting degrees of freedom: 
more DFT coefficients need to be quantized in source coding with erasures, 
less information symbols can be transmitted over a MIMO channel with ``0/1 fading'',
or less users can be active in non-orthogonal multiple access.     

In this chapter we focus on the former effect, 
i.e., on the advantage of a more rectangular subset.
Specifically, 
we show a monotonic improvement of the
sub-frame performance measures defined in 
Chapter~\ref{section-Intro-performance-measures},
$L^*_{Shannon}(n,m,k)$, $L^*_{MSE}(n,m,k)$,
$L^*_{Shannon}(\gamma, \beta)$ and $L^*_{MSE}(\gamma, \beta)$, 
as the subset aspect ratio $\beta = k/m$ deviates from 1.
%
Our analysis builds upon {\em information theoretic ``subset'' inequalities} for entropy and Fisher Information, 
and their implications to determinants and inverse-traces of positive semi-definite matrices 
\cite{CoverBook,dembo1991information,khina2020monotonicity}.
%



The monotonic behavior (in $k/m$) of the optimum performance over all frames 
(\ref{GAMR*}) and (\ref{AHMR*})
follows from a similar behavior that holds for any {\em fixed} frame.
For some $m \times n$ frame $F$,
recall the average subset log-determinant (Shannon transform) 
defined in (\ref{GAMR}) for $m \leq k \leq n$:
\begin{equation}
\label{GAMR-specific}
L_{\rm Shannon}(F,k) 
\Ddef
\frac{1}{{n \choose k}} \sum_{S: |S|=k}
 \frac{1}{m} \cdot \log \left( \det \left( \frac{m}{k} \cdot F_S \cdot F_S^\dagger \right) \right)  
 \end{equation}
%
%
(which is a non-positive quantity),
and recall the average subset trace-inverse (MSE goodness) 
defined in (\ref{AHMR}) for $1 \leq k \leq m$:
%
%
\begin{equation}
\label{AHMR-specific}
L_{\rm MSE}(F,k) 
\Ddef
\frac{1}{{n \choose k}} \sum_{S: |S|=k}
\log \left( \frac{1}{k} \cdot \tr{(F_S^\dagger \cdot F_S)^{-1}} 
\right).  
\end{equation}
%
(which is a non-negative quantity).

%
%
%
%
%

\vspace{5mm}

\begin{theorem}
\label{Thm-frame-shannon-monotonicity}
For any $m \times n$ unit-norm frame $F$, 
the average subset Shannon transform (\ref{GAMR}) is monotonically non-decreasing with the subset size $k$:
\begin{equation}
    \label{Inequalities-shannon-monotonicity}
L_{\rm Shannon}(F,k) 
\leq 
L_{\rm Shannon}(F,k+1)
\end{equation}
for $m \leq k \leq n$.
\end{theorem}


{\em Proof:}
This is a variation on the subset determinant inequality for increasing subsets in 
\cite{dembo1991information}. 
Unlike in \cite{dembo1991information}, 
the size of the (positive definite) sub-matrices $F_S F_S^\dagger$, for $|S|= m, m+1, \ldots, n$, is \underline{constant} ($m \times m$), 
and only the {\em inner} dimension $k$ is increasing. 
See the details of the proof in
Appendix~\ref{section-inequalities-II}.
$\Box$


\vspace{5mm}

\begin{theorem}
\label{Thm-frame-MSE-monotonicity}
For any $m \times n$ unit-norm frame $F$, 
the average logarithmic noise amplification (\ref{AHMR}) is monotonically non-decreasing with the subset size $k$:
\begin{equation}
    \label{Inequalities-MSE-monotonicity}
L_{\rm MSE}(F,k) 
\leq 
L_{\rm MSE}(F,k+1) 
\end{equation}
for $1 \leq k \leq m$.
\end{theorem}


{\em Proof:}
The proof is based on a trace-inverse inequality for increasing subsets of a general covariance matrix,
\cite{khina2020monotonicity}. 
See the details in Appendix~\ref{section-inequalities-II}.
$\Box$


Theorems~\ref{Thm-frame-shannon-monotonicity} and~\ref{Thm-frame-MSE-monotonicity} 
immediately imply the same monotonic behavior for the best 
attainable performance over all $m \times n$ frames in (\ref{GAMR*}) and (\ref{AHMR*}),
because for each $k$ we can select a {\em different} frame to further improve the goodness.\footnote{For example,
if $F_k^*$ is the $m \times n$ frame achieving $L_{Shannon}^*(n,m,k)$, then 
$L_{Shannon}^*(n,m,k) = L_{Shannon}(F_k^*,k) \leq L_{Shannon}(F_k^*,k+1) \leq L_{Shannon}^*(n,m,k+1)$.
} 


\begin{cor}
\label{Thm-OPTA-monotonicity}
The optimum $(n,m,k)$ Shannon transform goodness (\ref{GAMR*}) 
is monotonically non-decreasing with the subset size $k$:
\begin{equation}
    \label{Inequalities-shannon*-monotonicity}
L^*_{\rm Shannon}(n,m,k) 
\leq \ldots \leq L^*_{\rm Shannon}(n,m,n)
\leq 0
\end{equation}
for $m \leq k \leq n$, 
and the optimum $(n,m,k)$ logarithmic noise amplification (\ref{AHMR*}) is monotonically non-decreasing with the subset size $k$:
\begin{equation}
    \label{Inequalities-MSE-monotonicity}
0 \leq
L^*_{\rm MSE}(n,m,k) 
\leq \ldots \leq L^*_{\rm MSE}(n,m,m)
\end{equation}
for $1 \leq k \leq m$.
A similar monotonic behavior as a function of $\beta$ (for a fixed $\gamma$) holds   
for the asymptotic measures $L^*_{Shannon}(\gamma, \beta)$ and $L^*_{MSE}(\gamma, \beta)$.
\end{cor}



{\bf Remarks:}
\begin{enumerate}
\item
The negativity and positivity of $L_{\rm Shannon}$ and $L_{\rm MSE}$, respectively, follows
because the arguments of the logarithms in their definition 
can be written as geometric-to-arithmetic and arithmetic-to-harmonic 
means ratios for eigenvalues, (\ref{signal-amplification-eigenvalues}) and (\ref{shannon-transform}).
%
\item 
The case $k=m$ is worst for both goodness criteria, and they both improve
($L_{\rm Shannon}$ increases / $L_{\rm MSE}$ decreases) as $k$ moves away from $m$,  
i.e., the sub-matrix $F_S$ becomes more rectangular.
%
%
\item
Without the normalization by $k$ 
in (\ref{AHMR-specific}) 
the MSE sub-frame inequality 
becomes trivial, 
because for every two nested subsets 
$S_1 \subset S_2 \subset [n]$, 
we have 
$\tr{ (F_{S_1}^\dagger \cdot F_{S_1})^{-1} } \leq \tr{ (F_{S_2}^\dagger \cdot F_{S_2})^{-1}  }$.
%
\item 
The right-hand side of (\ref{Inequalities-shannon-monotonicity}) 
is reminiscent of the ``determinant criterion'', 
$\det(F \cdot F^\dagger)$,
for space-time coding over a Rayleigh fading channel 
\cite{tarokh1998space,tse2005fundamentals}
(viewing the full frame $F$ as the difference between two
codewords in the space-time code).
See Section~\ref{section-Apps-STC}. 
%
%
\item 
In Corollary~\ref{Thm-OPTA-monotonicity} above the frame dimensions $n$ and $m$ are held fixed while the subset size $k$ varies.
To fully justify the role of the design parameter $m$ (or $\beta$) in the applications, we need to study the behavior when $k$ and $n$ are held fixed while $m$ varies. 
This means that the frame $F$ must vary with the subset aspect ratio $\beta$  
while keeping $k/n = p = \gamma \beta$ fixed. 
    Empirically, \underline{for a fixed frame family}, the dependence of the goodness measures on 
    $\beta$ is much stronger than their dependence on the frame aspect ratio $\gamma$.
    Hence a similar behavior should be expected 
    for $L^*_{Shannon}(\gamma=p/\beta, \beta)$ and $L^*_{MSE}(\gamma=p/\beta, \beta)$,
    for a fixed $p$, as $\beta$ deviates from 1.
    See, e.g., Figure~\ref{fig-maya-slamovich}.
    %
    A more delicate analysis of this dependence 
    is left for further study.
\end{enumerate}


%
%




\chapter{Applications}
\label{section-Apps}
%
In this chapter we elaborate on some of the applications of analog frame codes in digital communication and signal processing, as specified in Table~\ref{table-nmk-applications} in the Introduction:
non-orthogonal code-division multiple access (NOMA/CDMA), discrete-Fourier transform (DFT) codes, space-time codes (STC)),
analog to digital (A/D) conversion, multiple-description (MD) source coding, the new application of {\em coded computing},
training of deep neural networks (DNN),
and phase transition in compressed sensing (CS).
We provide numerical examples that clearly demonstrate that ETFs outperform other common frame codes.

\section{NOMA/CDMA}
Consider the multiple-access channel with partially active 
users:
$m$ channel resources are allocated in a fixed (non-orthogonal) 
manner between $n \geq m$ users, 
where at any given time only $k = p n$ out of the $n$ users are active
\cite{verdu-shamai1999spectral,stoica2019massively,zaidel2001multicell,zaidel2018sparse,polyanskiy2017perspective}.
Here $m$ can be thought of as the time-bandwidth product of the channel,
and $\beta = k/m$ is the effective ``loading factor'' of the system.
The resource allocation is done via CDMA, i.e.,
\begin{equation}
    \label{CDMA0}
    \by = \sum_{i =1}^n x_i \bff_i + \bz = F \bx + \bz, 
    \end{equation}
if all the users are active, and
\begin{equation}
    \label{CDMA1}
    \by = \sum_{i \in S} x_i \bff_i + \bz = F_S \bx_S + \bz, 
    \end{equation}
if only a $k$-subset $S \subset [n]$ of the users is active,
where 
$\bx = (x_1,\ldots,x_n)^\dagger$ represents the information symbols of the $n$ users,
$\bff_i$ is the length-$m$ spreading sequence of user $i$,
$\bz = (z_1,\ldots,z_m)^\dagger$ is additive white Gaussian noise (AWGN),
and
$\by = (y_1,\ldots,y_m)^\dagger$ is the input to the (joint) receiver.
The $m \times n$ frame $F = [\bff_1 | \cdots | \bff_n]$, (\ref{frame}), defines the mapping 
from user symbols to channel signals,
and the $m \times k$ subframe $F_S$ and the subvector $\bx_S$ correspond to the $k$ active users.

The relevant fidelity criterion depends on the type of the front-end receiver:
matched-filter (MF), minimum mean-squared error (MMSE), 
or maximum likelihood (ML) 
\cite{verdu-shamai1999spectral,zaidel2001multicell}.
The simplest one, the matched-filter receiver, gives rise to the total interference power (average squared cross-correlation) criterion,
which is independent of the activity number $k$,
and {\em tightness} is sufficient for frame optimization;  
see the Welch bound (\ref{WB}) and \cite{rupf1994optimum}.
The other two receivers give rise to $k$-subset criteria: 
MMSE (\ref{AHMR}) and log-determinant (Shannon transform) (\ref{GAMR}),
which call for much stronger ``$(n,m,k)$'' frame symmetry conditions. 

\begin{figure}[htbp] 
	\begin{center}
		\includegraphics[width=3.5in]{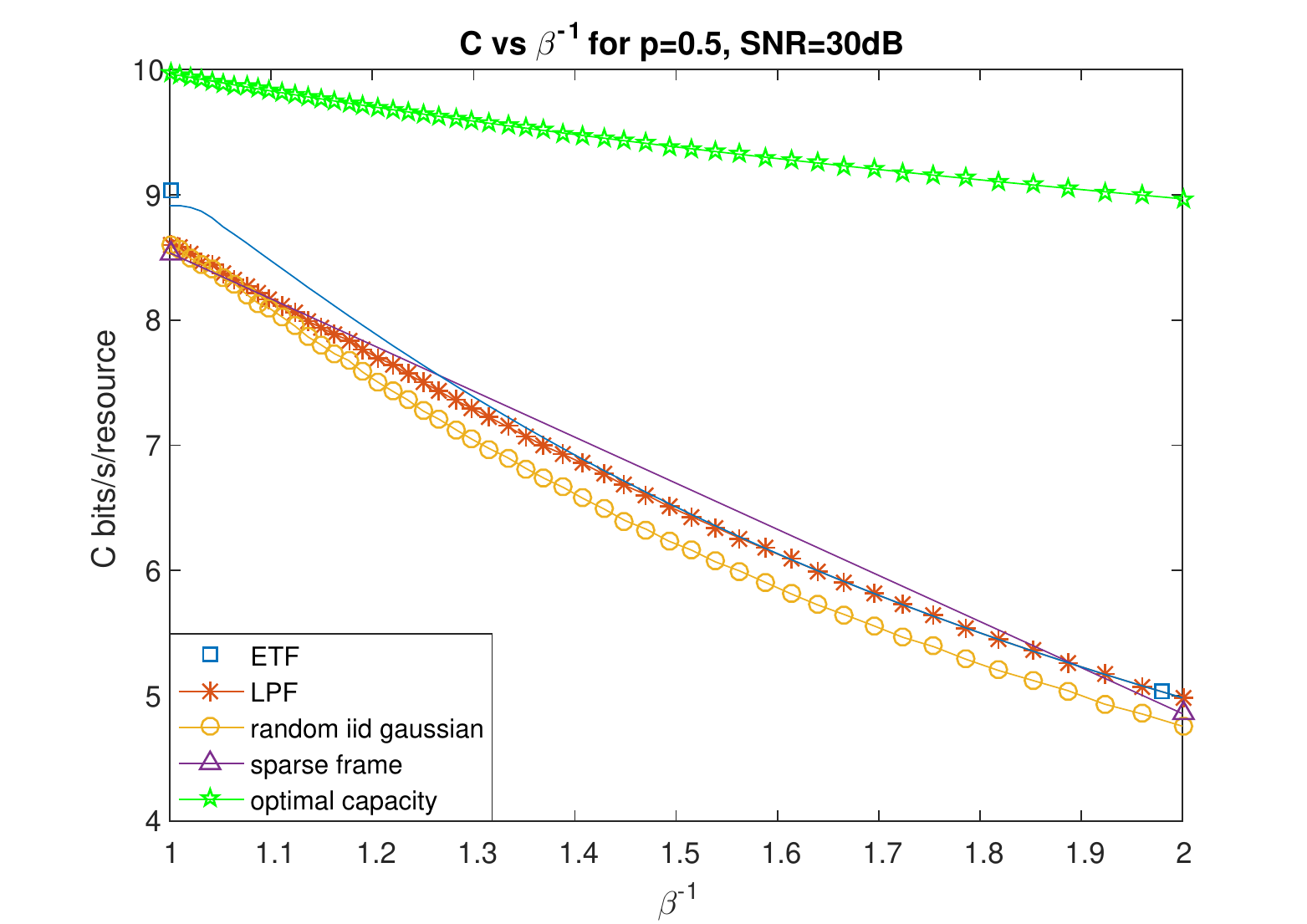}
		\includegraphics[width=3.5in]{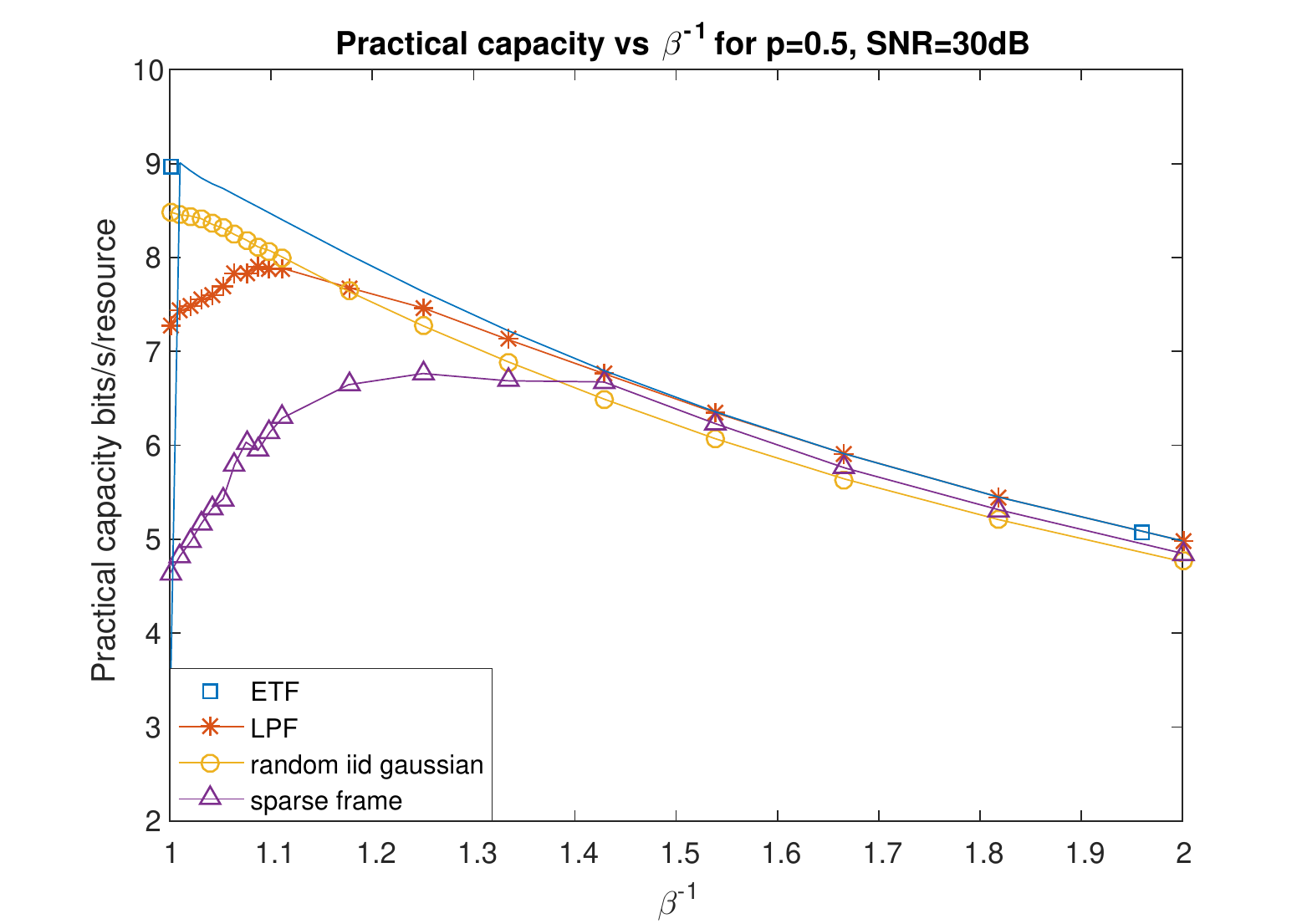}
	\caption{Average capacities per resource of a NOMA/CDMA system 
	for a fixed activity level $p = \gamma \beta = k/n = 0.5$, as a function of 
	the loading factor $\beta = k/m$, for several frames: 
	random (\ref{MP_density}), LPF (\ref{LPF-frame}), Steiner ETF \cite{fickus2012steiner}, and sparse \cite{zaidel2018sparse}.  
	All frame dimensions vary around $100 \times 200$.  
	(Upper) regular capacity (\ref{cdma-noma}).
	(Lower) practical capacity.
	For $\beta^{-1} \approx 1$, 
	the capacity gain of ETF is $\sim 0.4$ bit with respect to random frame, and almost 2 bits with respect to LPF (for practical capacity).    
	} 
		\label{fig-maya-slamovich}
	\end{center}
\end{figure}

Let us focus on the (optimal) maximum-likelihood receiver.
For a fixed $F_S$, the multiple access (MAC) sum capacity $C_S$ (assuming a white
Gaussian input \cite{CoverBook}) is given by 
$\log(\det[ I_m + {\rm SNR} \cdot F_S F_S^\dagger ])$
bits, where SNR is the signal-to-noise ratio per user.
Since in a practical system the 
loading factor $\beta=k/m$ is smaller than or equal to one,
it is preferable to rewrite $C_S$ with respect to the smaller dimension $k$, i.e., as
$\log(\det[ I_k + {\rm SNR} \cdot F_S^\dagger F_S ])$
(the two expressions are equal since the nonzero eigenvalues of $F_S F_S^\dagger$ and $F_S^\dagger F_S$ are the same).
The average capacity per resource (i.e., normalized by $m$) 
with respect to a uniform selection of the subset $S$ is thus given by 
\begin{equation}
\label{cdma-noma}
    C = \frac{1}{{n \choose k}} \sum_{S: |S|=k} 
    \frac{1}{m} \log(\det[ I_k + {\rm SNR} \cdot F_S^\dagger F_S ]) 
\end{equation}
bits per resource.
Note that for fixed $n$ and $k$ (equivalently fixed activity level $p=k/n$), 
one may wish to optimize the number of resources $m$, for $n \geq m \geq k$
(equivalently the loading factor $k/n \leq \beta \leq 1$) 
that maximizes the capacity per resource.
Figure~\ref{fig-maya-slamovich}, taken from \cite{MayaProject}, 
shows the average capacity $C$ as a function of $\beta$ for several frame families.
ETF clearly outperforms all other frames at its optimal $\beta$ operation point.
Figure~\ref{fig-maya-slamovich}-B shows the ``practical capacity'' 
(a quantity related to the concept of ``lattice decoding'' \cite{urbanke1998lattice,zamir2014lattice}),
which is equal to (\ref{cdma-noma}) without the ``$I_k$'' term inside the logarithm.
This quantity - similar to the Shannon-transform performance measure in (\ref{GAMR}) -
is more sensitive to eigenvalues of $F_S^\dagger F_S$ near zero; 
it thus demonstrates more clearly the gaps between the different frames
and the optimal $\beta$ operation point for each frame 
\cite{MayaProject}.

A similar behavior was observed in ``irregular pulse-shape'' optimization for faster-than-Nyquist TDMA signaling with FFE-DFE equalization \cite{DadushProject}.


%
%
%

\section{DFT codes}
Linear transformation can also be used to introduce redundancy
against fading (or erasures or impulses) plus noise.  
Say, $m$ information symbols $\bx = (x_1,\ldots,x_m)^\dagger$ are transformed into a larger
$n$-vector signal $F^\dagger \bx$, and the receiver gets a random $k$ subset $S$ corrupted with noise:
\begin{equation}
    \label{STC1}
    \by = F_S^\dagger \bx + \bz
\end{equation}
where $\by = (y_1,\ldots,y_k)^\dagger$ is the erased receiver input,
and $F^\dagger$ plays the role of the ``coding'' transformation.
Note that in contrast to the partially-active multiple access channel (\ref{CDMA1}), 
here $F^\dagger$ expands the dimension $m:n$, 
and the interesting regime where the channel can be inverted is $n \geq k \geq m$.
Wolf \cite{wolf1983redundancy} proposed the concept of {\em DFT codes}, where $F$ is a LPF frame,
as a means to mitigate the effect of impulse noise which acts similarly to the erasures
in (\ref{STC1}).
The redundancy introduced by the frame $F$ allows the receiver 
to estimate the information symbols $\bx$ after the $n-k$ impulses were detected and removed (erased).
A later work by Marshall \cite{marshall1984coding} hinted to the fact that 
irregular selection of DFT frequencies leads to better estimation
performance.
See also \cite{tulino2007gaussian,tulino2010capacity,zanko2014iterative}. 


%


\section{Space-time codes (STC)}
\label{section-Apps-STC}
\begin{figure}[htbp] 
	\begin{center}
		\includegraphics[width=3.5in]{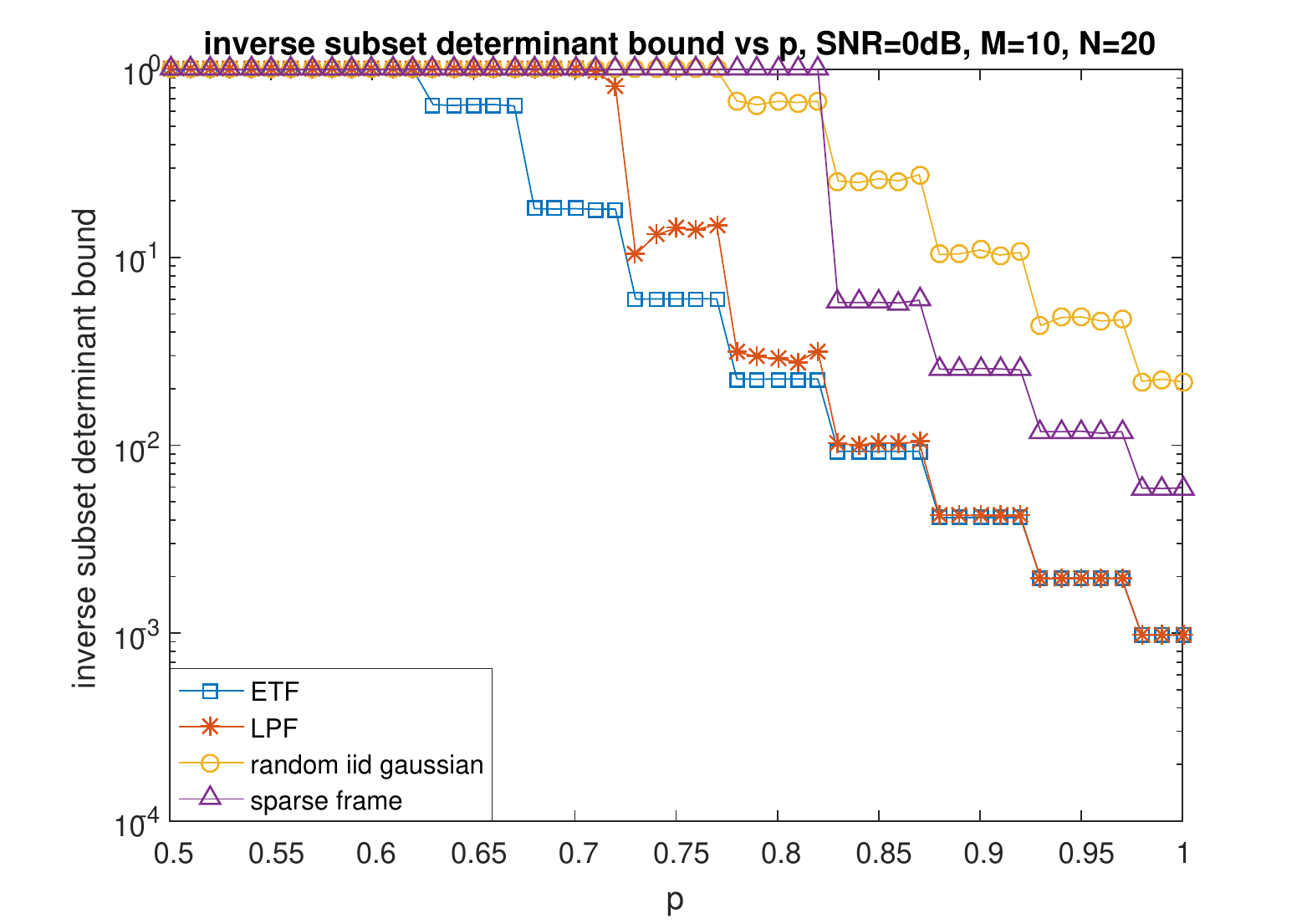}
		\includegraphics[width=3.5in]{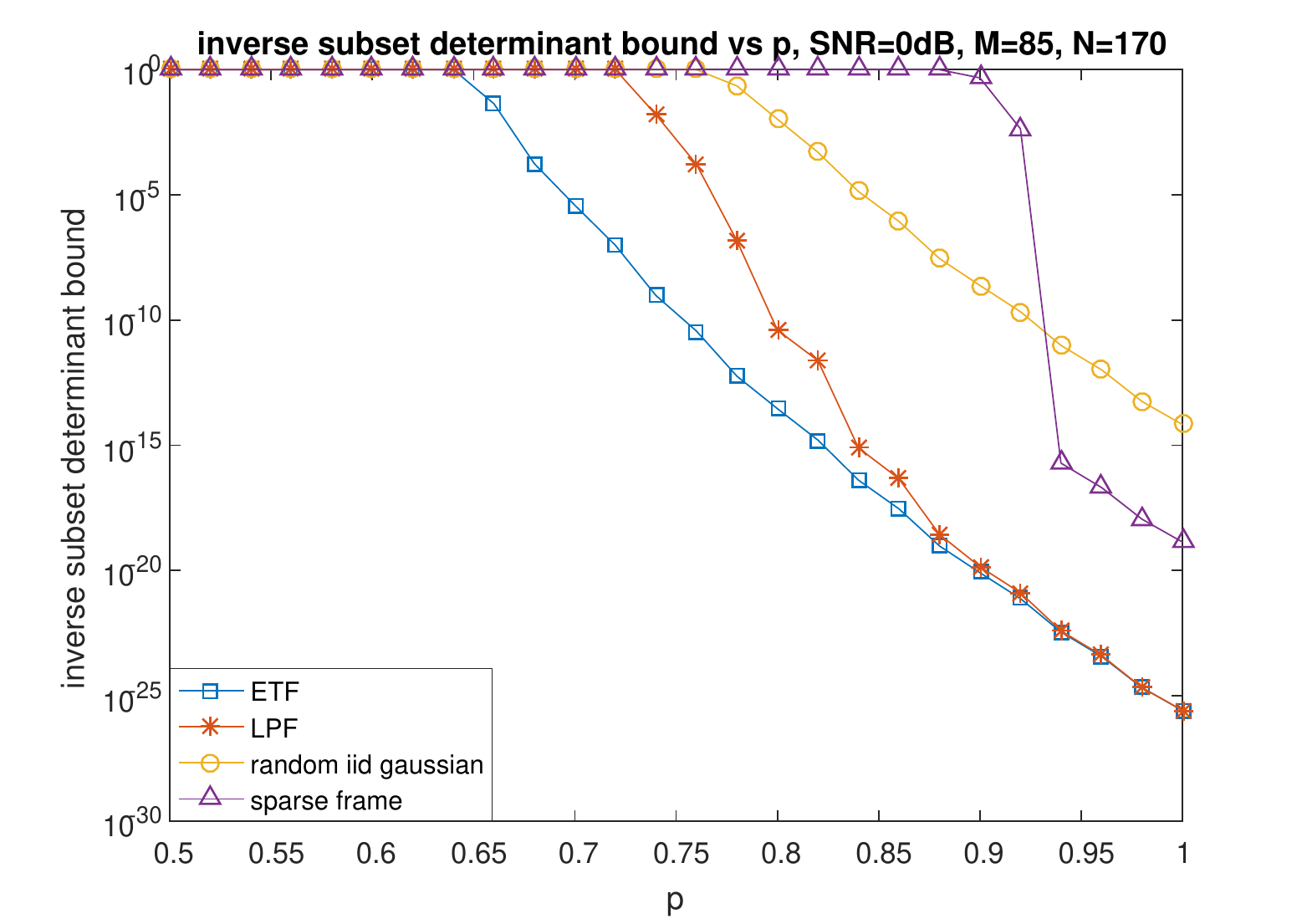}
	\caption{The ``inverse subset determinant'' bound (\ref{STC2}) of space-time codes, 
	for a fixed $\gamma = 0.5$ as a function of the not-erased probability $p = k/n$,
	for several frames: 
	random (\ref{MP_density}), LPF (\ref{LPF-frame}), Paley ETF \cite{fickus2012steiner}, and sparse \cite{zaidel2018sparse}.  
	Frame dimensions $m \times n =$: (A) $10 \times 20$. (B) $85 \times 170$.} 
		\label{fig-maya-slamovich2}
	\end{center}
\end{figure}

A space-time code 
can be viewed as a two-dimensional - space/time - analog modulation scheme, 
tailored to a vector-input block-fading channel,
that introduces diversity against random fading in the spatial dimension.
%
Codes for fading Gaussian channels need to preserve a sufficient Hamming
distance after the coordinates were randomly attenuated by fading \cite{tse2005fundamentals}.
Rayleigh fading gives rise to the ``product measure'' 
\cite[sec.~3.2.2]{tse2005fundamentals},
which becomes the ``determinant criterion'' for STC 
\cite{tarokh1998space}.
More specifically, 
STC transforms a message $A$ into an $m \times n$ codword $F_A$ of $m$ channel uses
in time over $n$ channel uses in space ($n$ transmit antennas).
The pairwise error between two such codewords $A$ and $B$ 
in the presence of Rayleigh fading is bounded by some (SNR-dependent) constant 
divided by the determinant $\det[(F_A - F_B) \cdot (F_A - F_B)^\dagger]$, \cite{tarokh1998space}, \cite[sec.~3.3.2]{tse2005fundamentals}.
If we take $B$ to be the zero codeword, then the goodness of the codeword $A$
is inverse proportional to $\det[F_A \cdot F_A^\dagger]$. 
In the presence of 0/1 spatial fading, the equivalent codword is $(F_A)_S$,
where $S \subset [n]$ is the set of surviving spatial coordinates.
%
%
If a random set of $k$ coordinates survive, where $k \geq m$, 
then we get that the error probability in the presence of a combination of 0/1 and Rayleigh fading is bounded by
\begin{equation}
 \label{STC2}
  P_e(A \rightarrow 0) 
  \leq
  \frac{1}{{n \choose k}} \sum_{S: |S|=k} 
    \frac{SNR^{-m}}{\det[(F_A)_S (F_A)_S^\dagger ]} .
    \end{equation}
The RHS of (\ref{STC2}) can be seen as 
a $k$-dependent criterion for STC design, which reduces to the usual 
``determinant criterion'' for $k=n$ ($p=1$).
Figure~\ref{fig-maya-slamovich2} illustrates this measure as a function of $p = k/n$ for several ``frames'' $F_A$. 
The clear advantage of ETFs for moderate $p$ gives rise to the question: can we choose multiple space-time codewords, such that all pairwise differences will be ``near ETF''?




%
%



\section{Over-sampling and $\Delta$-$\Sigma$ modulation}
%
Sampling of a band-limited signal beyond its Nyquist rate 
allows to trade off quantization resolution (distortion) for time resolution (rate) \cite{zamir1995rate}. 
The distortion enhancement follows since 
the desired signal has a low-pass spectrum in the over-sampled domain, while the quantization noise is approximately white and can be filtered out
at the reconstruction. 
$\Delta$-$\Sigma$ modulation enhances this effect using a ``noise shaping'' feedback filter around the quantizer, that creates a high-pass quantization noise spectrum,
hence lower reconstruction distortion for the same
quantization resolution 
\cite{candy1992oversampling,ordentlich2019performance}.
Interestingly, it was recognized that the mapping between the original (Nyquist-sampled) signal to the $\Delta$-$\Sigma$ modulated signal can be viewed as a frame expansion \cite{benedetto2004sigma}.
This opens the door to more general $\Delta$-$\Sigma$ modulation schemes, that are robust against various forms of interference and erasures
\cite{goyal2001quantized,strohmer2003grassmannian,eldar2003geometrically,boufounos2008causal}. 


\section{Multiple-description (MD) source coding}
%
When a source is transmitted over a lossy packet network, where ACK/NAK feedback cannot be assumed,
one may wish for graceful degradation as a function of the number of erased/received packets.
\cite{gamal1982achievable,goyal2001multiple}. 
In MD coding, the source is split into $n$ ``descriptions'', 
where a priori unknown subset 
of the descriptions arrive to the receiver;
it is convenient to assume  
that the receiver gets either a subset of size $k$, 
or all the $n$ descriptions \cite{pradhan2004n}.

A simple solution based on $\Delta$-$\Sigma$ modulation for the case ($n=2$, $k=1$) was proposed in \cite{ostergaard2009multiple}.
The $\Delta$-$\Sigma$-two-description coding scheme 
uses 1:2 oversampling ratio, where the two descriptions are created by splitting into even and odd samples, and where the noise shaping loop controls the distortion ratio between the $k=1$ case (even or odd samples) 
and the $k=2$ case (all samples).
It is shown in \cite{ostergaard2009multiple} that this solution achieves the optimal 
MD rate-distortion region for a Gaussian source under quadratic distortion if we
use an optimal dithered (lattice) quantizer
\cite{ozarow1980source}.
Naturally, one may wish to extend this solution 
to a general ($n,k$) pair using $m:n$ oversampling ratio,
for some $m \leq k$, with periodic allocation of the samples to the $n$ descriptions.  
This extension suffers, however, from {\em noise amplification}
whenever the received descriptions do not form a uniform sampling pattern
\cite{mashiach2010multiple,mashiach2013sampling},
which is exactly the same as the {\em signal amplification} problem in
band-limited interpolation for ``source coding with erasures''
(Chapter~\ref{section-Intro-motivations-IT}).

	\begin{figure}[h]
		\centering
		\includegraphics[trim={0.5cm 0 1cm 1cm},clip,width=3.4in]{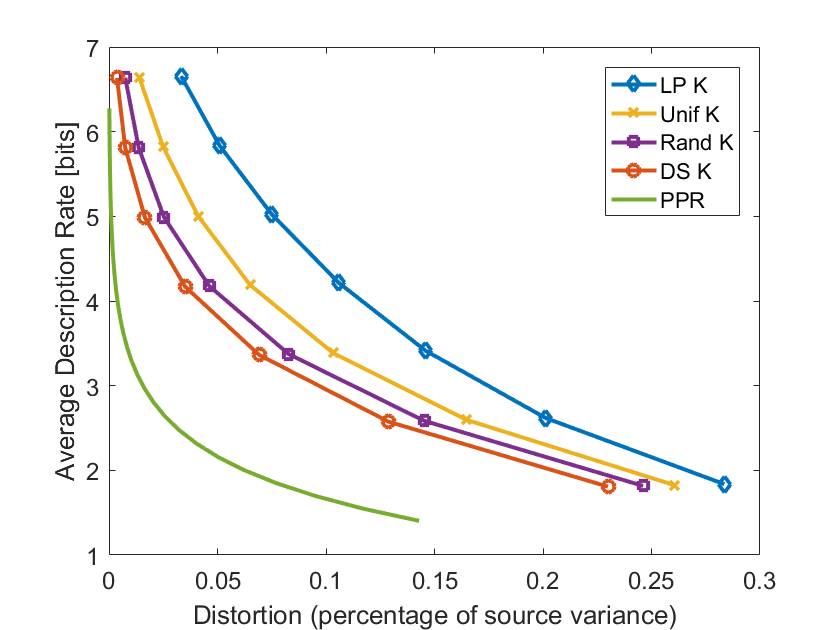}
		\caption{Rate-distortion performance of $\Delta$-$\Sigma$ MD coding with over-sampling ratio $n:m = 31:6$, where $k=6$ descriptions are received out of $n=31$ transmitted descriptions, for several interpolation filters:  
		low pass (LP), difference-set (DS), uniform (Unif), and random (rand) spectrum interpolation filter, compared to the ``PPR bound'' \cite{pradhan2004n}.
		}
		\label{fig-mor-goren}
	\end{figure}

The fact that ETFs -  which are highly robust against random erasures - 
correspond to ``irregular spectrum'' (see the difference-set DFT construction in Section~\ref{section-difference-set}),
inspired the idea to modify the interpolation filter in the MD $\Delta$-$\Sigma$ modulation scheme to be an ``irregular interpolator'' 
\cite{goren2021diversity}.
Figure~\ref{fig-mor-goren}, taken from 
\cite{goren2021diversity},
demonstrates this idea in MD image (JPEG-based) coding. 
See also \cite{palgy2010multiple}.

\section{Coded distributed computation (CDC)}
%
Analog MDS (Reed-Solomon) codes were recently proposed to protect against straggling worker nodes in distributed computation of a linear function; e.g., of the gradient in training a deep neural network
\cite{lee2018speeding,raviv2020gradient}.
Specifically, to obtain the matrix multiplication $X A$ with $K$ ideal workers,
the master node breaks $A$ into $K$ pieces $[A_1,...,A_K]$, 
and distribute the partial multiplications $X A_1, ..., X A_K$.
If there are more workers but some are stragglers, say, only $K$ out of $N$ compute their task in time,
the master worker can use the generator matrix of a $K:N$ ``analog erasure-correction code'' to expand 
$A_1,...,A_K$ into $\tilde{A}_1,...,\tilde{A}_N$, 
so $X A$ can be computed from any $K$ subset of $X A_1, ..., X A_N$.
For example, for $K=2, N=3$ 
the code is simply $\tilde{A}_1=A_1, \tilde{A}_2=A_2$ and $\tilde{A}_3=A_1+A_2$,
and analog versions of BCH and Reed Solomon codes provide simple encoding and
decoding schemes for more general $(K,N)$ pairs
\cite{lee2018speeding,li2020coded,raviv2020gradient}.
%
It is insightful to think of the expansion as multiplying $A$ by a frame $F$ with an aspect ratio $N/K$, i.e., 
$\tilde{A} = A F$,
where $F$ is composed of $K \times N$ square blocks, each being a scaled identity matrix
$g_{i,j} I$, for $i \in [K]$, $j \in [N]$.
For example, in the 2:3 example above all the $g$'s are 0/1, and
\[
F = \left[ 
\begin{array}{ccc}
I & 0 & I
\\
0 & I & I 
\end{array}
\right].
\] 
Now suppose that beyond the stragglers problem, we also face noise due to finite word-length computation.
That is, the master node needs to reconstruct $XA$ from
\[
Y_S = (X A F)_S + Z
\]
where $S$ denotes the subset of non straggling workers and $Z$ denotes the computation noise.
Our goal is to find coding frames $F$ which are resilient to noise, i.e., numerically stable 
\cite{fahim2019numerically,ramamoorthy2019numerically}.
The challenge though is to keep the simplicity of encoding and decoding of analog
codes inspired from digital erasure correction codes \cite{babadi2011spectral,calderbank2010construction,raviv2020gradient}, yet enjoy the better robustness to analog noise of ETFs.  	


%
%


\begin{figure} [ht]
\center
\includegraphics[width=\columnwidth, height=0.30\textheight]{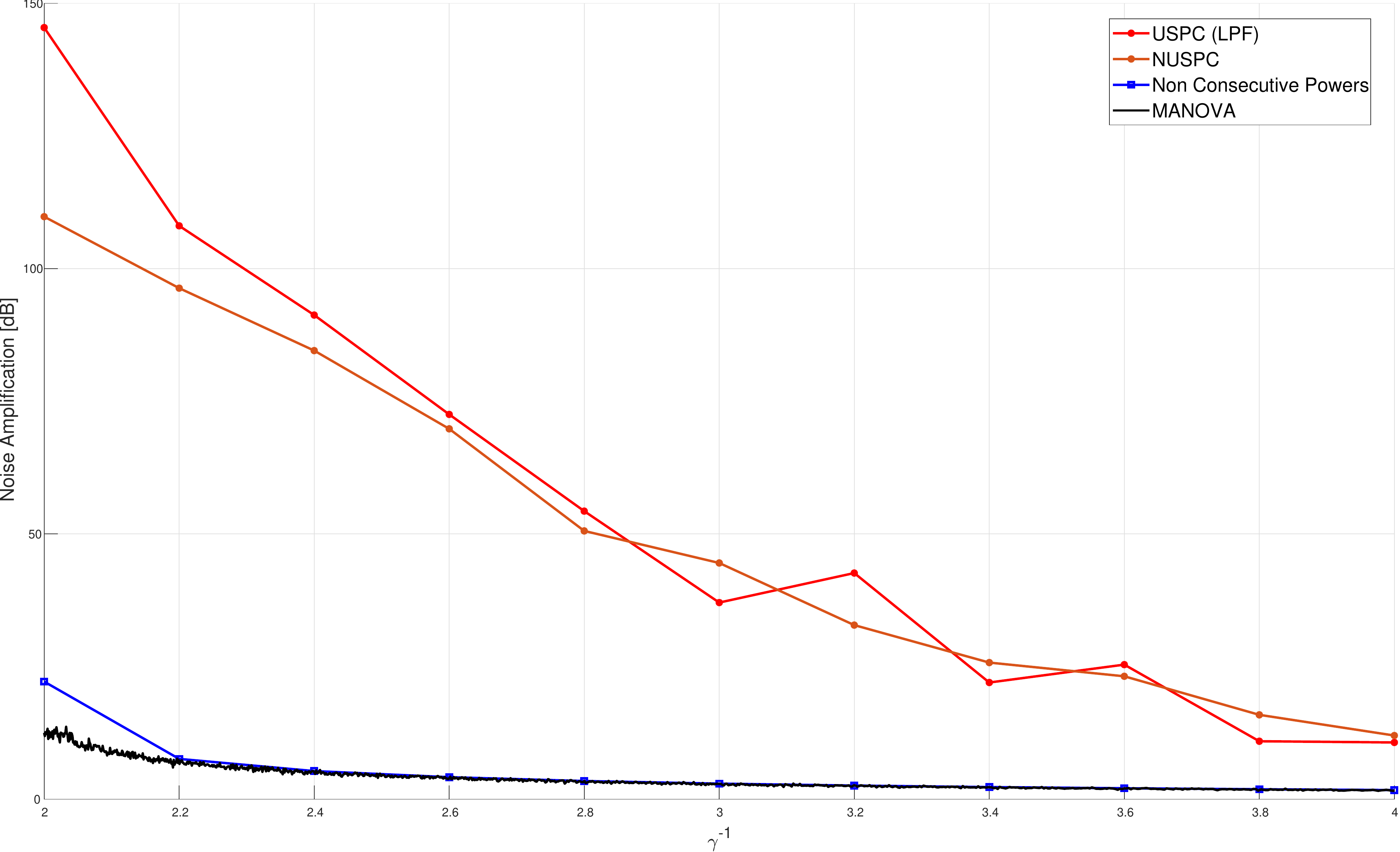}
\caption{Comparison of noise amplification Vs. the frame redundancy $\gamma^{-1}$ for several polynomial codes: uniform sampled unit circle (USPC), non-uniform sampled unit circle (NUSPC), and non-consecutive (difference-set) powers, compared to the MANOVA curve. 
Figure is taken from \cite{RoyeeYosibashPaper}.
} 
\label{fig-royee-yosibash}
\end{figure}


\section{Neural networks}
%
The robustness of ETF subsets, revealed in \cite{haikin2016analog,haikin2017random}, inspired the work of
Bank and Girayes \cite{bank2018relationship,bank2020ETF} on training 
of deep neural networks with Dropout.
Their scheme regularizes the training of the parameters in each iteration, so that
they form a set of vectors with a \underline{low coherence}, just like a subset from an ETF.

\vspace{10mm}

\section{Frames in compressed sensing: phase transition}
%
%
Pose the compressed sensing measurement as $x\mapsto Ax=y$. We observe $y$ and wish to recover $x$ for $A$ known.
Denote our chosen signal reconstruction scheme by $\mathcal{R}:t\mapsto x$. For example,
$\mathcal{R}$ can be $\mathcal{R}:y\mapsto \text{argmin}_{y=Ax}\norm{x}_1$.
The basic question in (noiseless) compressed sensing is - for which {\em (matrix,signal) } pair
$(A,x)$ do we have  $\mathcal{R}(y)=x$? 
When the underlying model includes randomness in the pair $(A,x)$, one can ask this question either 
in the worst-case sense  (asking to characterize those cases $(A,x)$ when $\mathcal{R}(y)=x$ w.p 1)
or average-case sense (asking to characterize those cases 
cases $(A,x)$ when $\mathcal{R}(y)=x$ with high probability). 

The Restricted Isometry Property became popular as it appeared in sufficient conditions for worst-case recovery in compressed sensing problems. However, these sufficient conditions are known to be far from tight, and indeed explain only a fraction of the cases $(A,x)$ in which recovery (using $\ell_1$ minimization, say) is known to hold. 

The situation is different when one considers average-case recovery. Donoho and Tanner 
\cite{donoho2009observed} 
developed a theory that places cases $(A,x)$ on a sparsity-undersampling phase plane. Using large-scale experiments they have identified - empiricially at first - an asymptotic
phase transition phenomenon, which holds approximately for finite $(A,x)$ and exactly in the large-system limit. An exact formula for the asymptotic phase transition was later identified for Gaussian i.i.d sensing matrices $A$ and a variety of signal models $x$.  In a definite sense, then, the average-case analysis is satisfying and complete. 

In \cite{monajemi2013deterministic} 
the authors reported a curious phenomenon: the same empirical apparatus that had been previously used to  establish the existence of the Donoho-Tanner phase transition for the case of Gaussian sensning matrices, has been applied to deterministic sensing matrices based on known frame constructions including ETFs. The authors reported overwhelming empirical evidence, based on a very large parallel computation, that most of the deterministic frames under study, when used as the sensing matrix $A$ in a recovery problem and when using a variety of reconstruction algorithms, exhibit a phase transition that agrees with that of Gaussian matrices. 

This observed universality, in which many different frame constructions exhibit a single, universal compressed sensing phase transition has not, to the best of our knowledge, proved rigorously. However in contrast with theory based on Restricted Isometry Property it provides a complete characterization, within a specified framework, of the success and failure of recovery of sparse vectors using $\ell_1$ minimization and related algorithms. 
The connection between this universal phenomenon and the new ETF-MANOVA relation is yet to be found.




 \chapter{ETF optimality conjecture} 
\label{section-ETF-super}
%
The examples of Chapter~\ref{section-Apps} (see also Figure~\ref{fig-goodness-comparison-Manova-MP-LPF}) showed that ETFs are superior to all other frame families that were investigated: LPF, random and code-based,
in terms of their performance as analog codes.
We are tempted to conjecture that ETFs outperform all other frames,
and achieve the optimal frame performance characteristics
(\ref{AHMR*}), (\ref{GAMR*}) and (\ref{Intro-StRIP})
defined in Chapter~\ref{section-Intro-performance-measures}.
In this chapter we provide some analytic support for the ``ETF superiority'' conjecture. 
We show that the Welch lower bound (\ref{WB}) naturally extends to random frame subsets and to higher-order moments (3rd and 4th moments in (\ref{moment_r})), and that the extended bound is achieved by (and sometimes only by) ETFs. 
%

%
%



\section{Erasure Welch bound}
\label{section-ETF-super--EWB}
\begin{theorem}[Erasure Welch Bound (EWB) of order $r = 2,3,4$ \cite{haikin2018frame,MarinaThesis}.] 
\label{th_bound_m1-m4}
	For any $m$-by-$n$ unit-norm frame $F$, selection probability $p$, and moment orders $r=2,3,4$, 
	the expected sub-frame $r$th moment \eqref{moment_r} is lower bounded by
	\begin{equation} 
	\label{moment d bound}
	m_r  \ge 
	m_r^{\rm ETF}
	\end{equation}
	where $m_r^{\rm ETF}$ is the corresponding value for an ETF given by Theorem~\ref{th_m_1_4} 
	(using the expressions in (\ref{Manova_moments})-(\ref{delta EWBd}), even if an ETF does not exist for the pair $(m,n)$).
 	Equality holds for $r=2,3$ if and only if $F$ is a tight frame, 
 	and for $r=4$ if and only if $F$ is an ETF.
\end{theorem}


The case ($r=2, p=1$) in Theorem~\ref{th_bound_m1-m4} is equivalent to the (mean-squared) Welch bound (\ref{WB}).
Thus, the EWB 
generalizes the Welch bound in two senses. 
First, it is a bound on average of subsets of $F$ rather than on the entire set of its pairwise moments.
Second, it is a bound on higher order ($r= 3, 4$) moments.\footnote{Our definition is different than the generalized Welch bound on the average $r$th power of the absolute cross-correlation \cite{welch1974lower}.}


\begin{proof}
For the case $r=2$, 
the equivalent form (\ref{WB_x}) of the mean-squared Welch bound \eqref{WB}, and the expression (\ref{m2a}) for $m_2$, imply 
%
%
%
\begin{equation} \label{m2_bound}
\begin{split}
&m_2 \ge  p+p^2x=m^{\rm MANOVA}(\gamma,p,2) = m_2^{\rm ETF} ,
\end{split}
\end{equation}
where the last two equations follow from (\ref{Manova_moments2}) and Theorem~\ref{th_m_1_4}.
The equality condition for the Welch bound implies that (\ref{m2_bound}) holds with equality for any tight frame.
The proof for $r=3,4$ is based on Jensen's inequality, 
and it appears in \cite{haikin2018frame,MarinaThesis}. 
\end{proof}

\section{Minimum empirical variance and kurtosis}
\label{section-ETF-super--ESV-ESK}
The EWB has an interesting implication on ETF having the most compact spectral distribution of a randomly selected sub-frame.  
By (\ref{m1}) and Theorem~\ref{th_bound_m1-m4}, 
all unit-norm frames share the same expected sub-frame 1st moment $m_1 = p$;
thus 
minimizing the expected sub-frame 2nd moment $m_2$ amounts to minimizing the 
{\em empirical spectral variance} 
\begin{equation}
    \label{empirical_spectral_variance}
    {\rm ESV} = m_2 - m_1^2
\end{equation}
%
of a {\em typical} sub-frame (typical in the sense that its moment matches the ensemble average), among all unit-norm frames.
Moreover, all unit-norm {\em tight} frames share the same expected sub-frame 2nd moment $m_2 = p + p^2 x$, and 3rd moment
$m_3 = p + 3x p^2 + (x^2-x) p^3$;
thus minimizing the expected sub-frame 4th moment $m_4$ amounts to minimizing the 
{\em empirical spectral kurtosis}
\begin{equation}
    \label{empirical_spectral_kurtosis}
        {\rm ESK} = \frac{m_4  -  4 m_3 m_1  +  6 m_2 m_1^2  -  3 m_1^4}{(m_2 - m_1^2)^2}
\end{equation}
of a typical sub-frame, among all unit-norm {\em tight} frames. 
We can thus 
use the MANOVA $1st$-to-$4th$ moments (\ref{Manova_moments}) to lower bound the variance and kurtosis of the limiting sub-frame ESD of any sequence of unit-norm (tight) frames. 



\vspace{5mm}

\begin{theorem}[Minimum sub-frame ESV and ESK]
\label{thm-ESV-ESK}
Consider any sequence 
of unit-norm frames with increasing dimensions, whose aspect ratio converges to $\gamma$, and whose (Bernoulli-$p$ selected) sub-frame ESD 
$\mu_{G_S}^{(n)}$
converges to some limiting law $\mu$ (see (\ref{ETF-MANOVA-conjecture})).
The asymptotic sub-frame ESV (\ref{empirical_spectral_variance}) 
is then lower bounded by that of a tight frame:
\begin{equation}
    \label{ESV_bound}
    {\rm ESV} \geq p + (x-1) p^2  
\end{equation}
where $x = 1/\gamma - 1$.
Moreover, if the sequence of frames is {\em tight}, then the asymptotic sub-frame ESK (\ref{empirical_spectral_kurtosis}) is lower bounded by that of an ETF:
\begin{equation}
    \label{ESK_bound}
        {\rm ESK} \geq \frac{p+p^2(6x-4)+p^3(6x^2-16x+6)+p^4(x^3-7x^2+11x-3)}{p^2(x^4+2x^2+1)-p^3(2x^2+2)+p^4} .
\end{equation}
\end{theorem}
%

Recalling that the variance and kurtosis are measures for the spread of a distribution about its mean, 
Theorem~\ref{thm-ESV-ESK} indicates that ETFs have the {\em most compact typical sub-frame ESD} among all unit-norm frames.
\begin{figure}[htbp]  
		\centering
		\includegraphics[width=2.7in]{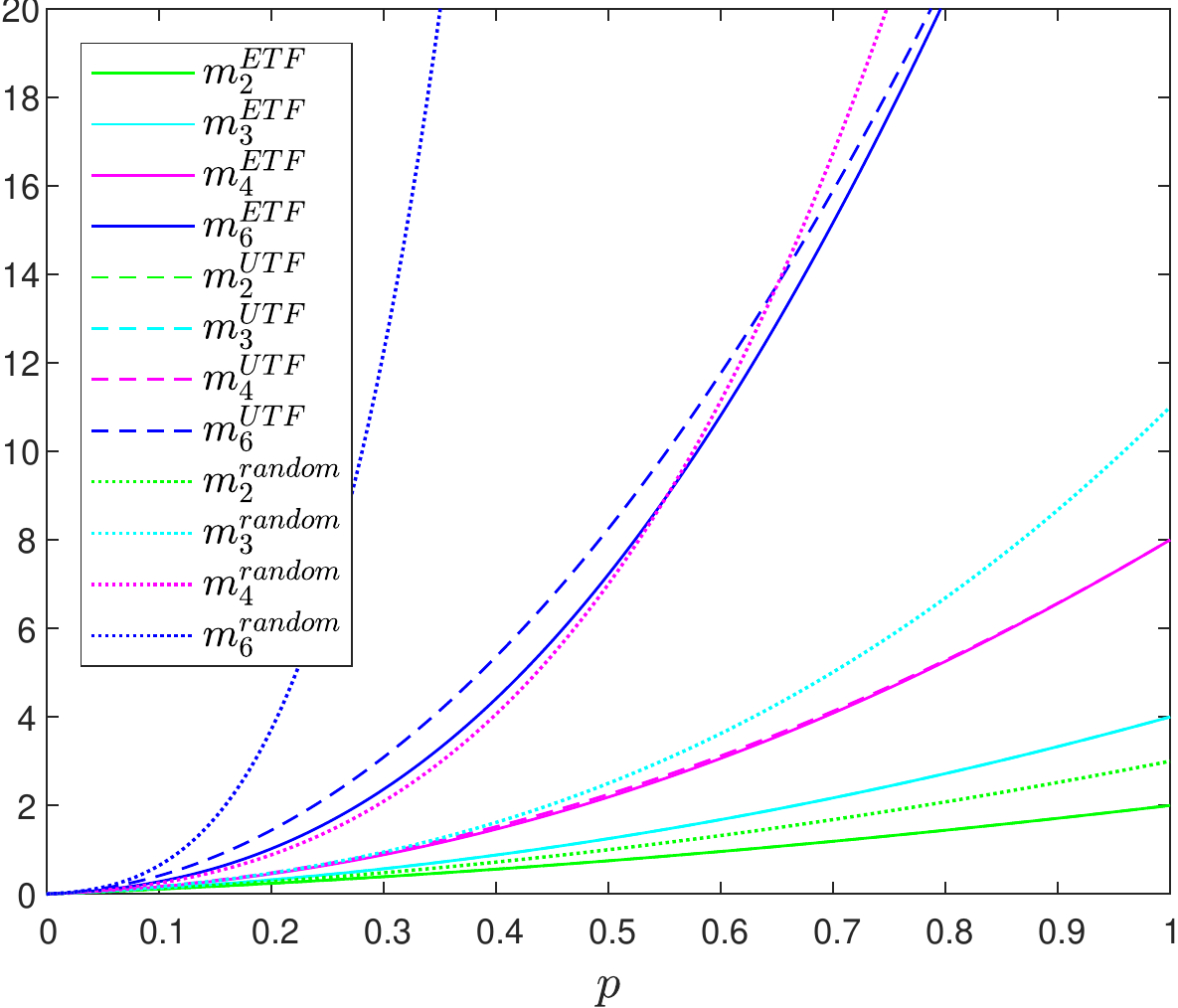}\\
		\includegraphics[width=2.3in]{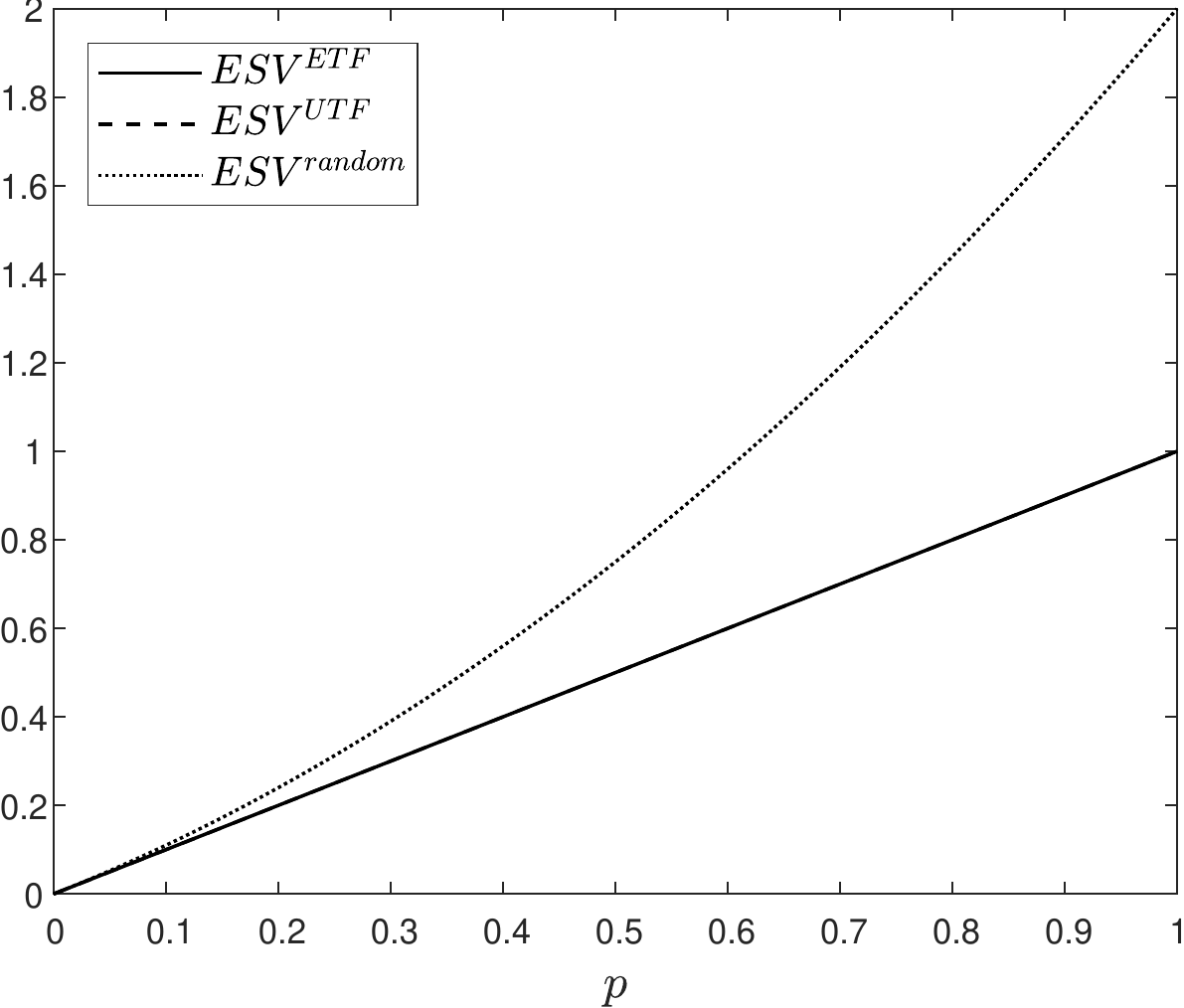}
		\includegraphics[width=2.3in]{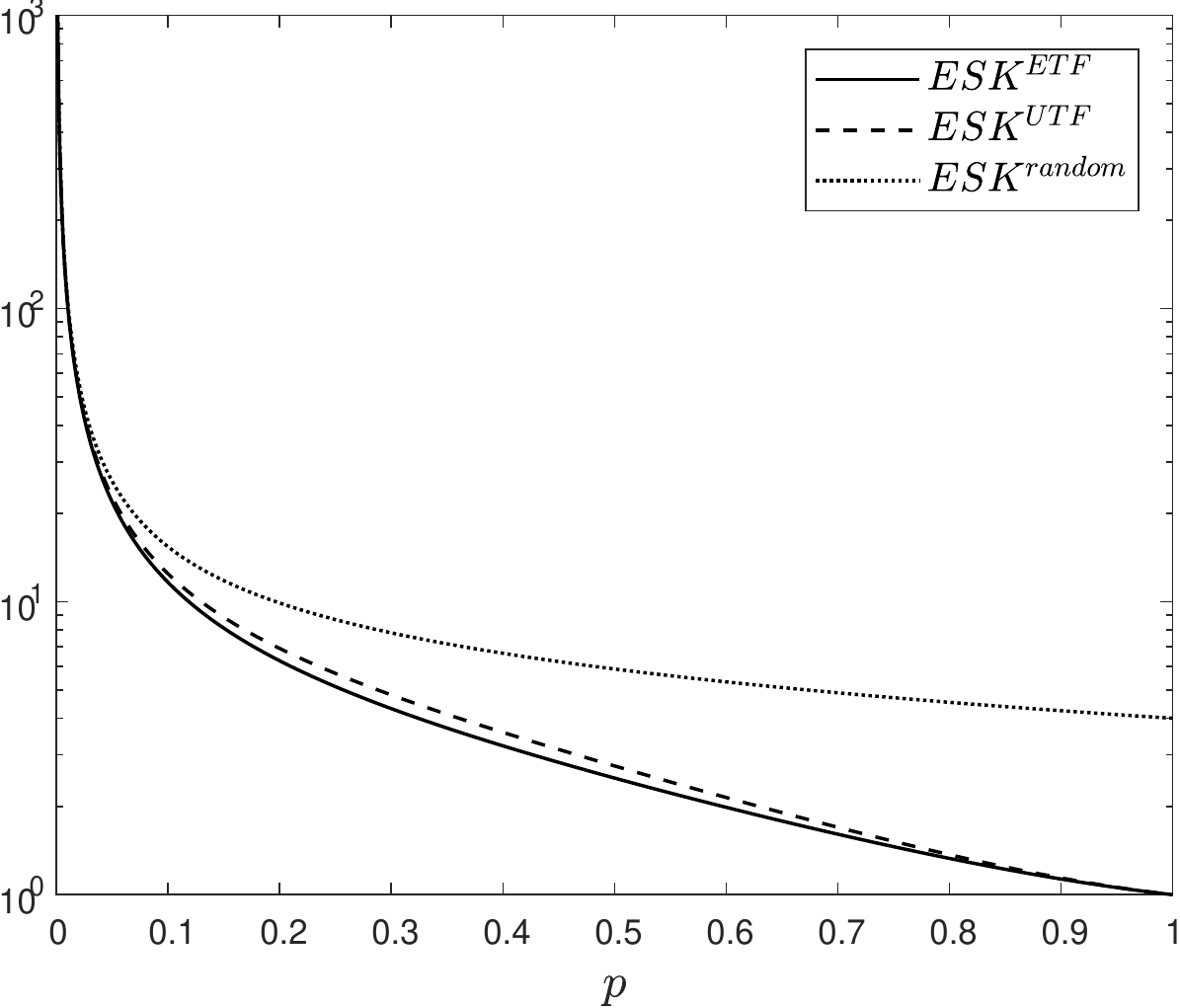}
		\caption{Comparison between a repetition (unit-norm tight) frame (UTF); random (i.i.d.) frame; and ETF, where all frames are with aspect ratio $\gamma=1/2$. 
		The $x$-axis shows the sub-frame selection probability $p$.  The $y$-axis shows the sub-frame moments 1-6 (top), the empirical spectral variance (bottom-left), and the empirical spectral kurtosis (bottom-right).
		}
		\label{fig-moments}
\end{figure}

Figure~\ref{fig-moments} shows a comparison of the sub-frame moments, and the corresponding variance and kurtosis, for several tight frames, as well as for a random (i.i.d.) frame.

\chapter{Conclusion and discussion}
\label{section-conclusion}
%

In this survey we tried to demonstrate the great potential of incorporating ideas from information theory and random matrix theory into frame theory.
We hope that it will stimulate future research along two main lines of open questions mentioned above. One line of investigation is practical, namely, design of robust analog codes. We raised the possibility that ETFs can be used as robust, and possibly optimally robust, analog codes in a variety of applications. The second line of investigation points down the ETF-MANOVA ``rabbit hole''.
It concerns formal mathematical analysis and explanation of the universality of the MANOVA distribution, as a limit of the spectrum of random subsets of frames in a class of well structured frames, and the universality of the rate of convergence to this limit.

\subsection*{Asymptotically robust analog codes} 

In all the applications we considered in Chapter~\ref{section-Apps}, exact mean-case performance is 
given by an expression of the form
\[
\mathbb{E}_S [\Psi(\lambda(G_S))]
\]
where $\Psi$ is a
functional of the empirical spectrum, which depends on the example. 
Here, $S$ denotes a subset of $[n]$, $F_S$ is the corresponding subset of the frame $F$, 
$G_S=F_S^\dagger F_S$ is the subset Gram matrix, $\lambda(G_S)$ is the spectrum of $G_S$, and $\mathbb{E}_S$ denotes expectation with respect to random selection of $S$. 
(Note that $\lambda(G_S)$ is equivalent to the empirical spectral distribution $\mu_{G_S}$ in (\ref{ESD}).)

A common property of the three performance measures $\Psi$ mentioned in Chapter~\ref{section-Intro-performance-measures}, is that a subframe $F_S$ which is ``more orthogonal'', i.e., whose spectrum is narrower and more removed from zero, gives better performance $\Psi(\lambda(G_S))$. 
%
%
From this perspective, ``good'' frames have the property that typical subframes have a compact and zero-removed spectrum support.  


\vspace{5mm}

\noindent {\bf Questions.} Generalizing from the examples we have seen, we arrive at fundamental new questions which span frame theory and random matrix theory:

\begin{enumerate}
    \item 
How does one calculate quantities of the form $\mathbb{E}_S [\Psi(\lambda(G_S))]$ for various functionals $\Psi$?
\item How can we design ``good frames'' - namely frames whose typical subframe has a spectrum with support as small and as zero-removed as possible?
\item Does there exist a ``best frame'', namely, a frame with ``best'' spectrum (with respect to one of the performance measures, or all of them)? If it exists, can we characterize it?
\end{enumerate}

\noindent {\bf Answers.} In this survey we review recent results which attempt to answer the questions above: 

\begin{enumerate}
    \item For a large variety of well known frames (Table~\ref{frames:tab}), and a large variety of functionals $\Psi$ (Chapter~\ref{section-Intro-performance-measures}), the quantity 
    $\mathbb{E}_S [\Psi(\lambda(G_S))]$ can be calculated to good and known precision (Chapter~\ref{section-RMT-meets-frames}).
    Specifically, 
    \[
    \mathbb{E}_S \left[ \Psi\left(\lambda (G_S)\right) \right]\approx 
    \Psi \left( f_{\beta,\gamma} \right )
    \]
    where $f_{\beta,\gamma}$ is a known cumulative distribution function that depends on the aspect ratios of the frame and subframe in question, and $\Psi(f)$ is defined in (\ref{functional_of_distribution}).
    
    \item Certain tight frames - of both deterministic and random design - are ``good frames'' in the sense that the spectrum of their typical subframe converges to a universal limiting spectral distribution $f_{\beta,\gamma}$ (Chapter~\ref{section-RMT}), whose support is compact and zero-removed. 
    In fact, this universal limiting 
    distribution function is the famous Wachter (Jacobi) astymptotic distribution familiar from random matrix theory 
    (Chapters~\ref{section-RMT-meets-frames} and~\ref{section-ETF-MANOVA-moments}).
    
    \item There is empirical and theoretic evidence 
    that suggests that equiangular tight frames in fact achieve the best performance, asymptotically as the frame size grows (Chapter~\ref{section-ETF-super}). 
\end{enumerate}




\subsection*{Mathematical analysis of the ETF-MANOVA relation}

More generally, this survey suggests that the spectrum of a typical subframe, taken from a frame with certain design properties, is a fascinating object - both from a theoretical perspective and a practical perspective. (Here, ``typical'' is taken to mean a subframe selected uniformly at random from the set of all subframes.)  Our results revolved around a fascinating universality property, whereby the spectrum of (the Gramian of) a typical subframe is indistinguishable from that of a matrix from the MANOVA (Jacobi) random matrix ensemble, both in finite-$n$ and asymptotically. 

Interestingly, we saw that this universality property holds for a certain kind of {\em deterministic} frames, just as it holds for certain random frames. This means that, curiously, a famous random matrix theory law arises in a deterministic and highly symmetric geometrical object. While one can find hints in the literature that this random-matrix phenomena can be suspected to arise asymptotically in these highly symmetric deterministic frames, the fact that universality holds for finite-$n$ as well is striking in our opinion. We highlight the many open questions and open problems that arise in this fascinating study of spectra of random subframes. 

We particularly note that, to the best of our knowledge, the observed universality of the {\em convergence rate} of the spectrum, and of functionals of the spectrum, to the universal limiting distribution, is a fascinating object for further investigation, whose proof could possibly require and stimulate new mathematical tools.




\section{Summary of asymptotic results for ETF subsets}
We conclude this survey by concise summaries of the asymptotic results and the open questions.

The ETF-MANOVA relation of Chapters~\ref{section-RMT-meets-frames} and~\ref{section-ETF-MANOVA-moments} implies that
we can compute the performance measures of large dimensional ETFs and near-ETFs as integrals of the MANOVA distribution: 
\begin{equation}
    \label{Manova-MMSE}
    L_{\rm MSE}(\mbox{ETF},\gamma,p) = 
    \log \left( \int_{\lambda_-}^{\lambda_+} \frac{1}{x} \cdot f_{\gamma,p}(x) dx \right) ,
\end{equation}
%
%
\begin{equation}
    \label{Manova-Shannon-Transform}
    L_{\rm Shannon}(\mbox{ETF},\gamma,p) = \int_{\lambda_-}^{\lambda_+} \log(x) \cdot f_{\gamma,p}(x) dx  ,
\end{equation}
and (by the convergence of the extreme eigenvalues\footnote{
\label{footnote-tracey-widom}
While the asymptotic results in Chapter~\ref{section-ETF-MANOVA-moments}
are only concerned with the bulk spectrum, our empirical results (Chapter~\ref{section-RMT-meets-frames}, \cite{haikin2017random}) demonstrate a ``Tracy-Widom'' kind of result 
(\cite{ZeitouniBook,feier2012methods})
regarding the high-probability convergence of the extreme eigenvalues to 
$\lambda_\pm$ of the MANOVA law.
})
\begin{equation}
    \label{Manova-StRIP}
    L_{\rm StRIP}(\mbox{ETF},\gamma,p,\delta) 
    = \left\{  
    \begin{array}{cc}
         1, & \mbox{if $\delta < \min\{1 - \lambda_-, \lambda_+-1\}$} \\
         0, & \mbox{otherwise,}
    \end{array}    
    \right.       
\end{equation}
where
$L_{\rm MSE}$, $L_{\rm Shannon}$ and $L_{\rm StRIP}$ 
are the performance measures (\ref{AHMR}), (\ref{shannon-transform}) and (\ref{Intro-StRIP}), respectively, 
and where the asymptotic MANOVA density $f_{\gamma,p}(x)$ and its support edges $(\lambda_-, \lambda_+)$ are given in (\ref{MANOVA_density}) (with $p = \beta \gamma$). 
The ETF superiority conjecture of Chapter~\ref{section-ETF-super} implies that the {\em optimal} measures 
(\ref{AHMR-asymptotic}), (\ref{GAMR*}) and (\ref{Intro-StRIP}) are given by:
\begin{equation}
    \label{Manova-MMSE-super}
    L_{\rm MSE}^*(\gamma,p) = L_{\rm MSE}(\mbox{ETF},\gamma,p) ,
\end{equation}
\begin{equation}
    \label{Manova-Shannon-Transform-super}
    L_{\rm Shannon}^*(\gamma,p) = L_{\rm Shannon}(\mbox{ETF},\gamma,p) ,
\end{equation}
and
\begin{equation}
    \label{Intro-StRIP-super}
    L_{\rm StRIP}^*(\gamma,p,\delta) = L_{\rm StRIP}(\mbox{ETF},\gamma,p,\delta) .
\end{equation}

The central conjecture of our work and main open question is whether ETFs are indeed the best possible frames
under random subset selection.

	

\section{Summary of open questions and conjectures}
\label{section-conclusion-open}
%
Finally, let us list some open questions 
and directions for future research.
%

\vspace{5mm}

\noindent
\underline{The ETF-MANOVA relation}
\begin{enumerate}
\item 
{\em Asymptotic ETF-MANOVA theorem:}
find a closed-form expression for the $r$th moment recursion formula (\ref{ETF-d-moment-non-centralized-formula}), in order to prove the conjectured identity (\ref{MarinaNew=Manova?}).

\item 
{\em Method of proof:}
can we find alternative proofs for the ETF-MANOVA relation via transform or free-probability methods? See
\cite{feier2012methods} and the result of \cite{raich2016eigenvalue}.

\item 
{\em Sub-frame selection:} 
compare the combinatoric ``$n$ choose $k$'' selection (assumed in the empirical experiments of Chapter~\ref{section-RMT-meets-frames}) 
with Bernoulli($p$) selection (assumed
in the asymptotic analysis of Chapter~\ref{section-ETF-MANOVA-moments}).
Is (and how much) the {\em non}-asymptotic behavior different?

\item
{\em Symmetries of ETF:} 
in the course of analyzing the $r$th-moment (\ref{moment_r}), we identified some invariant properties of small chains of products, which seem to hold for various ETF constructions, and may be useful to characterize the {\em non}-asymptotic behavior of the moments
\cite{MarinaThesis}.

\item
{\em Near-ETF:} how far can we relax the equi-angular and tightness conditions, without changing the asymptotic MANOVA behavior?

\item
{\em Non-asymptotic regime:} 
is the expected $r$th moment of a finite frame identical for all ETFs, 
or dependent on the specific ETF construction?
(Section~\ref{section-ETF-MANOVA-moments-NewNew}).

\item 
{\em Universality of convergence rate:} 
can we analyze the {\em non}-asymptotic behavior of ETF (and near-ETF) subsets? 
can we prove the universal convergence rate?
(Section~\ref{section-RMT-ETF-findings}) 
%

 \item
{\em Emergence of MANOVA ensemble: } 
what makes a typical ETF sub-frame 
similar to the ratio of i.i.d. matrices (\ref{MANOVA_ensemble}) in the MANOVA definition?

\item
{\em  Beyond the limiting spectrum:}
find other RMT-like properties of a typical ETF sub-frame; 
e.g., isotropic behavior of eigen-vectors;  
convergence of the extreme eigenvalues (see footnote~\ref{footnote-tracey-widom} in this chapter); 
diagonal of the Gram matrix, etc. 

\end{enumerate}


\vspace{5mm}

\noindent
\underline{General frame codes: asymptotic theory and bounds}

\begin{enumerate}
\setcounter{enumi}{9}

\item
{\em Low-pass frames:} 
do randomly-selected sub-frames of LPF have a limiting ESD?

\item
{\em Spherical codes:} find similarities and distinction between maximal distance spherical codes and ETFs.


\item
{\em Sub-frame rectangularity:} 
do the performance inequalities in Chapter~\ref{section-inequalities} continue to hold if we fix $p$ 
and let both $\beta$ and $\gamma$ vary accordingly?

\item
{\em RMT-like behavior of frames:}
are there other families of deterministic frames (for example, non-ETF harmonic frames) that exhibit a (possibly different) limiting ESD for the subframe eigenvalues? 
If so, what is the set of attainable distributions for fixed $\gamma$ and $\beta$? 
Is this set convex (closed under mixtures of frame families)?

\item
{\em Erasure Welch bound:}
how does the EWB (\ref{moment d bound}) connect to Welch's generalized bound for higher-order moments of the cross-correlations \cite{welch1974lower}; can we derive one from the other?
 
\item
{\em ETF superiority:} find a sharper definition (beyond variance, kurtosis and functional (``$\specstat$'') measures).

\item
{\em Analog versus digital:} are analog codes inherently inferior to digital (Shannon) codes? 

\end{enumerate}


\vspace{5mm}

\noindent
\underline{Applications for analog coding}

\begin{enumerate}
\setcounter{enumi}{16}

\item
{\em Coded distributed computing (CDC):}
modify the noise amplification performance measure (\ref{AHMR}) to the case where one needs to recover only the input sum (or one projection or subspace), rather than full inversion
\cite{RoyeeYosibashPaper}.

\item
{\em  Source with erasures:}
can we obtain the full MMSE gain, and eliminate the ``$+1$'' penalty term in (\ref{Ranalog1}) for a general $\beta$? 

\item
{\em Binary analog codes:}
can we approximate the superior ETF performance using {\em binary}
frame codes (similar to \cite{babadi2011spectral})?    

\end{enumerate}


It is our hope that these questions, and others implicitly raised in this survey, will stimulate research in this fascinating area at the intersection of information theory, random matrices, geometry and combinatorics. 

\begin{acknowledgements}
Matan Gavish was partially supported by the Israel Science Foundation, grant 1523/16. 

\noindent
Dustin G. Mixon was partially supported by AFOSR FA9550-18-1-0107 and NSF DMS 1829955. 

\noindent
Ram Zamir was partially supported by the Israel Science Foundation,
grants {676/15} and {2623/20}.

We thank Ofer Zeitouni for suggesting the moment approach to prove the
ETF-MANOVA theorem.
\end{acknowledgements}


\appendix
%
\chapter{Entropy-coded dithered quantization}
\label{section-ECDQ}
To asses the rate-distortion performance (\ref{Ranalog1}) of the analog coding scheme,
we shall adopt the additive-noise ``test channel'' of entropy-coded (subtractive) dithered quantization (ECDQ) 
\cite{zamir1992universal}.
In this model, 
$Q^{({\rm dither})}(\tilde{\bx}) = \tilde{\bx} + \bZ$,
where $\bZ$ is an independent uniform or Gaussian noise,
and the quantizer entropy is given by the mutual information 
$H(Q^{({\rm dither})}(\tilde{\bX})) = I(\tilde{\bX}; \tilde{\bX} + \bZ)$.\footnote{To be more precise, 
the quantizer operation is given by
$Q(\tilde{\bx} + {\rm dither}) - {\rm dither}$, where ${\rm dither}$ is uniform over the fundamental quantizer cell. For good high-dimensional lattice quantizers, this uniform dither distribution tends to be white Gaussian.  
See \cite{zamir2014lattice}.
}
The equivalent error in the important samples,  
$\bE_s = \hat{\bX}_s - \bX_s$ 
is thus, by (\ref{xtilde}), given by 
$\bE_s = F_S^\dagger \cdot (\tilde{\bX} + \bZ) - \bX_s = F_S^\dagger \cdot \bZ$,
implying that the mean-squared distortion per important sample (\ref{distortion}) is
\begin{equation}
    \label{appendix-Danalog}
    D = \frac{1}{k} \mathbb{E}\left\{ \|\bE_s\|^2 \right\}
    = \frac{\sigma_z^2}{k} \cdot \mbox{trace} \{ F_S^\dagger F_S \} 
    = \sigma_z^2 ,
\end{equation}
where $\sigma_z^2$ is the quantizer mean-squared error, 
and the last equation follows since $F_S$ has $k$ unit-norm columns.\footnote{Further improvement can be obtained using minimum mean-squared error ``Wiener'' estimation, \cite{haikin2016analog},
but this is negligible when $\sigma_x^2 \gg D$.}
%
Since the decoder is blind to the location of important samples (that affect the correlation
of the vector $\tilde{\bX}$),
the coding rate is equal to that of a {\em white} Gaussian vector $\tilde{\bW}$ with the same power 
as $\tilde{\bX}$, \cite{lapidoth1997role}.
Assuming the LS estimator (\ref{LS}), we have
$E \|\tilde{\bX} \|^2 = \sigma_x^2 \cdot \mbox{trace} \{ (F_S^\dagger F_S)^{-1} \}$,
so the ECDQ mutual information formula (with a Gaussian dither) becomes:
%
\begin{align}
    \label{appendix-Ranalog}
R_{\rm analog} 
& = \frac{1}{n} I(\tilde{\bW}; \tilde{\bW} + \bZ)   \\
& = \frac{m}{2n} \cdot \log 
\left( 1 + \frac{\frac{1}{m} \mathbb{E} \|\tilde{\bX} \|^2}{\sigma_z^2} \right)
\\
& = \beta \cdot \frac{p}{2} \cdot \log 
\left( 1 + \frac{\sigma_x^2}{D} \cdot \frac{\mbox{trace} \{ (F_S^\dagger F_S)^{-1} \}}{m} \right)
\end{align}
where $\beta \Ddef k/m$, and we used (\ref{distortion}) with $k = pn$. 



\chapter{Information theoretic proofs for sub-frame inequalities}
\label{section-inequalities-II}
%
%
In proving Theorems~\ref{Thm-frame-shannon-monotonicity} and~\ref{Thm-frame-MSE-monotonicity} we use the following two lemmas.
\begin{lemma}
\label{Inequalities-lemma1}
For any $m \times n$ frame $F$, and $k < n$,
\begin{equation}
\label{Inequalities-lemma1.1}
\frac{1}{{n \choose k}} \sum_{S: |S|=k}  F_S \cdot F_S^\dagger 
= \frac{k}{n} \cdot F \cdot F^\dagger  ,
\end{equation}
where the average in the left hand side 
is over all $k$-subsets of $\{1,\ldots,n\}$
as in (\ref{GAMR})-(\ref{AHMR}). 
\end{lemma}

{\em Remark:}
For Bernoulli($p$) selection,
i.e., when each index in $\{1,\ldots,n\}$ belongs to $S$ with probability $p$ independently of the other indices, 
the lemma becomes
$\mathbb{E}_S \{ F_S \cdot F_S^\dagger \} = p \cdot F \cdot F^\dagger$,
where $\mathbb{E}_S \{ \cdot \}$ denotes expectation with respect to the selection of $S$.


{\em Proof:} 
The case $k=1$ of (\ref{Inequalities-lemma1}) is the standard expansion 
$\sum_{i=1}^n \bff_i \cdot \bff_i^\dagger = F \cdot F^\dagger$ (multiplied by $1/n$), 
where the $\bff_i$'s are the columns of $F$. 
For a general $k$, let
$P$ denote a random diagonal matrix, whose $\{0,1\}$ diagonal elements 
correspond to the selection of the subset $S$. 
That is, $F_S \cdot F_S^\dagger = F \cdot P \cdot F^\dagger$,
where the diagonal of $P$ is uniform over all $n$ choose $k$ binary
vectors with $k$ ones and $n-k$ zeroes. 
Thus
$\mathbb{E}_S \{ F_S \cdot F_S^\dagger \} = \mathbb{E}_S\{ F \cdot P \cdot F^\dagger \} 
= F \cdot \mathbb{E}_S\{P\} \cdot F^\dagger = F \cdot (k/n \cdot I_n) \cdot F^\dagger$,
which is the right-hand side of (\ref{Inequalities-lemma1.1}).
$\Box$


\begin{lemma}
\label{Inequalities-lemma2}
For $k_1 < k_2 < n$, 
if $S_2$ is uniformly drawn from all $k_2$ subsets of $\{1,\ldots,n\}$, 
and $S_1$ is uniformly drawn from all $k_1$ subsets of $S_2$,
then $S_1$ is uniform on all $k_1$ subsets of $[n]$.
\end{lemma}

{\em Proof:}
Since the distribution of $S_1$ is invariant under the action of the symmetric group of $[n]$, it is necessarily uniform.
$\Box$


\vspace{5mm}

\underline{\em Proof of Theorem~\ref{Thm-frame-shannon-monotonicity}:}
We first prove the inequality with respect to the ``edge'':
$L(F,k) \leq L(F,n)$, and then extend to any $k_1 < k_2$.
%
%
The Ky Fan inequality 
\cite{CoverBook},
says that if $K_1$ and $K_2$ are $m \times m$ PSD matrices, then
\begin{equation}
    \label{KyFan}
\det[a \cdot K_1 + (1-a) \cdot K_2]  \geq  \det[K_1]^a \cdot \det[K_2]^{1-a}
\end{equation}
for any $0<a<1$, 
which by taking logarithm implies that $\log \det[K]$ is concave \cite{CoverBook}.
Thus, starting from (\ref{GAMR}),
\begin{align}
\label{KyFan2}
 L_{\rm Shannon}(F,k) & = \mathbb{E}_S\{ \log[ \sqrt[m]{ \det(F_S \cdot F_S^\dagger) } / (k/m) ]\}  \\
               & \leq \log[ \sqrt[m]{ \det(\mathbb{E}_S\{ F_S \cdot F_S^\dagger\}) } / (k/m) ]  \\
               & =  \log[ \sqrt[m]{ \det(k/n \cdot F \cdot F^\dagger) } / (k/m) ]   \\
               & =  \log[ \sqrt[m]{ \det(F \cdot F^\dagger) } / (n/m) ]   \\
               \label{KyFan3}
               & =  L_{\rm Shannon}(F,n)    
\end{align}
%
where the second line is by Jensen's inequality and the log concavity of the determinant (\ref{KyFan}), 
and the third line is by the identity in Lemma~\ref{Inequalities-lemma1}.
Turning to the general case, 
%
for $k_1 < k_2 < n$, let $S_1 \subset S_2$ denote a $k_1$-subset of a $k_2$-subset $S_2$ of $\{1,\ldots,n\}$.
By the definition of the average subset Shannon transform of a frame (\ref{GAMR}), 
we have 
\begin{align}
 L_{\rm Shannon}(F,k_1) 
 & = \mathbb{E}_{S_1} \{ \specstat_{\rm Shannon}(F_{S_1}) \} \\ 
 & = \mathbb{E}_{S_2}\{ E_{S_1|S_2}\{ \specstat_{\rm Shannon}(F_{S_1}) \} \\
 & = \mathbb{E}_{S_2}\{ L_{\rm Shannon}(F_{S_2},k_1) \} \\
 & \leq \mathbb{E}_{S_2}\{ L_{\rm Shannon}(F_{S_2},k_2) \} \\
 & = \mathbb{E}_{S_2}\{ \specstat_{\rm Shannon}(F_{S_2}) \} \\
 & = L_{\rm Shannon}(F,k_2)      
\end{align}
%
where $\mathbb{E}_{S_1|S_2}\{\cdot\}$
denotes expectation over a uniform distribution on all $k_1$-subsets of a given $S_2$,
and $\mathbb{E}_{S_2}\{\cdot\}$ denotes expectation over a uniform distribution on all $k_2$-subsets of $\{1,\ldots,n\}$.
The second line follows from Lemma~\ref{Inequalities-lemma2} by iterated expectation;
the third line follows by viewing $F_{S_2}$ as the full frame in the inner expectation;
the inequality follows from the first part of the proof (\ref{KyFan2})-(\ref{KyFan3})
setting $k = k_1$ and $n = k_2$;
and the last two lines are again by the definition (\ref{GAMR}).
$\Box$


\vspace{5mm}

\underline{\em Proof of Theorem~\ref{Thm-frame-MSE-monotonicity}:}
For an $n \times n$ covariance matrix $K$ (i.e., $K$ is a non-negative matrix), 
and a $k$-subset $S \subset \{1,\ldots,n\}$, 
let $K_S$ denote the corresponding $k \times k$ sub-matrix of $K$.
Define the {\em subset trace-inverse} of the covariance $K$ as
the average of $\tr{K_S^{-1}} / k$ over all $k$-subsets:
\begin{equation}
    \label{trace-inverse}
    M_k^{(n)} =  
    \frac{1}{{n \choose k}} \sum_{S: |S|=k} \frac{1}{k} \tr{K_S^{-1}} ,
\end{equation}
for $k = 1, \ldots, n$.
For example, the two extremes are 
$M_1^{(n)} = 1/n \sum_{i=1}^n (K_{ii})^{-1}$,
and
$M_n^{(n)} = 1/n \sum_{i=1}^n (K^{-1})_{ii}$.
It is shown in \cite{khina2020monotonicity} 
that the sequence of subset trace inverses 
is monotonically non decreasing, i.e., 
\begin{equation}
    \label{Khina-Yeredor-Zamir}
    M_1^{(n)} \leq M_2^{(n)} \leq \ldots \leq M_n^{(n)}  
\end{equation}
with equality if and only if the covariance matrix $K$ is proportional to identity.\footnote{
If the matrix $K$ is Toeplitz (corresponding to stationary vector), 
then the inequality between the two edges, $M_1^{(n)} \leq M_n^{(n)}$, follows from
the arithmetic-to-harmonic means inequality applied to the eigenvalues  
$\lambda_1,\ldots,\lambda_n$ of $K$,
and equality holds if and only if the eigenvalues are constant, i.e., $K= \lambda I$.
(This is because by the Toeplitz property 
$K_{ii} = \sigma^2$ for all $i$,
and therefore $M_1^{(n)} = 1/\sigma^2$;
furthermore,  
$\sigma^2 = \tr{K} /n = (\lambda_1 +\ldots + \lambda_n)/n$,
while the eigenvalues of $K^{-1}$ are the reciprocals of the eigenvalues of $K$, 
so $M_n^{(n)} = (1/\lambda_1 +\ldots + 1/\lambda_n)/n$.)
}
This inequality 
can be thought of as the ``MSE counterpart'' 
of the subset determinant inequality in \cite{CoverBook} 
and \cite{dembo1991information}.
Using similar tools as in \cite{CoverBook}, 
we can show the same inequality also for {\em subset \underline{log} trace-inverse}. 
Namely, letting
\begin{equation}
    \label{log trace-inverse}
    L_k^{(n)} =  
    \frac{1}{{n \choose k}} \sum_{S: |S|=k} \log \left( \frac{1}{k} \tr{K_S^{-1}} \right),
\end{equation}
we have 
\begin{equation}
    \label{log Khina-Yeredor-Zamir}
    L_1^{(n)} \leq L_2^{(n)} \leq \ldots \leq L_n^{(n)} . 
\end{equation}
%


{\em Proof:}
Starting with the right edge, 
noting that ${n \choose n} = 1$ and ${n \choose n-1} = n$, 
we have 
\begin{align}
\label{Inequalities-proof-of-thm2}
    L_{n}^{(n)} & = \log \left( \frac{1}{n} \tr{K^{-1}} \right) \nonumber \\
    & = \log \left( M_{n}^{(n)} \right)  \nonumber \\
    & \geq \log \left( M_{n-1}^{(n)} \right)  \nonumber \\
    & = \log \left( \frac{1}{n} \sum_{S: |S|=n-1} \frac{1}{n-1} \tr{K_S^{-1}} \right) \nonumber  \\
    & \geq \frac{1}{n} \sum_{S: |S|=n-1} \log \left( \frac{1}{n-1} \tr{K_S^{-1}} \right)  \nonumber \\
    & = L_{n-1}^{(n)} , 
\end{align}
where the equality lines are by definition;
the first inequality follows from 
(\ref{Khina-Yeredor-Zamir});
and the second inequality follows by Jensen.
We then continue similarly to the second part of the proof of Theorem~\ref{Thm-frame-shannon-monotonicity}
to prove that $L_k^{(n)} \leq L_{k+1}^{(n)}$ for any $k$.
Specifically, by Lemma~\ref{Inequalities-lemma2}, 
when computing $L_{k}^{(n)}$ we first condition on a specific $(k+1)$-subset $S_2$ and average over all its
$k$-subsets, and then average over all $(k+1)$-subsets $S_2$ of $\{1,\ldots,n\}$.
By (\ref{Inequalities-proof-of-thm2}), the inner average is upper bounded by
the log trace-inverse of $K_{S_2}$, which  becomes $L_{k+1}^{(n)}$ after taking the outer average.
$\Box$



Theorem~\ref{Thm-frame-MSE-monotonicity} now follows from (\ref{log Khina-Yeredor-Zamir})
by viewing the $F^\dagger \cdot F$ as the covariance matrix $K$,
so
$L_{\rm MSE}(F,k) = L_k^{(n)}$.
(Note that for $k>m$ the matrix $F_S^\dagger F_S$ is singular, so $L_k^{(n)}$ is infinite.)



\chapter{Variance of $1st$ and $2nd$ moments (proof of Theorem~\ref{th_var_1_2})}
\label{section-variance-proof}
\vspace{10mm}
Using the moment-variance formula (\ref{moment_variance}),
\begin{align*}
V_r 
& \triangleq 
{\rm Var}\left[\frac{1}{n}\mbox{trace} \left((FPF^\dagger)^r\right)\right]     
\\
& =
\mathbb{E} \left[ \left( \frac{1}{n}\mbox{trace} \left((FPF^\dagger)^r\right) \right)^2 \right]
    - m_r^2
\end{align*}
we obtain for $r=1$,
\begin{align} 
\nonumber
V_1^{\rm ETF} 
&= 
\mathbb{E}\left[\left(\frac{1}{n}\mbox{trace} \left(FPF^\dagger\right)\right)^2\right]-\mathbb{E}\left[\frac{1}{n}\mbox{trace} \left(FPF^\dagger\right)\right]^2 \\
\nonumber
&= \mathbb{E}\left[\left(\frac{1}{n}\mbox{trace} \left(FPF^\dagger\right)\right)^2\right] - m_1^2 \\
\nonumber
&= \frac{1}{n^2}\mathbb{E}\left[ \left(\sum_{i = 1}^{n}\langle \bff_i, \bff_i \rangle P_i \right)\left(\sum_{j = 1}^{n}\langle \bff_j, \bff_j \rangle P_j \right)\right] - p^2\\
\nonumber
&= \frac{1}{n^2}\mathbb{E}\left[ \left(\sum_{i = 1}^{n} P_i \right)\left(\sum_{j = 1}^{n} P_j \right)\right] - p^2\\
\nonumber
&= \frac{1}{n^2}\left[ np + n(n-1)p^2\right] - p^2\\
\label{var_moment_1}
&= \frac{1}{n}\left[ p - p^2\right]. 
\end{align}
For $r=2$, we obtain
\begin{align} 
\nonumber
V_2^{\rm ETF} 
&= 
\mathbb{E}\left[\left(\frac{1}{n}\mbox{trace} \left((FPF^\dagger)^2\right)\right)^2\right]-\mathbb{E}\left[\frac{1}{n}\mbox{trace} \left((FPF^\dagger)^2\right)\right]^2 \\
\nonumber
&= \mathbb{E}\left[\left(\frac{1}{n}\mbox{trace} \left((FPF^\dagger)^2\right)\right)^2\right] - m_2^2 \\
\label{var_moment_2}
&= \frac{1}{n^2}\mathbb{E}\left[ \left(\sum_{i,j = 1}^{n}|\langle \bff_i, \bff_j \rangle|^2 P_i P_j\right)\left(\sum_{k,m = 1}^{n}|\langle \bff_k, \bff_m \rangle|^2 P_k P_m\right)\right] - (p+p^2x)^2.
\end{align}
If $F$ is an ETF, then according to the Welch bound $|\langle \bff_i, \bff_j \rangle|^2 = \frac{x}{n-1}$ for $i\neq j$. Similarly to the computation of the moments, we split to cases by number of distinct values of $\{i,j,k,m\}$.

\begin{align} 
\nonumber
&\mathbb{E}\left[ \left(\sum_{i,j = 1}^{n}|\langle \bff_i, \bff_j \rangle|^2 P_i P_j\right)\left(\sum_{k,m = 1}^{n}|\langle \bff_k, \bff_m \rangle|^2 P_k P_m\right)\right]  = np \\
\nonumber
 & + \left[ n(n-1)+4n(n-1)\frac{x}{n-1}+2n(n-1)\frac{x^2}{(n-1)^2}\right]p^2\\
\nonumber
& + \left[ 2n(n-1)(n-2)\frac{x}{n-1}+4n(n-1)(n-2)\frac{x^2}{(n-1)^2}\right]p^3\\
\label{var_moment_2_proof}
& + n(n-1)(n-2)(n-3)\frac{x^2}{(n-1)^2}p^4.
\end{align}
It follows that
\begin{align}
\nonumber
V_2 = \frac{1}{n}\Bigl[ p &+ \left( -1 + 4x + \frac{2x^2}{n-1} \right)p^2 +\left( -4x + \frac{4(n-2)x^2}{n-1} \right)p^3\\ 
\label{var_moment_2_proof2}
&+\left( \frac{(6-4n)x^2}{n-1} \right)p^4\Bigr]. 
\end{align}


\backmatter  

\printbibliography

\end{document}